\documentclass[onecolumn]{aa}
\usepackage{graphicx}
\usepackage{longtable}
\begin{document}
   \title{The H$\alpha$ Galaxy Survey \thanks{
Based on observations made with the Jacobus Kapteyn Telescope operated 
on the island of La Palma by the Isaac Newton Group in the Spanish 
Observatorio del Roque de los Muchachos of the Instituto de Astrof\'\i sica 
de Canarias
   }}
   \subtitle{I. The galaxy sample, H$\alpha$
   narrow-band observations\\ 
   and star formation 
   parameters for 334 galaxies}
   \author{P.A. James
          \inst{1},
          N.S. Shane
          \inst{1},
          J.E. Beckman
          \inst{2},
          A. Cardwell
          \inst{2},
          C.A. Collins
          \inst{1},
          J. Etherton
          \inst{1},
          R.S. de Jong
          \inst{3},\\
          K. Fathi
          \inst{4},
          J.H. Knapen
          \inst{5,6}
          R.F. Peletier
          \inst{4},
          S.M. Percival
          \inst{1},
          D.L. Pollacco
          \inst{7},
          M.S. Seigar
          \inst{8},
          S. Stedman
          \inst{5},
          I.A. Steele
          \inst{1}
          } \offprints{P.A. James} 
          \institute{Astrophysics Research
	  Institute, Liverpool John Moores University, Twelve Quays
	  House, Egerton Wharf, Birkenhead CH41 1LD, UK \\
	  \email{nss,paj,cac,je,smp,ias@astro.livjm.ac.uk} 
          \and Instituto de Astrof\'\i sica de Canarias,
          C. V\'\i a L\'actea s/n 38200-La Laguna,
          Tenerife, Spain\\
          \email{jeb,cardwell@ll.iac.es}
          \and Space Telescope Science Institute,
          3700 San Martin Drive, Baltimore, MD 21218, USA\\
          \email{dejong@stsci.edu}
          \and School of Physics and Astronomy,
          University of Nottingham, 
          Nottingham NG7 2RD, UK\\
          \email{ppxkf,Reynier.Peletier@nottingham.ac.uk}
          \and Department of Physical Sciences,
          University of Hertfordshire, Hatfield, 
          Hertfordshire, AL10 9AB, UK\\
          \email{knapen, stedman@star.herts.ac.uk}
          \and Isaac Newton Group of Telescopes,
          Apartado 321, E-38700 Santa Cruz de La Palma, Spain
          \and Department of Pure and Applied Physics,
          Queen's University Belfast, University Road,
          Belfast BT7 1NN, UK\\
          \email{D.Pollacco@qub.ac.uk}
          \and UKIRT, Joint Astronomy Centre,
          660 North A'ohoku Place, University Park,
          Hilo, HI96720, USA\\ 
          \email{m.seigar@jach.hawaii.edu} } 
          \date{Received ; accepted }
   \abstract{ We discuss the selection and observations of a large
sample of nearby galaxies, which we are using to quantify the star
formation activity in the local Universe.  The sample consists of 334
galaxies across all Hubble types from S0/a to Im and with recession
velocities of between 0 and 3000 km~s$^{-1}$.  The basic data for each
galaxy are narrow band H$\alpha +$[N{\sc ii}] and $R$-band imaging, from which we
derive star formation rates, H$\alpha +$[N{\sc ii}] equivalent widths and surface
brightnesses, and $R$-band total magnitudes.  A strong correlation is found 
between total star formation rate and Hubble type, with the strongest star 
formation in isolated galaxies occurring in Sc and Sbc types.  More 
surprisingly, no significant trend is found between H$\alpha +$[N{\sc ii}] equivalent 
width and galaxy $R$-band luminosity.  More detailed analyses of the data 
set presented here will be described in subsequent papers.
   }
\authorrunning{James et al.}
\titlerunning{H$\alpha$ Galaxy Survey}
\maketitle
%
\section{Introduction}
Knowledge of the star formation histories of both the Universe and of
individual galaxies provides the foundation for our understanding of
the evolution of the Universe we see today.  Substantial advances have
been made in our understanding of high-redshift star formation, (e.g.
Madau et al.
\cite{madau96}; Madau et al. \cite{madau98}; Steidel et al.
\cite{stei}; Hopkins et al. \cite{hopk}; Lanzetta et al. \cite{lanz}).
This has resulted in an anomalous situation in which the star
formation history of the Universe appears to be more fully quantified
at high redshift than it is locally.  This results at least partly
from the relative ease with which a representative volume of the
Universe can be observed at high redshift, since only a small area of
the sky need be observed. Star formation rates have been determined
for many local galaxies using a range of different techniques, e.g.
emission-line fluxes (e.g., Kennicutt \& Kent \cite{ken83}; Young et al.
\cite{young}; Gallego et al. \cite{gallego95}; Gallego et al.
\cite{gallego96}), far-infrared luminosities (e.g., Kennicutt et al.
\cite{ken87}), radio luminosities (e.g., Condon \cite{condon}, Cram et
al. \cite{cram}) or direct ultraviolet emission from hot stars (e.g.
Bell \& Kennicutt
\cite{bell}). However, in the main these studies have looked at the
brightest and most rapidly star-forming galaxies, and {\bf most}
spectroscopic studies are biased towards galaxies with large
equivalent width (EW) in the line concerned.  In this study, one
focus will be on the star formation properties across the full range
of the numerically dominant dwarf galaxies, and the global star
formation rate from this population. A further consequence of the
observational limitations on areal coverage is that galaxy clusters
are easier to study than the field environment, and hence local field
galaxy star formation rates are relatively poorly constrained compared
with their cluster counterparts.

In the present study, we attempt to quantify star formation activity
across the full range of star-forming galaxies, down to the faintest
dwarf irregular types, by using narrow-band imaging through filters
centred on the redshifted Balmer H$\alpha$ line.  The advantages of
this technique are that it is sensitive to low levels of star
formation even in faint, low surface brightness galaxies, and that it
traces high mass stars, and hence recent star formation.  Given
suitable assumptions, principally about extinction and the stellar
initial mass function (IMF), it yields quantitative measurements of
the star formation rate.  The method can be applied to a sensitive level
using relatively modest integration times on small telescopes, which
is an important consideration given that this project must be done one
object at a time, to ensure that each target galaxy is observed with
the correct filter.  The resulting data not only give estimates of the
total star formation rate for each galaxy, but also detailed
information on the star formation distribution, enabling, for example,
the separation of nuclear and disk activity in spiral galaxies.
Finally, the large format of current CCDs gives a reasonable
probability of detecting nearby star-forming companion galaxies which
lie in the same field as target galaxies, and which have similar
recession velocities.

The principal drawbacks of narrow-band H$\alpha$ imaging are the need for
large and uncertain extinction corrections; the need to assume an IMF to
extrapolate from the quantity of high mass stars, responsible for the
ionising flux, to the total mass of the young stellar population;
contamination in most of the filters used by the [N{\sc ii}] 6548 and
6584~\AA\ lines, necessitating a further correction to the derived star
formation rates; and possible contributions to the line emission by
central active galactic nuclei (AGN). All of these assumptions and
corrections have been investigated in detail, principally by Kennicutt
and his collaborators (Kennicutt \& Kent
\cite{ken83}; Kennicutt \cite{ken98}); we will re-examine some of the
corrections they derived in Paper II of this series (James et al.
\cite{jame}).  The final disadvantage of using narrow-band filters is
that a comprehensive survey of a contiguous area (c.f. Sloan, IRAS)
cannot be completed in a reasonable amount of time, due to the small
recession velocity coverage of each of the narrow-band filters, and
hence a pre-existing galaxy catalogue must be used to provide a target
list for specific pointed observations with the appropriate filters.

Any discussion of previous work in this area must first acknowledge
the extensive studies undertaken by Kennicutt and collaborators in
defining the techniques for H$\alpha$ measurement and deriving star
formation rates from such measurements (Kennicutt \& Kent
\cite{ken83}), and in applying these measurements to studies of bright
spiral and irregular galaxies (e.g., Kennicutt et al. \cite{ken94}) and
interacting galaxies (Kennicutt et al. \cite{ken87}). All of this work
was comprehensively reviewed by Kennicutt (\cite{ken98}). Ryder \&
Dopita made a detailed study of 34 nearby southern spirals in
H$\alpha$, $V$ and $I$ band emission, demonstrating that star formation
can be much more asymmetric than the underlying stellar distribution
(Ryder \& Dopita \cite{ryde93}) and that the H$\alpha$ scale length
tends to be longer than that of the stellar light distribution (Ryder
\& Dopita \cite{ryde94}).  These observations were used to derive a
form for the star formation law in spiral galaxies (Dopita \& Ryder
\cite{dopi94}). Young et al. (\cite{young}) undertook a major study of
120 spiral galaxies using similar techniques to those used in the
present work, to look at trends in H$\alpha$ surface brightness with
Hubble type and at the dependence of star formation rate on gas mass and
interactions. Koopman et al. (\cite{koopman}) looked at the H$\alpha$
total emission and light profiles for 63 bright spiral galaxies in the
Virgo cluster, again using narrow-band imaging techniques.

The Universidad Complutense de Madrid (UCM) emission-line survey (Gallego
et al. \cite{gallego95}, Gallego et al. \cite{gallego96}) adopted a
different and in many ways complementary strategy of searching areas of
sky for all emission-line galaxies over a wide range of recession
velocities, using the objective prism technique to detect candidates, and
follow-up spectroscopy for confirmation and derivation of detailed
spectroscopic parameters.  Gallego et al. (\cite {gallego96}) present
H$\alpha$ and H$\beta$ fluxes and EWs for over 200 galaxies detected in
this way, and Gallego et al.  (\cite{gallego95}) use this dataset to
determine the total star formation rate in the Local Universe.  One of
the aims of the present study is to rederive this parameter using a
sample with very different selection criteria which are not dependent on
H$\alpha$ line strengths.  This will be presented in a later paper in
this series.  Finally, an important recent study with which the current
work should be compared is that of Charlot and collaborators (e.g.
Charlot et al. \cite{char02}), who used spectroscopic data from the
Stromlo-APM survey to study a representative sample of star-forming
galaxies with H$\alpha$ EW down to 0.2~nm, compared with a limit of
1.0~nm for Gallego et al. (\cite{gallego95}). The corresponding limit for
the current survey is about 0.4~nm EW, although a few detections of line
emission below this level are made.

The sample selection for the H$\alpha$ Galaxy Survey is described in
Sect. 2 of this paper.  The observational strategy and the data
reduction process are detailed in Sections 3 and 4 respectively.
Sect. 5 contains the data for the 334 galaxies, and Sect. 6 a
discussion of some of the first results of this study.
Sect. 7 contains our conclusions and a summary of planned future
publications.


\section{Sample selection}

The sample was selected using the Uppsala Galaxy Catalogue (Nilson
\cite{nilson}, henceforth UGC) as the parent catalogue.  The UGC was
chosen because of the uniform selection criteria used for its
compilation (all galaxies to a limiting diameter of 1\farcm0
and/or to a limiting apparent magnitude of 14.5 on the blue prints of
the Palomar Observatory Sky Survey), its uniform coverage of the
entire northern sky, its inclusion of galaxies of all Hubble types
(including faint and low-surface-brightness dwarfs), and because it
provides consistent diameters and classifications.  The principal
drawback of the UGC is that it does not contain recession velocities
for many galaxies.  This deficiency has largely been filled by studies
since the publication of the UGC, and a search using the NASA
Extragalactic Database (NED) shows that at least 85\% of all UGC
galaxies now have measured recession velocities.  However, the
remaining 15\% cannot be selected for observation in our study, and
this represents a possible source of bias which should be kept in mind
when interpreting the results.

UGC galaxies were selected, using the NED `Advanced All-Sky Search For
Objects By Parameters' facility, within 5 recession velocity shells:
0--1000~km~s$^{-1}$, 1000--1500~km~s$^{-1}$, 1500--2000~km~s$^{-1}$,
2000--2500~km~s$^{-1}$, and 2500--3000~km~s$^{-1}$. The galaxies were
required to be spiral or irregular galaxies, with Hubble types from S0/a to
Im inclusive, and to have $D_{25}$ diameters of 1\farcm7--6\farcm0.  This
last criterion ensures that all galaxies will fit on the field of the
CCD camera used, and the different shells effectively sample different
parts of the galaxy diameter function.  Thus the central shell is
dominated by dwarf Im and Sm galaxies, whilst the outer shells sample
the rarer S0/a-Sc galaxies.  The well-defined selection criteria, and
the large total number of galaxies observed (334), mean that it is
possible to combine the data for all shells such that galaxies with a
wide range of luminosities, diameters and surface brightnesses are
represented.


\section{Observations}

The primary data for this study are narrow-band H$\alpha +$[N{\sc ii}] and Johnson
$R$ band imaging.  These observations were made using the 1.0~metre
Jacobus Kapteyn Telescope (JKT), operated by the Isaac Newton Group of
Telescopes (ING) situated on La Palma in the Canary Islands.  This
project was allocated 100 nights of observing time on the JKT, between
February 2000 and January 2002.  Of these nights, 78 produced usable
data for the project and 52 were photometric. The instrument used was
the facility 2048$\times$2048 pixel SITe CCD camera, with
0\farcs33 pixels, giving a total field of view of over
11$^{\prime}$$\times$11$^{\prime}$, of which the central area of
10$^{\prime}$$\times$10$^{\prime}$ is unvignetted.  The CCD has a good quantum
efficiency ($>$60\%) in the $R$ band, and a read noise of about 7
electrons.

The filters used for this project are listed in Table~\ref{filters}.
The relative throughputs are given, normalised to the $R$-band filter.
The redshifted H$\alpha$ filters and the Harris $R$ filter are from
the standard ING filter set, whereas the H$\alpha$Cont filter is an
off-the-shelf item purchased for this project. The benefit of using
this filter is that it accurately samples the galaxy continuum flux
close to the H$\alpha$ line, but has a broader bandpass than the
narrow H$\alpha$ filters, thus reducing the time overhead in taking
continuum observations compared with using `off-line' narrow H$\alpha$
filters.  The H$\alpha$Cont filter proved useful in bright sky
conditions (moonlight or dark twilight), but for fully dark skies it
was found that scaled $R$-band exposures gave excellent continuum
subtraction.  The much greater speed of using the broad $R$ filter
more than offsets the small loss in effective throughput to H$\alpha$
light that results from having the H$\alpha$ line within the passband
of the $R$ filter.  Standard exposure times used were 3$\times$1200
seconds in the appropriate narrow band filter, chosen to maximise
throughput at the wavelength of the redshifted H$\alpha$ line; 300
seconds at $R$; and 3$\times$600 seconds in the H$\alpha$Cont filter
if the $R$ image was not obtained in fully dark sky conditions.
Additional calibrating $R$-band exposures were taken during a
photometric night if there were any doubts about the sky conditions
for the original observations.  The very narrow [N{\sc ii}] filter
centred on 6584~\AA\ was not used for general survey observations, but
was useful for isolating just the H$\alpha$ line, and {\em excluding}
[N{\sc ii}] emission, for galaxies with recession velocities of
approximately 1000~km~s$^{-1}$.  Such images will be used in a later
paper for an examination of the effects of [N{\sc ii}] contamination
in our H$\alpha$ imaging, but are not used in the present paper.  Hence
all line fluxes presented here are for H$\alpha +$[N{\sc ii}].

   \begin{table}
      \caption[]{Filters used. Filter throughput curves can be found on the ING webpages.}
         \label{filters}
      
          $$
          \begin{array}{lccc}
            \hline
            \noalign{\smallskip}
            $Filter name$ & $Central wavelength (\AA)$ & $Passband
            width (\AA)$ & $Normalised throughput$\\
            \noalign{\smallskip}
            \hline
            \noalign{\smallskip}
           $Harris R$ & 6373 & 1491 & 1    \\
           $H$\alpha$6570$   & 6570 &   55    & 0.019         \\
           $H$\alpha$6594$   & 6594 &   44    & 0.018         \\
           $H$\alpha$6607$   & 6607 &   50    & 0.021         \\
           $H$\alpha$6626$   & 6626 &   44    & 0.020         \\
           $N{\sc ii}6584$    & 6584 &   21       & 0.0084        \\
           $H$\alpha$Cont$   & 6471 &  115    & 0.050         \\
            \noalign{\smallskip}
            \hline
         \end{array}
         $$
   \end{table}

The photometric calibration strategy was to observe at least one
spectrophotometric standard from the ING standards list in all filters
at the start and end of the night to check for any changes in filter
transmission. Changes in sky transparency through the night were
monitored by regular observations of standard stars selected from the
lists of Landolt (\cite{landolt}) in the $R$ filter only.

The other calibration observations were the usual bias frames, taken
at the beginning and end of each night, and twilight sky flat fields
in all filters used, again taken at the beginning and end of each
night whenever possible.  Occasionally, weather conditions precluded
twilight observations, and flats from other nights were found to be
satisfactory, and in every case gave better flat fielding than dome
flats, so the latter were never used.  All galaxy observations were
autoguided.

\begin{figure} 
   \centering
   \rotatebox{-90}{
      \includegraphics[height=14cm,width=15cm]{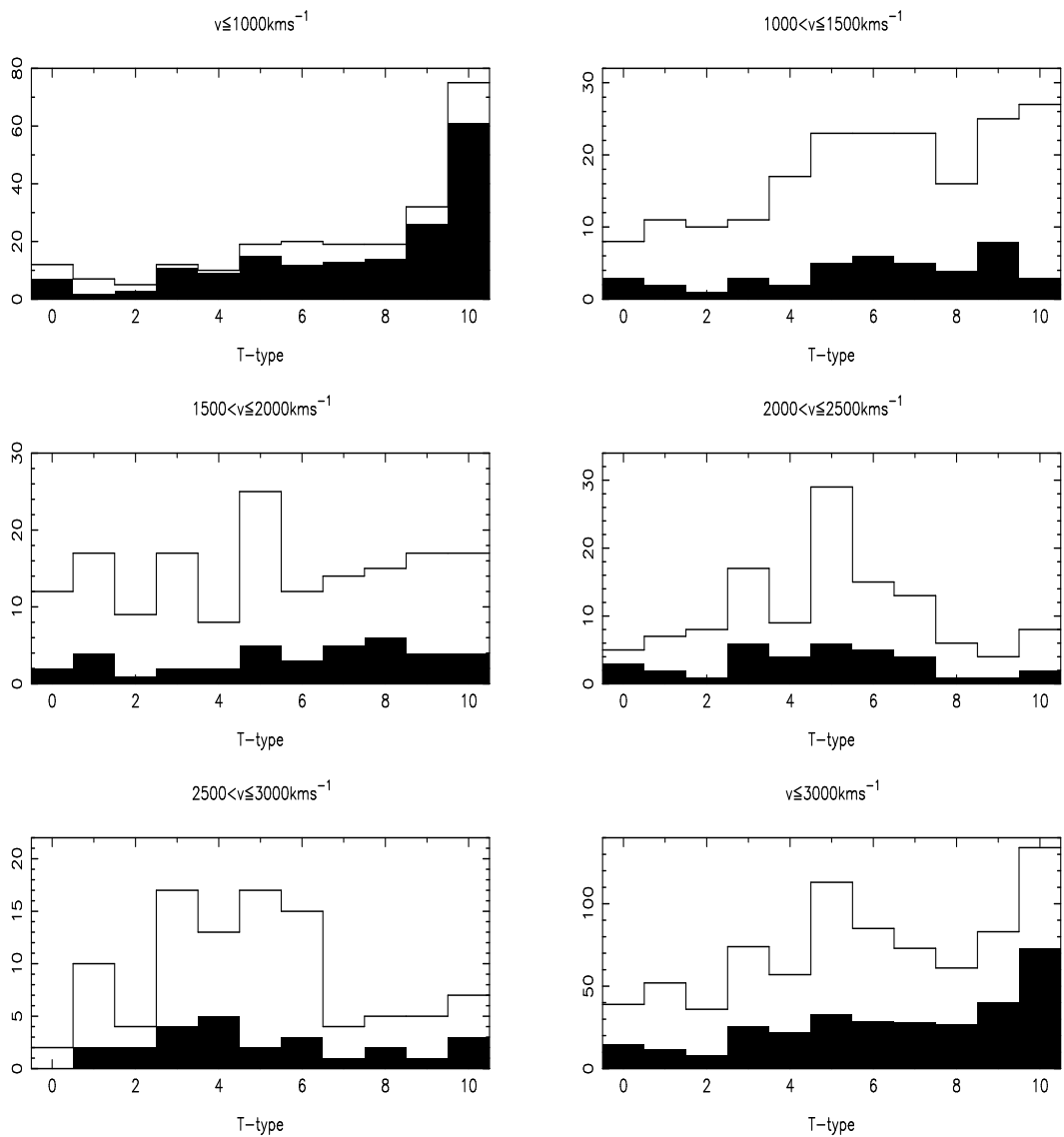}
      }
      \caption{The morphological make up of the parent sample (empty
      histograms) and the observed sample (filled histograms). See
      Sect. 3 for details.}
      \label{T_hist} 
      \end{figure}

Figure \ref{T_hist} illustrates the morphological makeup of both the
observed sample (filled histograms) and the parent sample (empty
histograms), where the latter is defined as all galaxies from the UGC
satisfying our selection criteria.  The x-axes display the galaxies'
T-types (where T is as defined such that T=0 represents an S0/a
galaxy, T=1 is an Sa galaxy, and so on up to T=10 for Im
classifications).  The first 5 plots show the breakdown for each
velocity shell. The final plot combines the data for the entire
sample.  The predominance of the Sm (T=9) and Im galaxies at low
redshift can clearly be seen, as can the emergence of the Sc (T=5)
class as the dominant detectable galaxy type at higher redshifts.  The
final plot in Fig. \ref{T_hist} shows that these Sc galaxies have been
undersampled in the observations.

In order to calculate intrinsic diameters and absolute magnitudes, it
is necessary to have reliable galaxy distances.  These were calculated
using NED heliocentric recession velocities and a Virgocentric infall
model for the local Hubble flow.  The model used assumes a global
Hubble constant of 75~kms$^{-1}$Mpc$^{-1}$, and accounted for Virgo
infall using the method of Schechter (\cite{schechter}). Calculated
distances were checked against those in the Nearby Galaxies Catalogue
(Tully \cite{tully}) with excellent agreement in almost all cases.
This catalogue was also used to resolve ambiguities in the
triple-valued region around the Virgo cluster, where we either
directly took the value preferred by Tully or, for galaxies not in his
catalogue, associated them with groups which he had identified.

\begin{figure} 
   \centering
   \rotatebox{-90}{
      \includegraphics[height=14cm,width=15cm]{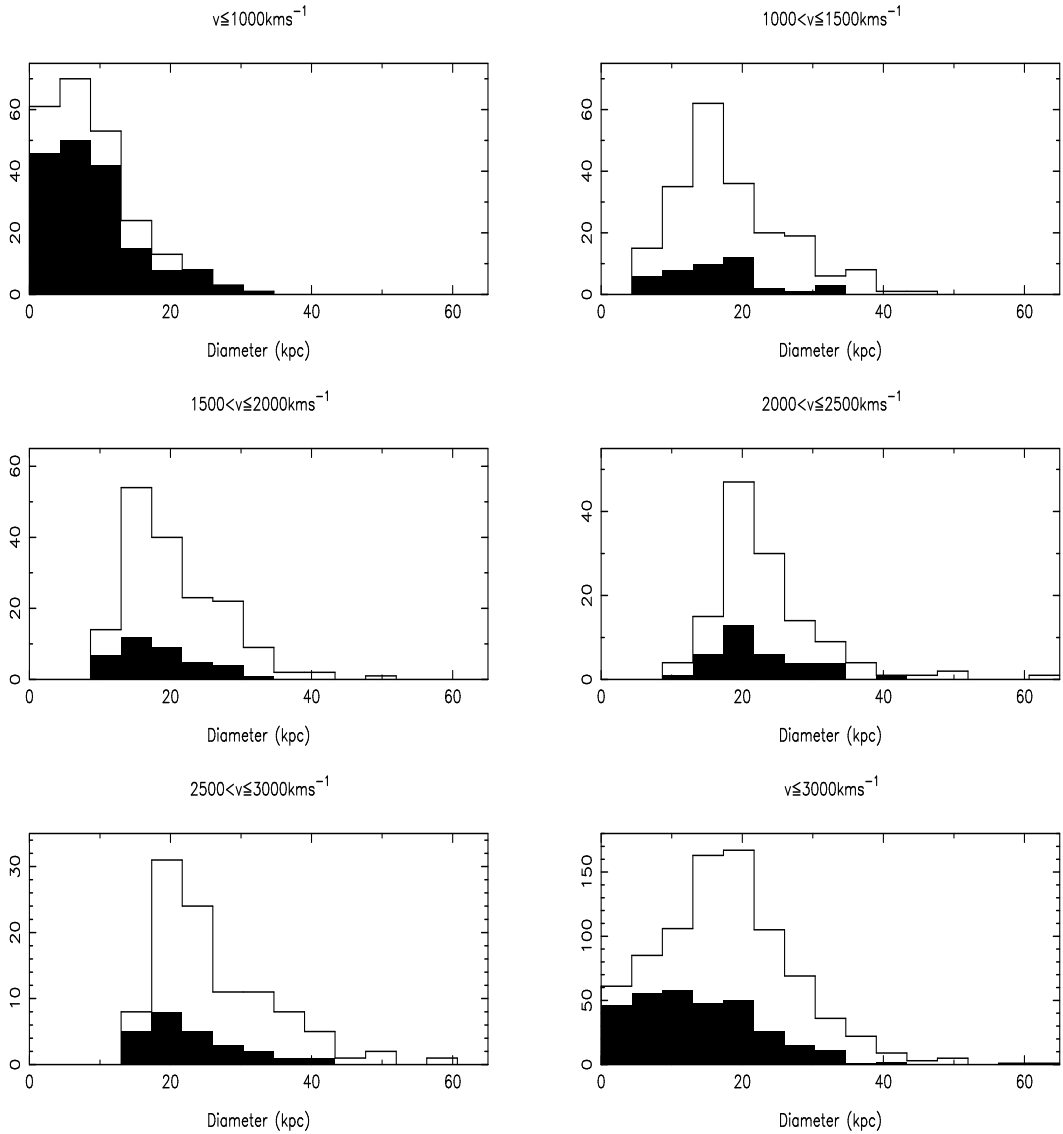}
      }
      \caption{The distribution of galaxy D$_{25}$ major-axis diameters in the parent
      sample (empty histograms) and the observed sample (filled
      histograms). See Sect. 3 for details.} 
      \label{size_hist} 
      \end{figure}

Figure \ref{size_hist} shows the distribution of galaxy diameters in
the original sample (empty histograms) and the observed sample (filled
histograms).  These diameters represent the intrinsic sizes of the
galaxies in kpc.  They are converted from the $D_{25}$ major axis
values quoted on NED (in arcminutes), using the Virgo-infall corrected
distances.  As expected, the lowest-redshift shell samples the
smallest galaxies, whereas the intrinsically largest objects are found
in the higher-redshift bins.  The final plot shows the distribution
for the entire sample and demonstrates that there is very good
coverage of the previously undersampled dwarf population.  The modal
galaxy sizes of $\sim$15-25~kpc are somewhat under-represented in the
observed sample.

\begin{figure}
   \centering
   \rotatebox{-90}{
   \includegraphics[height=14cm,width=15cm]{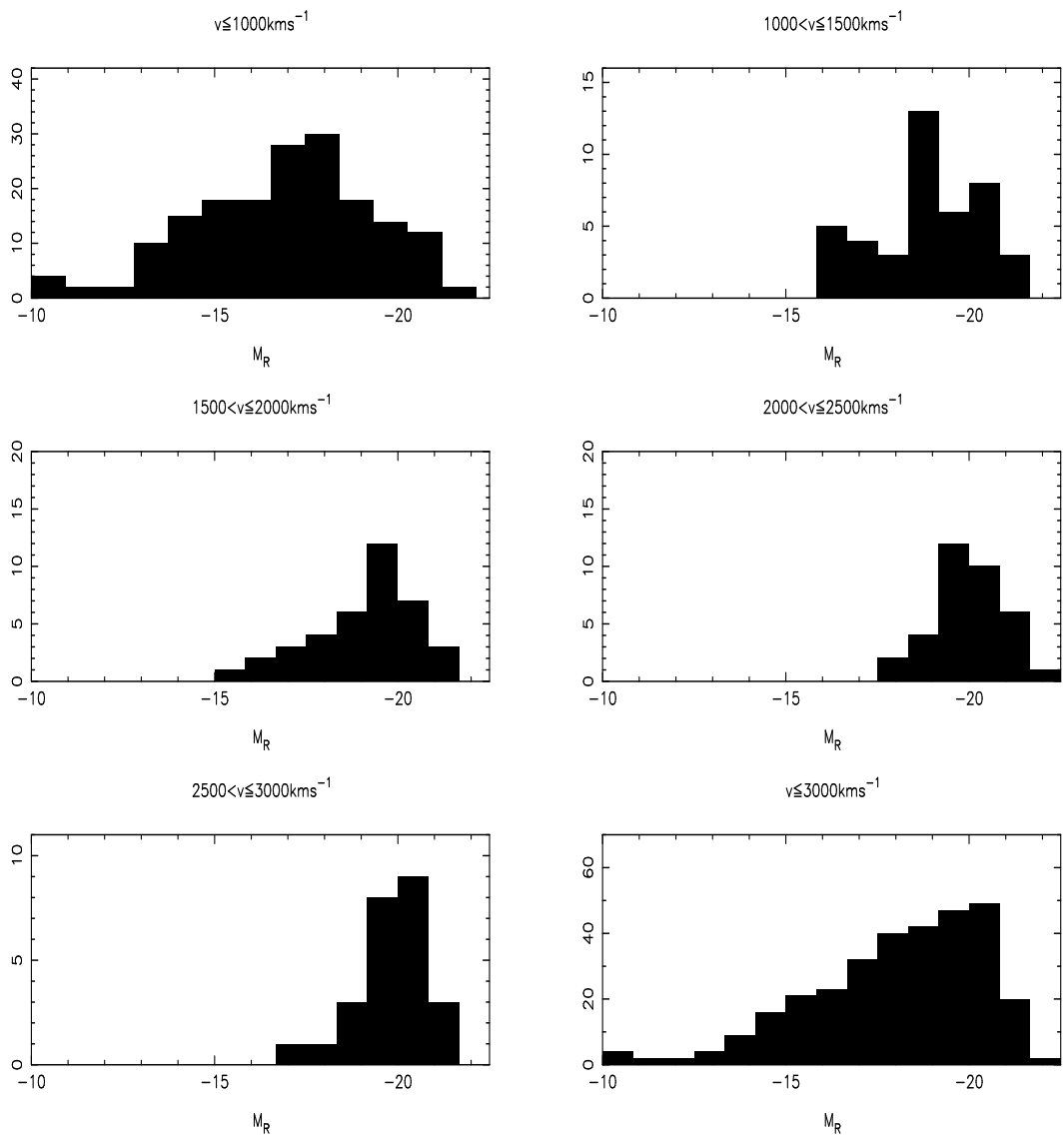}
   }
      \caption{The distribution of total $R$-band absolute magnitudes for the 
observed galaxies.  See Sect. 3 for details.
              }
         \label{MR_hist}
   \end{figure}

The distributions of absolute $R$-band magnitudes are shown in Fig.
\ref{MR_hist} for the observed galaxy sample.  These
magnitudes are calculated from the measured $R$-band fluxes, again using
Virgo-infall corrected distances.  The corresponding
histograms for the parent sample, thus, cannot be shown.  The $R$-band
fluxes have been measured in a consistent way, unlike the $B$-band
magnitudes quoted on NED, and are thus preferable.  As expected, only
intrinsically-bright galaxies were selected in the high-redshift bins.
The vast majority of the observed galaxies with magnitudes fainter
than --17.5 are to be found in the innermost shell.  These data show
that a wide range of luminosities is represented in this survey.

\section{Data reduction}

\subsection{Flux measurements}

The majority of the data reduction was performed using the {\it
Starlink} package {\sc CCDPACK}, with the rest making use of the {\it
Starlink} {\sc KAPPA} and {\sc FIGARO} packages.  The relevant commands were
assembled into a set of executable scripts, streamlining the reduction
process and ensuring that all data were treated in an objective and
reproducible way.  {\sc CCDPACK} was used to perform bias subtraction,
flat fielding by a median sky flat taken through the appropriate
filter, image registration and co-addition (by median stacking to
remove cosmic rays) for the multiple narrow-band images, and alignment
of images taken in different filters, to remove systematic offsets
which were found between different filters.  All galaxy and standard
star images were scaled to an effective integration time of 1~second
to simplify photometric calculations.  

One of the most critical stages of data reduction is the removal of
the continuum contribution to the flux in the images taken through the
narrow-band H$\alpha$ filters.  The key parameter is the scaling
factor applied to the image used for continuum subtraction (either an
$R$-band or a H$\alpha$Cont image), which depends on the relative
effective throughputs of the narrow-band and continuum filters to
continuum light.  These scaling factors were estimated in three ways.
The first was to integrate numerically under the scanned, digitised
filter profiles, which were already available for all the ING broad-
and narrow-band filters; the ING kindly scanned the profile of the new
H$\alpha$Cont filter in the same way.  The ratio of these integrals
for the narrow and continuum filters used for a specific galaxy
observation then gives the scaling factor to be applied to the
continuum image before it is subtracted from the narrow-band image.
The second method was to use photometry of standard spectrophotometric
stars through the pairs of filters in question, and using the ratio of
counts per second detected to give the scaling factor.  The third
method was to calculate the ratio from foreground stars in the
narrow-band and continuum images of each galaxy, thereby deriving a
scaling factor which forces the cancellation of stellar images, at
least in a statistical sense, in the continuum-subtracted images.
This latter method has the advantage of accommodating any changes in
the sky transparency between the two images, and can make use of
several stars giving a statistical improvement in precision compared
with using the standard stars.  The drawback of this final method is
that the colours of the stars used are not known, and may well be
significantly bluer on average than the galaxy continuum, which would
lead to a systematic error in the scaling factor.  However, in
practice it was found that the three methods gave very consistent
values for the scaling factors, and the differences between them give
good estimates of the likely errors in the continuum subtraction
process.  For non-photometric data, the final method (stars in the
galaxy fields) was used to allow for sky transparency changes, but in
photometric conditions standard ratios, calculated for every
narrow/continuum filter pair from a combination of standard stars and
field stars, were used, with the integrated filter profiles giving a
strong consistency check.  Good estimates of the scaling factors can
be quickly obtained from the throughputs of the filters listed in the
final column of Table~\ref{filters}, where the values are quoted as
fractions of the $R$-band filter throughput.

Galactic extinction corrections were derived with the aid of NED, which
uses data and methods from Schlegel et al. (\cite{schlegel}) and Cardelli
et al. (\cite{cardelli}).  These corrections were applied to the H$\alpha
+$[N{\sc ii}] fluxes prior to calculating star formation rates, but have
not been applied to fluxes and magnitudes presented in this paper.

There is a known problem of light leaks with the JKT.  This was only
found to give significant problems when the moon was above the
horizon, but did lead to some minor problems with gradients in the 
sky background.  These were removed using a 2-dimensional polynomial
fit to the background light, after removal of all stars and
galaxies, using the KAPPA routine {\sc surfit}.  The fitted function
was then subtracted from the original image, achieving both gradient
removal and sky subtraction.

An extensive study was made into the dominant sources of errors on
derived line fluxes and equivalent widths.
The typical errors in total H$\alpha +$[N{\sc ii}] fluxes due to
background sky subtraction are 1\%, with a worst-case error of 10\%;
filter throughput uncertainties are 9\% (typical) or 15\%
(worst-case); continuum subtraction 10\% (typical) or 35\%
(worst-case); and photometric errors from standard star observations
1.5\% (typical) or 5\% (worst-case).  In calculating errors on 
H$\alpha +$[N{\sc ii}] fluxes on a galaxy by galaxy basis, it was found
that only the continuum subtraction and filter throughput errors
contributed significantly to the totals, and given the importance of the
first of these, the errors are strongly dependent on the EW
of the H$\alpha +$[N{\sc ii}] emission.  For high EW
($>$2~nm) the total flux errors were calculated to be 10\% or 15\%,
with the lower value being for the lower redshift galaxies.  For low
EW galaxies ($<$2~nm) the corresponding errors were
calculated to be 25\% and 35\% respectively.  The same percentage errors
were allocated the the EW measurements, as in all cases the
fractional errors on the continuum flux levels are much smaller than the
fractional errors on line fluxes.  The minimum detectable H$\alpha
+$[N{\sc ii}] EW is 0.2~nm, and the minimum line flux is
10$^{-17}$W~m$^{-2}$.

Photometric calibration for the $R$-band images was quite
straightforward, with the observations of Landolt standards being used
to define zero-points and airmass corrections for each night in the
usual way.  Airmass corrections were generally very small, as most of
the galaxy observations were made at airmass $<$1.5.  Photometric
calibration of the narrow-band continuum subtracted images is
inevitably more involved, and the interested reader is referred to
Shane (\cite{shane}) for full details.  In outline, the procedure was
to tie the calibration of the H$\alpha$ fluxes to the Landolt
standards observed on the same night, taking into account the
transmission profiles of the filters used to both normalise the
response to continuum sources through the $R$ filter, and to normalise
the throughput of the narrow-band filter to the H$\alpha$ line.  This
latter is a function of the recession velocity of the galaxy, and it
was assumed that all the line emission occurred at a wavelength
corresponding to 6563~\AA\ redshifted by the systemic recession
velocity of the galaxy as listed in NED.  This value will be slightly
in error for the satellite [N{\sc ii}] lines, and for specific H{\sc
ii} regions due to galaxy rotation, but such effects are small.  This
calculation thus accounts for overall transmission differences between
filters used, and for the effect of the transmission profile on the
specific line of interest.  The latter has not been done in several
previous studies, due to the lack of sufficiently accurate filter
transmission data.  The derived value for H$\alpha$ transmission was
corrected, where appropriate, for the effective subtraction of a small
part of the line flux along with the continuum, when the $R$-band filter
was used to derive the continuum.  The effect of this was minor,
reducing the throughput to H$\alpha$ by about 3\%.  All line fluxes
quoted in this paper have been corrected for this effect.

The net result of the calibration process is that the detected
counts~second$^{-1}$ in the $R$-band frames can be converted into either
$R$-band magnitudes or flux densities in W~m$^{-2}$~nm$^{-1}$, and in the
H$\alpha$ continuum subtracted frames they can be directly converted to
line fluxes in W~m$^{-2}$.  Both calibrations include all airmass,
telescope, instrumental and filter transmission corrections.

The errors given on apparent $R$ magnitudes in this paper include
contributions from the photometric zero point error for the given night,
calculated from the scatter in standard star magnitudes, and from the
error in determining the sky level on each image.  This latter
effect gives an error that varies as a function of galaxy magnitude and
surface brightness. As a result we find that the total errors range from
a of 0.04~mag for the brightest galaxies to 0.10~mag for
low-surface-brightness dwarfs.  Even for these latter galaxies, the
errors due to photon shot noise are negligible and were not included in
the error estimation.

Initial galaxy photometry was obtained within the {\it Starlink} {\sc
GAIA} package, using a set of between 30 and 60 concentric apertures
ranging from 3\farcs3 up to between 100$^{\prime \prime}$
and 200$^{\prime\prime}$ in radius/semi-major axis, with the upper limit
depending on the angular size of the galaxy. All apertures used, and
hence all profiles calculated, are centred on the R-band galaxy centres.
For irregular galaxies and face-on spiral galaxies, circular apertures
were used, and for inclined spiral galaxies, elliptical apertures were
used to obtain the data presented here.  The ellipse parameters
(ellipticity and major axis) were taken from the UGC, and checked by eye
to ensure that they were a good fit to the present images.  From this
photometry, growth curves were constructed, both in $R$-band and H$\alpha
+$[N{\sc ii}] light.  Dividing the H$\alpha +$[N{\sc ii}] aperture fluxes
by the corresponding $R$-band flux densities enables an estimate to be
made of the H$\alpha +$[N{\sc ii}] EW, assuming that the average
continuum level within the $R$ filter is equal to the continuum level at
6563~\AA.

Line fluxes and R-band magnitudes quoted in this paper are total values,
where the apertures were set by requiring that the enclosed flux varied
by less than 0.5\% over 3 consecutive points in the H$\alpha +$[N{\sc
ii}] growth curves. Inspection of the apertures thus defined showed this
to be a conservative criterion which encompassed all visible H{\sc ii}
regions; examples of these apertures are shown in figures later in this
paper.

\subsection{Comparison with literature measurements}

   \begin{table}
      \caption[]{Repeat observations; asterisks indicate the adopted
measurements for these galaxies.}
         \label{repeats}
      
          $$
          \begin{array}{lcc|lcc}
            \hline
            \noalign{\smallskip}
            ${\bf Name}$ &  &  & ${\bf Name}$ &  & \\
            $Obs. date $ & $H$\alpha +$[N{\sc ii}] flux$ & $H$\alpha +$[N{\sc ii}] EW$ &
            $Obs. date $ & $H$\alpha +$[N{\sc ii}] flux$ & $H$\alpha +$[N{\sc ii}] EW$\\
                         & $(10$^{-16}$~W~m$^{-2}$)$     & $(nm)$          &
                         & $(10$^{-16}$~W~m$^{-2}$)$     & $(nm)$          \\
            \noalign{\smallskip}
            \hline
            \noalign{\smallskip}
       ${\bf UGC~2023}$ & & & ${\bf UGC~4115}$ & & \\
       22/11/00*  &  3.440  &  3.7 & 19/11/00*  &  2.450  &  3.4\\
       18/12/00   &  2.610  &  2.4 & 20/11/00   &  2.299  &  3.4\\
       \hline
       ${\bf UGC~4173}$ & & & ${\bf UGC~4469}$ & & \\
       22/11/00*  &  2.041  &  5.8 & 17/02/01   &  8.948  &  2.7\\
       17/12/00   &  1.339  &  4.0 & 20/10/01*  &  9.771  &  4.5\\
       \hline
       ${\bf UGC~4484}$ & & & ${\bf UGC~6251}$ & & \\
       14/02/01   &  8.781  &  2.0 & 28/03/01*  &  1.795  &  4.7\\
       22/10/01*  &  6.057  &  1.7 & 11/05/01   &  1.391  &  4.6\\
       \hline
       ${\bf UGC~7232}$ & & & ${\bf UGC~8188}$ & & \\
       19/01/01* &   4.685  &  2.0 & 09/05/00* &   8.479  &  4.0\\
       03/04/01  &   7.004  &  3.2 & 28/03/01  &   5.657  &  2.0\\
       \hline
       ${\bf UGC~11331}$ & & & ${\bf UGC~11332}$ & & \\
       18/05/01  &   1.013  &  2.7 & 18/05/01  &   10.34  &  6.2\\
       22/10/01* &   0.927  &  2.9 & 22/10/01* &   10.57  &  5.0\\
       \hline
       ${\bf UGC~12294}$ & & & ${\bf NGC2604b}$ & & \\
       04/08/01  &   25.57  &  4.2 & 14/02/01  &   0.490  &  1.4\\
       23/10/01* &   31.63  &  4.4 & 20/10/01* &   0.682  &  2.4\\
            \noalign{\smallskip}
            \hline
         \end{array}
         $$
   \end{table}

In order to test our photometry, extensive internal and external
comparison tests were performed.  Table~\ref{repeats} shows the total
measured H$\alpha +$[N{\sc ii}] fluxes and line EWs for galaxies observed by us on
more than one occasion, as a test of the internal reliability of the
photometry in this study.  These repeats give a conservative
indication of the internal errors in the photometry for this study, as
many of the measurements in the table were repeated due to doubts
about the observing conditions for the first set of observations.
UGC~2023 (18/12/00 observation) and UGC~7232 (03/07/01) both suffer from
an unexplained `creased' pattern in the narrow-band filter images.
The observations of UGC~8188 (28/03/01), UGC~11331 (18/05/01) and
UGC~11332 (18/05/01) for various reasons only resulted in two usable
H$\alpha$ integrations, instead of the three needed to achieve a good
signal-to-noise ratio and to remove cosmic-ray contamination.  The
mean variations shown by the 12 pairs of repeats are 29\% in H$\alpha +$[N{\sc ii}]
flux, and 38\% in EW.  For the three galaxies which were observed on
two photometric nights with no major problems (UGC~4115, UGC~6251 and
UGC~12294), the agreements between the two sets of measurements are
much better (a mean difference of 17\% in the H$\alpha +$[N{\sc ii}] flux, and of
only 2\% in the EW).

   \begin{figure}
   \centering
   \includegraphics[angle=-90,width=10cm]{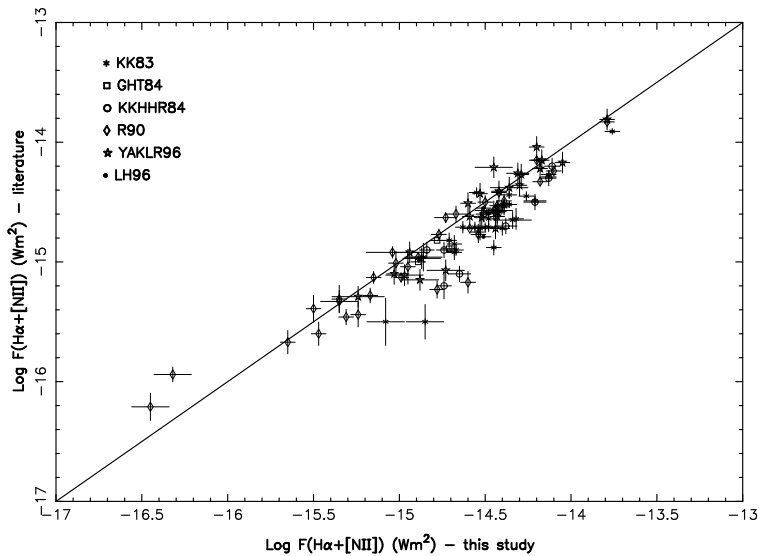}
      \caption{Comparison of literature values for total galaxy
H$\alpha +$[N{\sc ii}] fluxes with values from the present study. 
              }
         \label{fhacomp}
   \end{figure}

   \begin{figure}
   \centering
   \includegraphics[angle=-90,width=10cm]{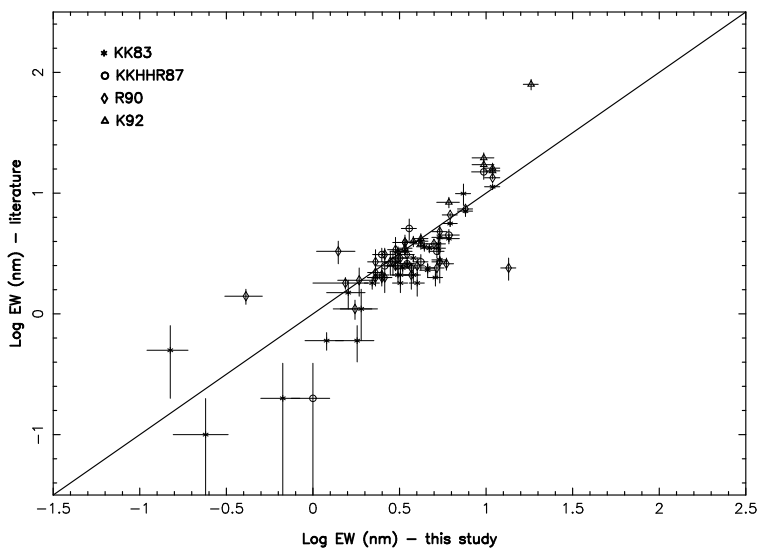}
      \caption{Comparison of literature values for total galaxy 
      H$\alpha +$[N{\sc ii}] EWs with values from the present study.
              }
         \label{ewcomp}
   \end{figure}

Figure~\ref{fhacomp} illustrates a comparison of 105 literature
measurements of H$\alpha +$[N{\sc ii}] fluxes with the equivalent values
from the present study for the same galaxies. The literature data are
taken from the following papers: Kennicutt \& Kent (\cite{ken83}, KK83);
Gallagher et al. (\cite{gal84}, GHT84); Kennicutt et al. (\cite{ken87},
KKHHR87); Kennicutt \cite{ken92}, K92; Romanishin (\cite{rom90}, R90);
Young et al. (\cite{young}, YAKLR96); and Lehnert
\& Heckman (\cite{lehn96}, LH96).  All of these studies quote H$\alpha
+$[N{\sc ii}] fluxes which are total or near total values using either
CCD photometry or large aperture spectrophotometry, with apertures in the
following ranges: KK83, most fluxes in a 3$^{\prime}$ aperture, remainder
2$^{\prime}$, 5$^{\prime}$ or 7$^{\prime}$; GHT84) fluxes in 2$^{\prime}$
apertures; KKHHR87, total fluxes within an observed field of
2\farcm3 square or 2\farcm7 square; K92, 4 objects in
0\farcm75 apertures, remainder in range 1--2$^{\prime}$; R90,
apertures from 1\farcm3 to 4\farcm8, with most close to
2$^{\prime}$; YAKLR96), apertures not quoted, but measured from CCD
images with a field of 6$^{\prime}$ by 6$^{\prime}$ or 7$^{\prime}$ by
4$^{\prime}$ ; and LH96, no apertures quoted but they state that these
are total fluxes, measured from CCD images. Where specific apertures are
quoted, these tend to be somewhat smaller than the software apertures
used in the present study for deriving total fluxes, and the CCD-based
studies use cameras with smaller fields than that of the JKT CCD.  This will tend to
offset points somewhat to the right of the one-to-one correspondence,
indicated by the diagonal solid line. Data from the present study shown
in Fig.~\ref{fhacomp} have been corrected for Galactic extinction,
where this was done in the comparison study (GHT84, KK83, KKHHR87 and
LH96), but not otherwise, and no corrections for extinction internal to
the galaxy concerned have
been applied. The mean offset in Fig.~\ref{fhacomp} is 0.11 dex (0.21 dex
relative to KK83 and KKHHR87; 0.055 dex relative to R90 and YAKLR96), in the
sense that this study finds higher fluxes than literature studies,
possibly due to the larger effective apertures used.
The plotted points have an RMS scatter of 0.16 dex about the best-fit
regression line, or 0.20 dex about the line representing perfect
agreement.  Given the above uncertainties and differences in reduction
procedures, the agreement found in Fig.~\ref{fhacomp} is generally good,
and uncertainties in the internal extinction corrections to H$\alpha$
fluxes probably dominate over photometric errors as the main error in
derived star formation rates.

Figure~\ref{ewcomp} shows a comparison of the H$\alpha +$[N{\sc ii}] EWs
for 88 galaxies from the present study with literature values for the
same galaxies, again showing reasonable agreement overall, but with
significant scatter, particularly for low EW galaxies.  These objects are
often early-type galaxies where uncertainties in H$\alpha +$[N{\sc ii}]
flux measurements are high due to low H$\alpha$ emission and the presence
of H$\alpha$ absorption.   The mean offset in Fig.~\ref{ewcomp} is 0.10
dex, in the sense that this study finds rather higher EWs
than literature studies. The plotted points have an RMS scatter of 0.21
dex about the best-fit regression line, or 0.36 dex about the line
representing perfect agreement.

\subsection{Calculation of star formation rates}

For the present paper, we calculate star formation rates from total
H$\alpha +$[N{\sc ii}] luminosities using the same relationship as Kennicutt et al.
(\cite{ken94}):

$$
{\rm SFR}(M_{\odot}~{\rm yr}^{-1}) = 7.94\times10^{-35}L_{{\rm
H\alpha}}({\rm W})
$$

This transformation is appropriate under the assumption of solar
abundances and a Salpeter initial mass function (Salpeter \cite{sal55})
over a range of stellar masses from 0.1--100~M$_{\odot}$, and does not
include the effect of dust attenuation on measured line fluxes.  The
H$\alpha$ luminosities in the present study were corrected for internal
extinction assuming a constant value of 1.1~mag for each galaxy,
independent of type and inclination.  This value was found to be
appropriate for local galaxies in studies by Kennicutt \& Kent
(\cite{ken83}) although Niklas et al. (\cite{nikl97}) found a rather
smaller value of A$_{\rm H\alpha}=$0.8~mag.  Corrections for contamination
by the [N{\sc ii}] lines which lie within the passband of our filters
were applied using the H$\alpha$/(H$\alpha+$[N{\sc ii}]) ratios derived
spectrophotometrically by Kennicutt \& Kent (\cite{ken83}): 0.75$\pm$0.12
for spiral galaxies and 0.93$\pm$0.05 for irregular galaxies.  However,
it should be noted that Tresse et al. (\cite{tres99}), in a very large
spectroscopic study of local galaxies, find that this ratio varies
systematically as a function of EW(H$\alpha+$[N{\sc ii}]), in the sense
that [N{\sc ii}] is less significant in high-EW galaxies. These
corrections can also be affected by the possible presence of
high-excitation emission-line regions around AGN.  The [N{\sc ii}] and
extinction corrections are applied only when calculating star formation
rates in the present paper; hence any H$\alpha$ EW values and line fluxes
presented here are in fact for H$\alpha+$[N{\sc ii}], uncorrected for
internal and Galactic extinction. All of these corrections will be
examined in more detail in a later paper, but initial analysis of our
data shows a good general agreement with the values obtained by Kennicutt
and coworkers.

The errors on the star formation rates were derived from line flux errors
discussed above, added in quadrature with errors resulting from distance
uncertainties, and in most cases the latter was the dominant
contribution.  The distance error was taken as $\pm$25\% for galaxies closer
than 8~Mpc (corresponding to a 56\% error in star formation rate), and as
$\pm$2~Mpc for galaxies outside this distance.


\section{Results}

The main results in this paper are listed in Table 3, which contains the
photometric data and derived star formation rates for the full sample of
334 galaxies.  Column 1 contains the number of each galaxy in the Uppsala
Galaxy Catalogue; col. 2 the Hubble type, taken from NED; col. 3 the
heliocentric recession velocity from NED; col. 4 the distance in Mpc of
the galaxy, assuming a Hubble constant of 75~km~s$^{-1}$Mpc$^{-1}$ and
after corrections from a Virgo infall model; col. 5 the galaxy major axis
in minutes of arc, from NED; col. 6 the major-to-minor axis ratio, from
NED;  col. 7 the total $R$ magnitude derived from this study, with errors
in brackets; col. 8 the total H$\alpha+$[N{\sc ii}] flux, after all
corrections described in the previous section except those for [N{\sc
ii}] contamination, Galactic extinction and internal extinction, in units
of 10$^{-16}$W~m$^{-2}$, and with errors given in brackets; col. 9 the EW
in nm of the H$\alpha +$[N{\sc ii}] lines, with errors in brackets; col.
10 contains the total star formation rate, based on the total measured
H$\alpha +$[N{\sc ii}] line flux corrected for [N{\sc ii}] contamination,
Galactic extinction and internal extinction, with the conversion factor
as described in Sect. 4.3, and errors in brackets; and col. 11 contains
H$\alpha +$[N{\sc ii}] surface brightnesses within Petrosian radii,
calculated as described in section 6.2, and with errors in brackets.  The
galaxies are listed in Right Ascension order, within each of the five
recession velocity shells, starting with the lowest velocity shell (up to
1000~km~s$^{-1}$).  Serendipitously discovered galaxies are listed at the
end of the table.  Data for the different recession velocity shells are
separated by a horizontal line in the table.

\setlongtables
   \begin{longtable}{llrrrrrrrrr}
      \caption*{{\bf Table 3.} Photometric, distance and star formation data for
        334 galaxies.}\\
            \hline
            UGC 
            & Class
            & \multicolumn{1}{c}{$v_{\rm{rec}}$} 
            & \multicolumn{1}{c}{Dist.} 
            & \multicolumn{1}{c}{Dia}
            & \multicolumn{1}{c}{$a/b$}
            & \multicolumn{1}{c}{$m_{R}$}  
            & \multicolumn{1}{c}{F(H$\alpha +$[N{\sc ii}])} 
            & \multicolumn{1}{c}{EW} 
            & \multicolumn{1}{c}{SFR} 
            & \multicolumn{1}{c}{SB$\times$10$^{-20}$} \\
            ~ 
            & ~ 
            & \multicolumn{1}{c}{kms$^{-1}$}
            & \multicolumn{1}{c}{Mpc}
            & \multicolumn{1}{c}{$^{\prime}$}
            & ~
            & \multicolumn{1}{c}{mag}
            & \multicolumn{1}{c}{10$^{-16}$W~m$^{-2}$} 
            & \multicolumn{1}{c}{nm}
            & \multicolumn{1}{c}{M$_{\odot}$~yr$^{-1}$}
            & \multicolumn{1}{c}{Wm$^{-2}$/$^{\prime\prime2}$}\\
            \hline
            \endhead
   17   &  Sm      &   878 &  9.9 & 2.5 & 1.5 & 14.49(0.06) &   2.1(0.21) &  6.0(0.6) & 0.0448(0.0204) &   2.4(0.34) \\
   75   &  IBm     &   865 &  9.7 & 2.8 & 1.3 & 12.33(0.04) &   4.2(1.05) &  1.6(0.4) & 0.1167(0.0606) &   5.2(1.40) \\
  655   &  Sm      &   836 & 10.6 & 2.5 & 1.0 & 14.05(0.06) &   0.9(0.22) &  1.6(0.4) & 0.0224(0.0108) &   1.1(0.29) \\
  891   &  SABm    &   643 &  7.2 & 2.3 & 2.3 & 14.05(0.06) &   0.5(0.13) &  1.0(0.3) & 0.0059(0.0036) &   0.6(0.17) \\
 1176   &  Im      &   633 &  7.3 & 4.6 & 1.3 & 14.29(0.06) &   1.8(0.18) &  4.3(0.4) & 0.0275(0.0156) &   1.3(0.19) \\
 1195   &  Im      &   774 &  8.8 & 3.4 & 3.1 & 13.35(0.05) &   1.7(0.43) &  1.8(0.4) & 0.0386(0.0218) &   1.8(0.47) \\
 1200   &  IBm     &   808 &  9.2 & 2.0 & 1.4 & 13.19(0.05) &   3.4(0.34) &  3.0(0.3) & 0.0795(0.0391) &   7.4(1.05) \\
 1865   &  Sm:     &   580 &  7.5 & 2.8 & 1.3 & 14.42(0.06) &   1.1(0.11) &  3.1(0.3) & 0.0149(0.0085) &   1.1(0.15) \\
 1983   &  SAb:    &   609 &  8.0 & 2.4 & 1.8 & 11.49(0.04) &  17.1(1.71) &  3.1(0.3) & 0.2471(0.1412) &  19.5(2.76) \\
 2002   &  Sdm:    &   597 &  7.8 & 2.3 & 1.6 & 12.00(0.04) &  10.3(1.03) &  3.0(0.3) & 0.1467(0.0834) &   6.8(0.96) \\
 2017   &  Im      &   985 & 12.1 & 2.3 & 1.4 & 15.27(0.10) &   0.2(0.10) &  1.4(0.4) & 0.0111(0.0060) &   0.3(0.15) \\
 2014   &  Im:     &   565 &  7.5 & 2.0 & 3.3 & 14.69(0.06) &   0.9(0.10) &  3.1(0.3) & 0.0138(0.0078) &   1.4(0.19) \\
 2023   &  Im:     &   603 &  7.8 & 1.7 & 1.0 & 13.42(0.05) &   3.5(0.35) &  3.8(0.4) & 0.0662(0.0377) &   2.3(0.33) \\
 2141   &  S0/a    &   987 & 12.2 & 2.5 & 2.3 & 12.02(0.04) &  15.6(1.56) &  4.6(0.5) & 0.6539(0.2410) &  10.9(1.54) \\
 2193   &  SAc     &   518 &  6.9 & 3.0 & 1.1 & 11.08(0.04) &  30.6(3.06) &  3.8(0.4) & 0.3330(0.1894) &  21.7(3.06) \\
 2455   &  IBm     &   375 &  4.9 & 3.3 & 1.3 & 11.91(0.04) &  27.6(2.76) &  7.4(0.7) & 0.2800(0.1593) &  36.8(5.20) \\
 2684   &  Im      &   350 &  4.6 & 1.8 & 2.0 & 16.19(0.10) &   0.2(0.10) &  2.9(0.3) & 0.0015(0.0010) &   0.4(0.20) \\
 2947   &  SBm     &   863 & 10.8 & 3.6 & 4.0 & 11.99(0.04) &  12.9(1.29) &  3.7(0.4) & 0.4413(0.1839) &   1.3(0.18) \\
 3174   &  IABm:   &   670 &  9.2 & 1.7 & 1.5 & 15.36(0.10) &   0.8(0.10) &  5.3(0.5) & 0.0214(0.0106) &   1.3(0.19) \\
 3371   &  Im:     &   816 & 13.3 & 4.6 & 1.3 & 14.73(0.06) &   0.7(0.10) &  2.4(0.2) & 0.0418(0.0141) &   0.6(0.10) \\
 3429   &  SBab    &   893 & 14.5 & 6.0 & 1.8 & 10.06(0.04) &  51.2(5.12) &  2.5(0.2) & 2.6773(0.8337) &  11.3(1.60) \\
 3711   &  IBm     &   436 &  7.6 & 2.2 & 1.3 & 12.29(0.04) &  13.4(1.34) &  5.1(0.5) & 0.2339(0.1330) &  26.1(3.69) \\
 3734   &  SAc:    &   974 & 15.9 & 1.7 & 1.0 & 11.60(0.04) &   3.4(0.85) &  0.7(0.2) & 0.2144(0.0785) &   2.9(0.77) \\
 3817   &  Im:     &   438 &  8.3 & 1.8 & 2.0 & 15.16(0.10) &   0.8(0.10) &  4.3(0.4) & 0.0175(0.0096) &   1.3(0.19) \\
 3847   &  IRR     &    70 &  2.9 & 1.7 & 1.5 & 14.75(0.06) &   9.2(0.92) & 33.9(3.4) & 0.0207(0.0118) &  19.7(2.78) \\
 3851   &  IBm     &   100 &  2.9 & 8.1 & 2.5 & 11.76(0.04) & 101.7(10.2) & 23.6(2.4) & 0.2277(0.1295) &  24.8(3.51) \\
 3876   &  SAd     &   860 & 14.5 & 2.2 & 1.7 & 12.97(0.04) &   3.5(0.35) &  2.5(0.3) & 0.1630(0.0508) &   2.6(0.37) \\
 3966   &  Im      &   361 &  6.2 & 1.7 & 1.0 & 14.51(0.06) &   0.8(0.10) &  2.4(0.2) & 0.0087(0.0050) &   0.9(0.12) \\
 4115   &  IAm     &   338 &  5.8 & 1.8 & 1.8 & 13.70(0.05) &   2.5(0.25) &  3.5(0.3) & 0.0221(0.0125) &   2.3(0.32) \\
 4165   &  SBd     &   514 &  9.0 & 2.9 & 1.1 & 11.51(0.04) &  20.0(2.00) &  3.7(0.4) & 0.3518(0.1773) &   9.9(1.40) \\
 4173   &  Im:     &   860 & 14.3 & 1.9 & 3.2 & 14.48(0.06) &   2.1(0.21) &  6.0(0.6) & 0.1122(0.0354) &   1.5(0.21) \\
 4274   &  SBm     &   447 &  7.7 & 1.7 & 1.1 & 11.40(0.04) &  24.7(2.47) &  4.1(0.4) & 0.3280(0.1866) &  43.5(6.15) \\
 4325   &  SAm     &   524 &  9.2 & 3.5 & 1.5 & 12.59(0.04) &   6.6(0.66) &  3.3(0.3) & 0.1222(0.0601) &   3.9(0.55) \\
 4426   &  Im:     &   397 &  6.7 & 2.0 & 2.0 & 14.72(0.06) &   0.8(0.10) &  3.0(0.3) & 0.0100(0.0057) &   0.9(0.12) \\
 4499   &  SABd    &   691 & 12.2 & 2.6 & 1.4 & 13.06(0.05) &   6.4(0.64) &  4.9(0.5) & 0.2047(0.0754) &   5.0(0.70) \\
 4514   &  SBcd    &   691 & 12.2 & 2.1 & 2.3 & 13.18(0.05) &   3.4(0.34) &  2.9(0.3) & 0.1072(0.0395) &   1.6(0.23) \\
 4645   &  SAB0/a  &   692 & 12.3 & 3.6 & 1.1 & 10.04(0.04) &  18.1(4.53) &  0.9(0.2) & 0.5675(0.2449) &  11.3(3.05) \\
 4879   &  IAm     &   600 & 10.5 & 1.7 & 1.3 & 13.18(0.05) &   0.2(0.10) &  0.2(0.2) & 0.0050(0.0024) &   0.1(0.05) \\
 5139   &  IABm    &   143 &  2.3 & 3.6 & 1.2 & 13.71(0.05) &   4.0(0.40) &  5.5(0.6) & 0.0058(0.0033) &   0.7(0.10) \\
 5221   &  SAc     &     3 &  2.1 & 5.9 & 2.2 & 10.21(0.04) &  62.4(6.24) &  3.5(0.3) & 0.0641(0.0365) &  12.8(1.82) \\
 5272   &  Im      &   520 &  7.7 & 2.1 & 2.6 & 13.78(0.05) &   4.1(0.41) &  6.2(0.6) & 0.0629(0.0358) &   4.5(0.64) \\
 5340   &  Im      &   503 &  7.2 & 2.7 & 2.7 & 14.14(0.06) &   2.6(0.26) &  5.3(0.5) & 0.0339(0.0193) &   2.6(0.36) \\
 5336   &  Im      &    46 &  3.4 & 2.5 & 1.3 & 13.88(0.05) &   1.1(0.27) &  1.8(0.4) & 0.0035(0.0022) &   0.4(0.10) \\
 5364   &  IBm     &    20 &  1.0 & 5.1 & 1.6 & 13.58(0.05) &   3.3(0.33) &  4.1(0.4) & 0.0008(0.0005) &   1.7(0.24) \\
 5373   &  Im      &   301 &  3.9 & 5.1 & 1.5 & 12.58(0.04) &   7.3(0.73) &  3.6(0.4) & 0.0291(0.0165) &   4.0(0.57) \\
 5398   &  I0      &    14 &  2.1 & 5.4 & 1.2 & 10.35(0.04) &  66.2(6.62) &  4.2(0.4) & 0.0838(0.0477) &  40.6(5.74) \\
 5414   &  IABm    &   603 &  9.5 & 3.2 & 1.5 & 13.07(0.05) &   6.4(0.64) &  4.9(0.5) & 0.1443(0.0687) &   4.0(0.56) \\
 5637   &  IBm     &   753 & 10.8 & 5.0 & 1.5 & 11.07(0.04) &  49.9(4.99) &  6.2(0.6) & 1.5315(0.6384) &  19.2(2.71) \\
 5672   &  Sab     &   531 &  6.8 & 1.8 & 3.6 & 13.72(0.05) &   1.5(0.15) &  2.2(0.2) & 0.0147(0.0083) &   1.3(0.18) \\
 5692   &  Sm:     &   180 &  2.9 & 3.2 & 1.8 & 13.67(0.05) &   1.8(0.18) &  2.5(0.2) & 0.0034(0.0019) &   1.9(0.27) \\
 5721   &  SABd    &   537 &  6.9 & 2.1 & 2.1 & 12.16(0.04) &  12.2(1.22) &  4.1(0.4) & 0.1213(0.0690) &   7.5(1.06) \\
 5719   &  SBdm:   &   941 & 17.1 & 2.9 & 2.4 & 13.14(0.05) &   7.1(0.71) &  5.9(0.6) & 0.4179(0.1116) &   2.5(0.35) \\
 5740   &  SABm    &   649 & 10.9 & 1.7 & 1.4 & 13.67(0.05) &   2.3(0.23) &  3.1(0.3) & 0.0558(0.0231) &   1.8(0.25) \\
 5761   &  SABdm   &   641 &  8.1 & 2.2 & 1.3 & 12.20(0.04) &   3.7(0.92) &  1.3(0.3) & 0.0506(0.0308) &   2.8(0.75) \\
 5764   &  IBm:    &   586 &  7.8 & 2.0 & 1.8 & 14.57(0.06) &   1.5(0.15) &  4.6(0.5) & 0.0234(0.0133) &   2.6(0.37) \\
 5786   &  SABbc   &   993 & 18.2 & 3.1 & 1.3 & 10.40(0.04) & 163.4(16.3) & 10.9(1.1) & 11.231(2.8359) & 477.6(7.54) \\
 5829   &  Im      &   629 &  8.6 & 4.7 & 1.1 & 13.13(0.05) &  15.0(1.50) & 12.3(1.2) & 0.2859(0.1512) &   3.7(0.53) \\
 5848   &  Sm:     &   822 & 14.9 & 2.1 & 2.1 & 14.09(0.06) &   1.4(0.14) &  2.8(0.3) & 0.0624(0.0189) &   1.3(0.18) \\
 5889   &  SABm    &   572 &  6.9 & 2.2 & 1.0 & 13.63(0.05) &   1.2(0.30) &  1.5(0.4) & 0.0120(0.0074) &   0.9(0.24) \\
 5918   &  Im:     &   340 &  5.4 & 2.4 & 1.0 & 14.52(0.06) &   0.7(0.10) &  2.0(0.2) & 0.0050(0.0029) &   0.5(0.07) \\
 6123   &  SBb     &   979 & 20.9 & 3.4 & 1.2 & 10.80(0.04) &  40.5(4.05) &  3.9(0.4) & 3.6838(0.8255) &  18.6(2.63) \\
 6161   &  SBdm    &   756 & 12.3 & 2.6 & 2.2 & 13.52(0.05) &   2.9(0.29) &  3.4(0.3) & 0.0881(0.0322) &   1.6(0.22) \\
 6251   &  SABm:   &   927 & 17.1 & 1.8 & 1.1 & 14.39(0.06) &   1.8(0.18) &  4.6(0.5) & 0.1050(0.0280) &   1.8(0.26) \\
 6272   &  SA0/a:  &   628 &  7.1 & 5.2 & 2.7 & 10.70(0.04) &  29.4(2.94) &  2.6(0.3) & 0.3052(0.1736) &  11.0(1.55) \\
 6399   &  Sm:     &   805 & 14.4 & 2.8 & 3.5 & 13.27(0.05) &   2.6(0.26) &  2.5(0.2) & 0.1122(0.0352) &   0.9(0.13) \\
 6439   &  SAb     &   770 & 12.2 & 5.9 & 1.9 & 10.08(0.04) &  35.2(8.80) &  1.8(0.4) & 1.0789(0.4682) &   7.0(1.90) \\
 6446   &  SAd     &   645 & 10.5 & 3.5 & 1.5 & 13.88(0.05) &   4.9(0.49) &  8.0(0.8) & 0.1095(0.0470) &   6.7(0.95) \\
 6565   &  Irr     &   229 &  3.1 & 2.5 & 1.3 & 11.68(0.04) &  16.1(1.61) &  3.5(0.3) & 0.0388(0.0221) &  17.3(2.45) \\
 6572   &  Im      &   229 &  2.9 & 2.0 & 1.8 & 13.77(0.05) &   4.4(0.44) &  6.6(0.7) & 0.0097(0.0055) &   8.0(1.14) \\
 6595   &  SBb:    &   732 & 11.8 & 3.1 & 3.1 & 11.69(0.04) &  13.3(1.33) &  2.9(0.3) & 0.3855(0.1469) &   5.3(0.75) \\
 6618   &  SABcd:  &   739 & 11.7 & 1.7 & 1.5 & 12.55(0.04) &  11.1(1.11) &  5.3(0.5) & 0.3120(0.1199) &   9.8(1.39) \\
 6628   &  SAm     &   850 & 15.2 & 2.9 & 1.0 & 12.38(0.04) &   5.8(0.58) &  2.4(0.2) & 0.2811(0.0837) &   2.1(0.29) \\
 6644   &  SAc     &   993 & 16.9 & 4.3 & 1.4 & 10.39(0.04) &  67.3(6.73) &  4.4(0.4) & 4.2056(1.1351) &  20.6(2.92) \\
 6670   &  IBm     &   922 & 17.6 & 3.0 & 3.3 & 12.63(0.04) &  12.0(1.20) &  6.2(0.6) & 0.9595(0.2496) &  13.6(1.92) \\
 6778   &  SABc:   &   977 & 18.5 & 4.5 & 1.6 & 10.35(0.04) &  79.4(7.94) &  5.0(0.5) & 5.6248(1.3999) &  30.2(4.27) \\
 6781   &  SB0/a:  &   905 & 18.5 & 1.4 & 1.4 & 13.05(0.05) &   2.0(0.51) &  1.5(0.4) & 0.1432(0.0484) &   4.4(1.17) \\
 6782   &  Im      &   525 &  5.7 & 2.0 & 1.0 & 15.71(0.10) &   0.6(0.10) &  4.9(0.5) & 0.0047(0.0027) &   1.1(0.15) \\
 6797   &  SBd     &   961 & 18.2 & 1.9 & 1.1 & 12.33(0.04) &   6.8(0.68) &  2.7(0.3) & 0.4688(0.1184) &   7.7(1.09) \\
 6813   &  SAd:    &   954 & 17.7 & 2.6 & 1.0 & 12.60(0.04) &   4.9(0.49) &  2.5(0.2) & 0.3105(0.0804) &   5.5(0.78) \\
 6815   &  SAcd:   &   968 & 18.1 & 5.1 & 3.9 & 11.58(0.04) &   7.6(1.89) &  1.5(0.4) & 0.5124(0.1752) &   0.4(0.12) \\
 6818   &  SBb     &   819 & 14.0 & 2.0 & 2.0 & 13.58(0.05) &   2.2(0.22) &  2.8(0.3) & 0.0913(0.0294) &   0.5(0.07) \\
 6817   &  Im      &   243 &  2.9 & 4.1 & 2.7 & 13.90(0.05) &   3.6(0.36) &  6.0(0.6) & 0.0078(0.0044) &   2.0(0.28) \\
 6824   &  S0/a    &   906 & 16.9 & 1.7 & 2.1 & 12.28(0.04) &   1.2(0.30) &  0.4(0.2) & 0.0710(0.0251) &   1.5(0.40) \\
 6833   &  SABc    &   919 & 17.6 & 3.2 & 1.3 & 12.67(0.04) &  25.0(2.50) & 13.5(1.4) & 1.6029(0.4170) &  13.7(1.94) \\
 6869   &  SAbc:   &   807 & 13.9 & 2.9 & 1.7 & 10.70(0.04) &  60.2(6.02) &  5.3(0.5) & 2.4053(0.7800) &  54.3(7.68) \\
 6900   &  Sd      &   590 &  6.8 & 2.1 & 1.6 & 13.93(0.05) &   0.6(0.15) &  1.0(0.3) & 0.0056(0.0034) &   0.4(0.11) \\
 6904   &  SAbc    &   842 & 15.4 & 3.9 & 3.5 & 11.94(0.04) &   9.8(0.98) &  2.7(0.3) & 0.4702(0.1383) &   1.1(0.16) \\
 6917   &  SBm     &   910 & 17.0 & 3.5 & 1.8 & 13.20(0.05) &   1.9(0.47) &  1.6(0.4) & 0.1143(0.0404) &   1.4(0.37) \\
 6930   &  SABd    &   778 & 13.2 & 4.4 & 1.6 & 12.46(0.04) &   8.6(0.86) &  3.8(0.4) & 0.3173(0.1082) &   5.3(0.75) \\
 6956   &  SBm     &   917 & 17.1 & 2.2 & 1.0 & 14.43(0.06) &   1.1(0.11) &  2.9(0.3) & 0.0642(0.0171) &   0.8(0.12) \\
 6955   &  IBm:    &   905 & 16.5 & 5.0 & 1.9 & 13.32(0.05) &   2.0(0.50) &  2.0(0.5) & 0.1372(0.0492) &   1.1(0.30) \\
 6962   &  SABcd   &   784 & 11.8 & 2.3 & 1.2 & 11.89(0.04) &  13.8(1.38) &  3.6(0.4) & 0.3969(0.1512) &   9.1(1.29) \\
 6973   &  Sab:    &   701 &  9.7 & 2.6 & 2.2 & 11.28(0.04) &  17.2(1.72) &  2.6(0.3) & 0.3340(0.1555) &  20.4(2.88) \\
 7002   &  SBb:    &   932 & 17.5 & 2.5 & 1.3 & 12.09(0.04) &   6.2(1.56) &  2.0(0.5) & 0.4029(0.1401) &   3.5(0.94) \\
 7007   &  Sm:     &   774 &  9.4 & 1.7 & 1.1 & 14.14(0.06) &   0.3(0.10) &  0.7(0.2) & 0.0063(0.0034) &   0.3(0.10) \\
 7030   &  SABbc   &   725 & 10.6 & 5.2 & 1.3 & 10.05(0.04) &  66.3(6.63) &  3.2(0.3) & 1.5093(0.6413) &  14.2(2.01) \\
 7047   &  IAm     &   210 &  2.8 & 3.3 & 1.9 & 12.59(0.04) &   9.0(0.90) &  4.5(0.5) & 0.0182(0.0103) &   5.7(0.80) \\
 7045   &  SAc     &   769 &  9.1 & 4.1 & 2.4 & 10.70(0.04) &  22.3(5.57) &  2.0(0.5) & 0.3850(0.2111) &   2.2(0.61) \\
 7054   &  SBa:    &   913 & 17.6 & 4.4 & 2.6 & 11.04(0.04) &  10.8(2.70) &  1.3(0.3) & 0.6917(0.2398) &   3.1(0.83) \\
 7075   &  SABc:   &   752 & 12.6 & 2.8 & 3.5 & 11.86(0.04) &  21.7(2.17) &  5.6(0.6) & 0.7076(0.2526) &   2.4(0.34) \\
 7081   &  SABbc   &   760 & 12.9 & 5.8 & 2.6 & 10.08(0.04) &  89.1(8.91) &  4.4(0.4) & 3.0566(1.0660) &  14.7(2.09) \\
 7096   &  SABb    &   837 & 15.2 & 3.0 & 1.8 & 10.47(0.04) &  36.5(3.65) &  2.6(0.3) & 1.7388(0.5178) &  22.5(3.18) \\
 7134   &  SABc    &   609 &  6.8 & 4.0 & 1.1 & 11.27(0.04) &  15.0(1.50) &  2.2(0.2) & 0.1423(0.0809) &   5.9(0.83) \\
 7151   &  SABcd   &   265 &  3.3 & 6.0 & 4.6 & 11.32(0.04) &  23.2(2.32) &  3.6(0.4) & 0.0514(0.0292) &   9.8(1.39) \\
 7199   &  IAm     &   165 &  1.9 & 1.8 & 1.1 & 12.87(0.04) &   1.3(0.32) &  0.8(0.2) & 0.0012(0.0007) &   1.6(0.42) \\
 7215   &  SBdm    &   378 &  3.9 & 5.1 & 2.8 & 11.09(0.04) &  37.0(3.70) &  4.6(0.5) & 0.1181(0.0672) &   2.5(0.35) \\
 7216   &  SBcd:   &  -183 & 17.4 & 1.9 & 2.4 & 13.69(0.05) &   2.5(0.25) &  3.5(0.3) & 0.1657(0.0436) &   1.6(0.23) \\
 7232   &  Im      &   228 &  2.6 & 1.7 & 1.1 & 12.42(0.04) &   4.8(0.48) &  2.0(0.2) & 0.0085(0.0048) &   5.2(0.74) \\
 7261   &  SBdm    &   861 &  9.0 & 3.6 & 1.2 & 12.43(0.04) &  16.4(1.64) &  7.1(0.7) & 0.2835(0.1428) &   5.1(0.72) \\
 7267   &  Sdm:    &   472 &  6.6 & 2.1 & 2.6 & 13.73(0.05) &   1.2(0.31) &  1.8(0.4) & 0.0111(0.0068) &   1.1(0.29) \\
 7271   &  SBd:    &   546 &  7.0 & 2.0 & 3.3 & 13.89(0.05) &   1.1(0.28) &  1.9(0.5) & 0.0112(0.0069) &   0.4(0.12) \\
 7315   &  SABbc   &   867 & 17.6 & 2.1 & 1.6 & 11.18(0.04) &  13.2(3.30) &  1.8(0.4) & 0.8649(0.2998) &  10.8(2.91) \\
 7323   &  SABdm   &   517 &  6.8 & 5.0 & 1.3 & 11.29(0.04) &  11.6(2.90) &  1.8(0.4) & 0.1088(0.0667) &   1.9(0.51) \\
 7326   &  Im:     &  -164 & 17.4 & 1.9 & 3.2 & 15.29(0.10) &   1.2(0.12) &  7.3(0.7) & 0.0963(0.0253) &   2.3(0.32) \\
 7328   &  SB0/a:  &   890 & 10.4 & 2.9 & 1.3 & 10.82(0.04) &   9.0(2.24) &  0.9(0.2) & 0.2001(0.0981) &   9.0(2.42) \\
 7405   &  SB0/a   &   893 & 17.6 & 5.6 & 2.2 & 10.25(0.04) &  11.6(2.89) &  0.7(0.2) & 0.7765(0.2692) &   2.2(0.59) \\
 7414   &  SABdm:  &   232 &  2.4 & 1.7 & 1.1 & 12.03(0.04) &  25.5(2.55) &  7.6(0.8) & 0.0312(0.0178) &  32.3(4.56) \\
 7523   &  SBb     &   922 & 17.5 & 3.6 & 1.1 & 10.52(0.04) &  13.6(3.40) &  1.0(0.3) & 0.8852(0.3078) &   5.3(1.42) \\
 7539   &  SAc     &   716 &  8.2 & 3.6 & 1.8 &  9.61(0.04) &  72.1(7.21) &  2.3(0.2) & 0.9977(0.5551) &  30.0(4.24) \\
 7559   &  IBm     &   218 &  2.5 & 3.2 & 1.6 & 13.62(0.05) &   4.0(0.40) &  5.1(0.5) & 0.0063(0.0036) &   2.1(0.29) \\
 7561   &  SBa:    &   439 &  4.6 & 3.6 & 2.0 & 11.80(0.04) &   9.4(0.94) &  2.3(0.2) & 0.0410(0.0233) &   7.3(1.04) \\
 7622   &  SB0/a   &   508 & 18.8 & 3.8 & 2.9 & 11.14(0.04) &   6.1(1.53) &  0.8(0.2) & 0.4439(0.1490) &   1.8(0.47) \\
 7690   &  Im:     &   537 &  6.8 & 1.7 & 1.1 & 12.45(0.04) &   5.6(0.56) &  2.5(0.2) & 0.0681(0.0387) &   5.8(0.82) \\
 7753   &  SBb     &   486 & 18.1 & 5.4 & 1.3 &  9.95(0.04) &  18.7(4.68) &  0.8(0.2) & 1.3205(0.4514) &   2.9(0.78) \\
 7826   &  SABb    &   631 & 18.0 & 1.7 & 1.5 & 11.80(0.04) &   7.7(1.92) &  1.9(0.5) & 0.5352(0.1835) &   8.7(2.35) \\
 7866   &  IABm    &   359 &  4.2 & 3.4 & 1.1 & 13.80(0.05) &   3.3(0.33) &  5.0(0.5) & 0.0148(0.0084) &   3.4(0.48) \\
 7874   &  SABdm:  &   291 &  3.0 & 2.1 & 2.3 & 13.10(0.05) &   4.9(0.49) &  3.9(0.4) & 0.0092(0.0053) &   2.4(0.34) \\
 7901   &  SAc     &   805 & 17.8 & 4.0 & 1.5 & 10.33(0.04) &  37.1(3.71) &  2.3(0.2) & 2.4632(0.6344) &  21.4(3.02) \\
 7971   &  Sm:     &   467 &  6.5 & 2.2 & 1.0 & 13.45(0.05) &   3.0(0.30) &  3.3(0.3) & 0.0255(0.0145) &   1.8(0.25) \\
 7985   &  SABd    &   652 &  6.8 & 2.7 & 1.6 & 11.29(0.04) &  41.0(4.10) &  6.2(0.6) & 0.3985(0.2267) &  24.4(3.45) \\
 8024   &  IBm     &   376 &  4.1 & 3.0 & 1.4 & 13.46(0.05) &   4.2(0.42) &  4.7(0.5) & 0.0177(0.0101) &   2.4(0.34) \\
 8034   &  Im      &   915 & 17.6 & 1.7 & 1.9 & 13.61(0.05) &   6.3(0.63) &  8.0(0.8) & 0.5142(0.1338) &  10.1(1.42) \\
 8054   &  SAcd:   &   778 & 20.7 & 2.8 & 2.5 & 11.46(0.04) &  30.5(3.05) &  5.4(0.5) & 2.8064(0.6340) &  19.3(2.73) \\
 8098   &  SBm:    &   847 & 11.0 & 4.0 & 2.7 & 12.35(0.04) &  45.6(4.57) & 18.2(1.8) & 1.1130(0.4553) &   6.2(0.88) \\
 8116   &  SBc     &   969 & 17.1 & 2.2 & 1.0 & 10.96(0.04) &  38.0(3.80) &  4.2(0.4) & 2.3140(0.6179) &  33.2(4.70) \\
 8188   &  SAm     &   321 &  3.7 & 6.0 & 1.1 & 12.53(0.04) &   8.7(0.87) &  4.1(0.4) & 0.0242(0.0138) &   1.2(0.17) \\
 8201   &  Im      &    37 &  2.8 & 3.5 & 1.8 & 13.06(0.05) &   1.3(0.32) &  1.0(0.2) & 0.0025(0.0015) &   0.6(0.16) \\
 8256   &  SABbc   &   946 & 18.9 & 5.8 & 2.1 &  9.46(0.04) &  44.1(11.0) &  1.2(0.3) & 3.1988(1.0713) &   8.4(2.25) \\
 8303   &  IABm    &   944 & 18.9 & 2.2 & 1.2 & 13.11(0.05) &   7.8(0.78) &  6.3(0.6) & 0.7073(0.1728) &   6.2(0.87) \\
 8313   &  SBc     &   625 &  8.3 & 1.7 & 4.3 & 13.52(0.05) &   3.3(0.33) &  3.9(0.4) & 0.0463(0.0254) &   2.2(0.30) \\
 8320   &  IBm     &   195 &  2.4 & 3.6 & 2.6 & 12.82(0.04) &   6.0(0.60) &  3.7(0.4) & 0.0087(0.0049) &   3.8(0.53) \\
 8331   &  IAm     &   260 &  3.3 & 2.7 & 3.0 & 13.79(0.05) &   1.7(0.17) &  2.6(0.3) & 0.0046(0.0026) &   1.7(0.24) \\
 8396   &  SBd     &   946 & 18.6 & 1.7 & 3.4 & 13.06(0.05) &  10.7(1.07) &  8.3(0.8) & 0.7517(0.1862) &   9.8(1.39) \\
 8403   &  SBcd    &   965 & 19.1 & 4.0 & 1.4 & 11.49(0.04) &  29.0(2.90) &  5.2(0.5) & 2.1315(0.5159) &   7.7(1.09) \\
 8490   &  SAm     &   201 &  2.8 & 5.0 & 1.7 & 11.44(0.04) &  31.3(3.13) &  5.4(0.5) & 0.0496(0.0282) &   7.1(1.01) \\
 8508   &  IAm     &    62 &  2.7 & 1.7 & 1.7 & 13.41(0.05) &   3.6(0.36) &  3.8(0.4) & 0.0065(0.0037) &   4.7(0.66) \\
 8565   &  SABdm   &   232 &  3.1 & 1.7 & 1.2 & 13.24(0.05) &   4.0(0.40) &  3.6(0.4) & 0.0077(0.0044) &   3.3(0.47) \\
 8760   &  Im      &   193 &  2.3 & 2.2 & 3.1 & 13.80(0.05) &   0.7(0.19) &  1.1(0.3) & 0.0010(0.0006) &   0.5(0.14) \\
 8837   &  IBm     &   144 &  3.9 & 4.3 & 3.3 & 12.07(0.04) &   4.5(1.12) &  1.4(0.3) & 0.0172(0.0105) &   1.4(0.36) \\
 8839   &  Im      &   957 & 20.5 & 4.0 & 1.5 & 14.01(0.06) &   0.2(0.10) &  0.5(0.2) & 0.0271(0.0140) &   2.6(1.30) \\
 9013   &  SAcd    &   273 &  3.8 & 4.8 & 1.1 & 10.80(0.04) &  22.8(2.28) &  2.2(0.2) & 0.0664(0.0378) &   5.6(0.79) \\
 9018   &  SAm     &   304 &  4.3 & 1.7 & 1.3 & 13.64(0.05) &   7.0(0.70) &  9.2(0.9) & 0.0261(0.0148) &   8.2(1.15) \\
 9128   &  Im      &   154 &  2.2 & 1.7 & 1.3 & 13.86(0.05) &   0.4(0.10) &  0.6(0.2) & 0.0005(0.0003) &   0.3(0.10) \\
 9179   &  SABd    &   305 &  4.5 & 5.8 & 1.6 & 11.21(0.04) &  28.5(2.86) &  4.0(0.4) & 0.1179(0.0671) &   5.3(0.75) \\
 9211   &  Im:     &   686 & 11.2 & 1.7 & 1.2 & 15.00(0.10) &   3.6(0.91) & ~~---~~   & 0.1129(0.0522) &  ~~---~~    \\
 9219   &  Im:     &   663 & 10.2 & 2.6 & 2.0 & 13.17(0.05) &   4.7(0.47) &  4.0(0.4) & 0.1220(0.0539) &   8.3(1.18) \\
 9240   &  IAm     &   150 &  3.7 & 1.8 & 1.0 & 12.98(0.04) &   3.7(0.37) &  2.7(0.3) & 0.0128(0.0073) &   3.5(0.49) \\
 9405   &  Im      &   222 &  3.2 & 1.7 & 2.8 & 14.24(0.06) &   0.8(0.20) &  1.8(0.4) & 0.0020(0.0012) &   0.9(0.23) \\
 9649   &  SBb     &   447 &  7.7 & 3.7 & 1.7 & 11.77(0.04) &   8.7(0.87) &  2.0(0.2) & 0.1083(0.0616) &   3.1(0.44) \\
 9753   &  SAbc:   &   772 & 13.8 & 4.2 & 3.2 & 11.42(0.04) &  16.7(1.67) &  2.8(0.3) & 0.6413(0.2094) &   8.9(1.26) \\
 9769   &  SABdm:  &   841 & 15.1 & 2.7 & 1.4 & 13.34(0.05) &   1.9(0.46) &  1.9(0.5) & 0.0861(0.0325) &   1.0(0.27) \\
 9866   &  SAbc    &   435 &  7.4 & 2.2 & 2.2 & 11.78(0.04) &   9.6(0.96) &  2.3(0.2) & 0.1089(0.0620) &   7.4(1.05) \\
 9906   &  Sc      &   656 & 11.6 & 3.3 & 1.3 & 12.18(0.04) &  13.0(1.30) &  4.4(0.4) & 0.3515(0.1363) &  49.7(7.03) \\
10075   &  SAcd    &   835 & 14.7 & 5.4 & 2.6 & 10.96(0.04) &  27.4(2.74) &  3.1(0.3) & 1.1993(0.3686) &   4.6(0.66) \\
10310   &  SBm     &   716 & 12.7 & 2.8 & 1.3 & 13.08(0.05) &   8.0(0.80) &  6.3(0.6) & 0.2610(0.0924) &   4.0(0.57) \\
10445   &  SBc     &   963 & 16.9 & 2.8 & 1.6 & 12.89(0.04) &   7.8(0.78) &  5.1(0.5) & 0.4732(0.1277) &  10.7(1.52) \\
10521   &  SAc     &   852 & 15.0 & 3.0 & 2.3 & 11.24(0.04) &  36.7(3.67) &  5.3(0.5) & 1.6847(0.5079) &  17.3(2.45) \\
10606   &  SBcd:   &   919 & 15.9 & 3.6 & 2.4 & 14.21(0.06) &   3.9(0.39) &  8.7(0.9) & 0.2017(0.0576) &   1.0(0.14) \\
10736   &  SABdm   &   490 &  8.6 & 3.1 & 3.4 & 13.92(0.05) &   1.0(0.24) &  1.6(0.4) & 0.0153(0.0088) &   0.6(0.16) \\
10806   &  SBdm    &   932 & 15.7 & 2.1 & 2.3 & 12.76(0.04) &   6.7(0.67) &  3.9(0.4) & 0.3469(0.1002) &   2.2(0.31) \\
11300   &  SBcd    &   490 &  8.4 & 3.8 & 2.9 & 12.29(0.04) &   7.6(0.76) &  2.9(0.3) & 0.1241(0.0673) &   1.1(0.16) \\
12048   &  IBm     &   986 & 12.2 & 2.1 & 1.2 & 12.15(0.04) &  16.6(1.66) &  5.6(0.6) & 0.7043(0.2596) &  16.2(2.30) \\
12082   &  Sm      &   802 & 10.1 & 2.6 & 1.2 & 13.47(0.05) &   2.7(0.27) &  3.1(0.3) & 0.0694(0.0310) &   1.5(0.21) \\
12101   &  SAd     &   786 &  9.9 & 2.2 & 2.0 & 12.63(0.04) &   7.2(0.72) &  3.7(0.4) & 0.1671(0.0762) &   7.0(0.99) \\
12613   &  Im      &  -183 &  1.0 & 5.0 & 1.9 & 14.89(0.06) &   0.1(0.10) &  0.4(0.2) & .00002(.00002) &   .01(0.01) \\
12732   &  Sm:     &   749 &  8.9 & 3.0 & 1.1 & 13.70(0.05) &   6.3(0.63) &  8.8(0.9) & 0.1220(0.0622) &   5.7(0.80) \\
12754   &  SBcd    &   751 &  8.9 & 4.4 & 1.5 & 11.45(0.04) &  26.5(2.64) &  4.6(0.5) & 0.4936(0.2516) &   6.9(0.98) \\
\hline							          	                                                       
12893   &  SAdm    &  1108 & 12.5 & 1.7 & 1.0 & 13.33(0.05) &   0.4(0.10) &  0.4(0.2) & 0.0121(0.0051) &   0.4(0.11) \\
  763   &  SABm    &  1162 & 12.7 & 5.1 & 1.4 & 11.38(0.04) &  19.4(1.94) &  3.2(0.3) & 0.6567(0.2326) &   9.9(1.40) \\
 2053   &  Im      &  1029 & 12.7 & 2.0 & 2.0 & 14.40(0.06) &   0.4(0.10) &  1.0(0.3) & 0.0205(0.0087) &   0.4(0.10) \\
 2210   &  SBc     &  1211 & 13.9 & 4.9 & 1.1 & 11.57(0.04) &  26.3(2.63) &  5.1(0.5) & 1.0953(0.3552) &   6.1(0.86) \\
 2275   &  Sm:     &  1025 & 11.9 & 7.5 & 1.4 & 17.52(0.10) &   0.2(0.10) & 11.1(1.1) & 0.0081(0.0031) &   0.2(0.10) \\
 2302   &  SBm:    &  1104 & 12.8 & 4.8 & 1.3 & 14.37(0.06) &   8.0(0.80) & 20.5(2.1) & 0.3119(0.1096) &   7.1(1.00) \\
 2855   &  SABc    &  1202 & 17.5 & 4.4 & 2.2 & 11.11(0.04) &  26.5(2.65) &  3.4(0.3) & 9.7830(2.5583) &   9.4(1.33) \\
 3384   &  Sm:     &  1089 & 17.0 & 1.7 & 1.0 & 14.68(0.06) &   1.3(0.13) &  4.6(0.5) & 0.1229(0.0330) &   1.8(0.25) \\
 3403   &  SBcd    &  1264 & 19.2 & 2.7 & 3.4 & 13.02(0.05) &   3.3(0.33) &  2.5(0.2) & 0.4179(0.1007) &   1.3(0.18) \\
 3439   &  SABc:   &  1494 & 22.3 & 3.3 & 2.7 & 12.81(0.04) &   2.8(0.71) &  1.7(0.4) & 0.3503(0.1094) &   1.1(0.31) \\
 3574   &  SAcd    &  1441 & 21.8 & 4.2 & 1.2 & 12.21(0.04) &   6.1(0.61) &  2.2(0.2) & 0.6487(0.1404) &   3.4(0.47) \\
 3580   &  SAa:    &  1201 & 18.8 & 3.6 & 2.3 & 12.44(0.04) &   6.7(0.67) &  2.9(0.3) & 0.5300(0.1301) &   6.5(0.92) \\
 4097   &  SAa     &  1442 & 22.5 & 1.9 & 1.3 & 11.33(0.04) &   4.6(1.15) &  0.7(0.2) & 0.5461(0.1701) &   8.6(2.31) \\
 4121   &  Sm:     &  1092 & 18.0 & 2.0 & 2.5 & 15.14(0.10) &   0.4(0.10) &  1.9(0.5) & 0.0243(0.0083) &   0.3(0.08) \\
 4637   &  SAB0/a  &  1404 & 21.8 & 4.9 & 1.2 & 11.20(0.04) &   2.0(0.49) &  0.3(0.2) & 0.1961(0.0618) &   ~~---~~   \\
 4781   &  Scd:    &  1443 & 24.3 & 1.7 & 2.8 & 14.40(0.06) &   1.4(0.14) &  3.7(0.4) & 0.1794(0.0356) &   1.2(0.17) \\
 4779   &  SAc:    &  1289 & 21.3 & 3.2 & 1.9 & 11.15(0.04) &  15.9(1.59) &  2.1(0.2) & 1.5585(0.3438) &   3.8(0.54) \\
 5349   &  Sdm:    &  1381 & 24.4 & 2.5 & 2.8 & 13.30(0.05) &   3.3(0.33) &  3.2(0.3) & 0.4020(0.0795) &   1.5(0.21) \\
 5393   &  SBdm:   &  1448 & 25.5 & 1.9 & 1.7 & 13.54(0.05) &   2.7(0.27) &  3.2(0.3) & 0.3498(0.0669) &   3.4(0.49) \\
 5589   &  SBcd    &  1154 & 20.3 & 3.0 & 1.6 & 12.28(0.04) &  10.1(1.01) &  3.8(0.4) & 0.8439(0.1938) &   3.7(0.52) \\
 5731   &  SAab    &  1408 & 25.9 & 1.9 & 1.1 & 11.36(0.04) &   4.6(1.15) &  0.7(0.2) & 0.6457(0.1918) &  11.0(2.96) \\
 6023   &  Sd      &  1334 & 25.5 & 1.9 & 2.4 & 12.49(0.04) &  11.3(1.13) &  5.1(0.5) & 1.5203(0.2908) &  14.2(2.00) \\
 6077   &  SBb:    &  1434 & 27.4 & 2.3 & 1.1 & 11.82(0.04) &  10.5(1.05) &  2.6(0.3) & 1.6135(0.2926) &  13.1(1.86) \\
 6112   &  Sd      &  1036 & 22.0 & 2.2 & 2.8 & 13.21(0.05) &   1.9(0.47) &  1.6(0.4) & 0.1854(0.0582) &   0.8(0.22) \\
 6923   &  Im:     &  1066 & 19.8 & 2.0 & 2.5 & 12.91(0.04) &   4.4(0.44) &  3.0(0.3) & 0.4532(0.1063) &   5.0(0.71) \\
 9036   &  SAm:    &  1390 & 24.0 & 1.5 & 1.7 & 12.94(0.04) &   5.0(0.50) &  3.5(0.3) & 0.5984(0.1199) &   4.3(0.60) \\
 9465   &  SABdm   &  1491 & 26.4 & 2.3 & 1.9 & 13.19(0.05) &   4.7(0.47) &  4.1(0.4) & 0.6677(0.1244) &   3.7(0.52) \\
 9645   &  SABb    &  1359 & 24.5 & 3.5 & 1.8 & 11.10(0.04) &  12.3(3.07) &  1.6(0.4) & 1.6406(0.4959) &   6.2(1.68) \\
 9824   &  SBbc    &  1480 & 25.6 & 4.5 & 1.1 & 10.79(0.04) &  23.6(2.36) &  2.2(0.2) & 3.3419(0.6372) &   9.3(1.31) \\
 9935   &  SBd     &  1447 & 24.8 & 4.5 & 1.3 & 12.10(0.04) &  11.6(1.16) &  3.7(0.4) & 1.6163(0.3157) &   3.4(0.48) \\
 9987   &  SBd:    &  1108 & 20.0 & 2.9 & 3.6 & 12.24(0.04) &   6.8(0.68) &  2.5(0.2) & 0.5963(0.1387) &   2.8(0.39) \\
10470   &  SBbc    &  1362 & 21.2 & 3.0 & 1.2 & 11.10(0.04) &  29.8(2.98) &  3.8(0.4) & 2.9252(0.6478) &  31.9(4.52) \\
10546   &  SABcd   &  1280 & 20.4 & 2.8 & 1.6 & 12.51(0.04) &   8.2(0.82) &  3.8(0.4) & 0.7296(0.1669) &   3.8(0.53) \\
10564   &  SBd     &  1129 & 18.4 & 2.6 & 2.2 & 13.27(0.05) &   5.3(0.53) &  4.9(0.5) & 0.3802(0.0951) &   2.6(0.37) \\
10762   &  SA0/a   &  1198 & 19.1 & 3.7 & 1.1 & 10.69(0.04) &   3.5(0.89) &  0.3(0.2) & 0.2869(0.0956) &   2.3(0.61) \\
10792   &  Im      &  1233 & 19.4 & 1.8 & 1.0 & 14.37(0.06) &   1.6(0.16) &  4.1(0.4) & 0.1592(0.0380) &   1.2(0.17) \\
10876   &  Scd:    &  1164 & 18.6 & 2.7 & 3.4 & 12.87(0.04) &   5.7(0.57) &  3.7(0.4) & 0.4271(0.1058) &   1.6(0.22) \\
10897   &  SAc     &  1324 & 20.5 & 2.5 & 1.1 & 11.40(0.04) &  19.1(1.91) &  3.2(0.3) & 1.7315(0.3944) &  12.1(1.71) \\
11218   &  SAc     &  1484 & 22.2 & 3.6 & 2.1 & 10.72(0.04) &  37.8(3.78) &  3.4(0.3) & 4.2349(0.9029) &  11.2(1.59) \\
11557   &  SABdm   &  1390 & 19.7 & 2.2 & 1.3 & 12.62(0.04) &   6.8(0.68) &  3.5(0.3) & 0.9242(0.2178) &   4.8(0.67) \\
11604   &  SABbc   &  1424 & 20.2 & 3.9 & 1.2 & 10.54(0.04) &  21.7(5.43) &  1.6(0.4) & 4.2698(1.3881) &   6.6(1.77) \\
11782   &  SBm     &  1112 & 13.4 & 2.3 & 1.6 & 13.53(0.05) &   2.6(0.26) &  3.1(0.3) & 0.1020(0.0343) &   1.3(0.18) \\
11820   &  Sm      &  1104 & 13.3 & 2.0 & 1.1 & 14.43(0.06) &   1.0(0.10) &  2.8(0.3) & 0.0479(0.0162) &   1.2(0.17) \\
11861   &  SABdm   &  1481 & 20.9 & 3.5 & 1.3 & 12.51(0.04) &  11.1(1.11) &  5.1(0.5) & 4.2181(0.9452) &   4.9(0.69) \\
11868   &  SBm     &  1093 & 13.2 & 1.9 & 1.2 & 13.10(0.05) &   2.3(0.57) &  1.8(0.5) & 0.0956(0.0393) &   1.4(0.37) \\
11872   &  SABb    &  1150 & 13.2 & 2.7 & 1.4 & 10.68(0.04) &  13.5(3.38) &  1.2(0.3) & 0.5739(0.2358) &  16.5(4.44) \\
12043   &  S0/a    &  1008 & 12.4 & 1.6 & 2.7 & 13.06(0.05) &   3.1(0.31) &  2.4(0.2) & 0.1084(0.0393) &   3.2(0.45) \\
\hline							          	                                                       
  806   &  SABcd:  &  1761 & 19.3 & 3.5 & 1.2 & 11.94(0.04) &  19.4(1.94) &  5.3(0.5) & 1.5643(0.3752) &   7.9(1.12) \\
 1356   &  SABa    &  1733 & 19.3 & 2.3 & 1.1 & 16.11(0.10) &   0.2(0.10) &  2.4(0.2) & 0.0150(0.0075) &   0.2(0.10) \\
 1670   &  Sm:     &  1601 & 18.1 & 2.2 & 1.0 & 14.43(0.06) &   1.1(0.11) &  3.0(0.3) & 0.0792(0.0201) &   0.9(0.13) \\
 1736   &  SABc    &  1562 & 17.6 & 5.2 & 1.4 & 11.00(0.04) &  19.6(1.96) &  2.3(0.2) & 1.3405(0.3488) &   6.1(0.86) \\
 1888   &  SABc:   &  1507 & 17.5 & 3.8 & 1.7 & 11.56(0.04) &   9.8(2.44) &  1.9(0.5) & 1.3707(0.4766) &   2.7(0.72) \\
 1954   &  SABc    &  1608 & 18.1 & 3.3 & 1.3 & 12.20(0.04) &   7.8(0.78) &  2.7(0.3) & 0.5464(0.1387) &   4.6(0.65) \\
 2045   &  Sab     &  1543 & 18.5 & 4.0 & 2.0 & 10.94(0.04) &  25.4(2.54) &  2.8(0.3) & 2.2182(0.5521) &  15.4(2.18) \\
 2183   &  Sa:     &  1545 & 18.6 & 1.9 & 1.3 & 11.93(0.04) &   8.5(0.85) &  2.3(0.2) & 0.8304(0.2057) &  19.1(2.70) \\
 2245   &  SABc    &  1519 & 17.3 & 3.8 & 1.4 & 10.78(0.04) &  50.2(5.02) &  4.7(0.5) & 3.2074(0.8475) &  21.5(3.04) \\
 2345   &  SBm:    &  1506 & 17.2 & 3.5 & 1.2 & 14.06(0.06) &   2.3(0.23) &  4.5(0.5) & 0.1615(0.0429) &   1.7(0.23) \\
 2392   &  Scd     &  1548 & 19.0 & 1.9 & 3.8 & 14.53(0.06) &   0.4(0.11) &  1.3(0.3) & 0.0404(0.0135) &   0.7(0.20) \\
 2729   &  S0/a    &  1940 & 26.1 & 3.5 & 1.4 & 13.24(0.05) &   1.5(0.38) &  1.4(0.3) & 1.3883(0.4114) &   0.7(0.19) \\
 3546   &  SBa:    &  1871 & 26.9 & 2.6 & 1.9 & 11.16(0.04) &   6.1(1.53) &  0.8(0.2) & 1.0344(0.3039) &   2.4(0.66) \\
 3496   &  Im:     &  1581 & 23.4 & 2.1 & 1.4 & 15.42(0.10) &   0.6(0.10) &  4.3(0.4) & 0.1062(0.0217) &   0.9(0.13) \\
 3598   &  IBm     &  1991 & 28.4 & 2.0 & 1.7 & 13.07(0.05) &   6.4(0.64) &  4.9(0.5) & 1.4924(0.2639) &   5.0(0.70) \\
 3685   &  SBb     &  1797 & 26.3 & 3.3 & 1.2 & 11.69(0.04) &   7.3(1.82) &  1.6(0.4) & 1.1441(0.3383) &   7.3(1.96) \\
 3826   &  SABd    &  1733 & 25.7 & 3.5 & 1.2 & 12.63(0.04) &   3.5(0.87) &  1.8(0.5) & 0.5355(0.1594) &   1.4(0.36) \\
 4238   &  SBd     &  1544 & 23.5 & 2.6 & 1.7 & 12.97(0.04) &   3.2(0.32) &  2.3(0.2) & 0.3783(0.0771) &   2.1(0.29) \\
 4533   &  Sdm     &  1939 & 29.8 & 1.9 & 2.7 & 12.93(0.04) &   3.6(0.36) &  2.5(0.2) & 0.6875(0.1176) &   3.4(0.47) \\
 4659   &  SAdm:   &  1756 & 27.6 & 1.7 & 3.4 & 14.59(0.06) &   0.4(0.11) &  1.3(0.3) & 0.0675(0.0197) &   0.4(0.10) \\
 4680   &  Sbc     &  1631 & 26.2 & 1.7 & 3.4 & 12.63(0.04) &   6.9(0.69) &  3.6(0.4) & 0.9718(0.1821) &   2.3(0.33) \\
 4708   &  SBb:    &  1815 & 28.5 & 3.0 & 2.0 & 11.66(0.04) &   9.6(0.96) &  2.0(0.2) & 1.6147(0.2848) &   4.1(0.59) \\
 4922   &  SAm     &  1991 & 30.7 & 3.5 & 1.8 & 13.18(0.05) &   0.5(0.12) &  0.4(0.2) & 0.0935(0.0266) &   0.2(0.05) \\
 5015   &  SABdm   &  1650 & 27.3 & 1.7 & 1.1 & 13.74(0.05) &   1.1(0.27) &  1.5(0.4) & 0.1789(0.0523) &   1.1(0.29) \\
 5688   &  SBm:    &  1920 & 29.2 & 3.5 & 1.8 & 13.57(0.05) &   3.5(0.35) &  4.3(0.4) & 0.6340(0.1099) &   1.2(0.17) \\
 5717   &  SABbc:  &  1686 & 26.8 & 2.0 & 2.0 & 12.50(0.04) &   6.2(0.62) &  2.9(0.3) & 0.9099(0.1677) &   9.6(1.35) \\
 6506   &  SBd     &  1580 & 29.1 & 1.7 & 3.0 & 14.74(0.06) &   0.5(0.12) &  1.7(0.4) & 0.0852(0.0245) &   0.4(0.11) \\
 7656   &  SAa     &  1774 & 32.3 & 1.7 & 1.1 & 11.50(0.04) &   3.0(0.76) &  0.6(0.2) & 0.6595(0.1851) &   2.1(0.55) \\
 9576   &  SABd    &  1567 & 27.4 & 3.0 & 1.2 & 12.09(0.04) &  16.6(1.66) &  5.2(0.5) & 2.7147(0.4924) &   5.8(0.82) \\
 9579   &  SBc     &  1681 & 28.8 & 4.2 & 4.2 & 10.71(0.04) &  27.8(2.78) &  2.5(0.2) & 5.0033(0.8760) &   9.5(1.34) \\
 9926   &  SAc     &  1958 & 31.2 & 2.8 & 1.4 & 10.97(0.04) &  29.1(2.92) &  3.3(0.3) & 6.3637(1.0554) &  20.9(2.96) \\
10805   &  SBm     &  1554 & 23.8 & 1.7 & 1.1 & 14.10(0.06) &   1.6(0.16) &  3.2(0.3) & 0.2331(0.0470) &   2.0(0.29) \\
11124   &  SBcd    &  1613 & 23.7 & 2.5 & 1.1 & 12.38(0.04) &   4.8(0.48) &  2.0(0.2) & 0.5879(0.1189) &   3.2(0.45) \\
11238   &  SB0/a   &  1821 & 26.2 & 2.9 & 1.3 & 11.36(0.04) &   1.0(0.24) &  0.2(0.2) & 0.1509(0.0447) &   0.5(0.15) \\
11283   &  SBdm    &  1959 & 27.5 & 1.8 & 1.2 & 12.79(0.04) &  10.6(1.06) &  6.4(0.6) & 1.8206(0.3293) &   9.9(1.40) \\
11331   &  Sm:     &  1554 & 22.9 & 1.5 & 1.3 & 14.58(0.06) &   0.9(0.10) &  3.0(0.3) & 0.1156(0.0240) &   2.3(0.33) \\
11332   &  SBd     &  1558 & 23.0 & 2.5 & 2.8 & 12.88(0.04) &  10.2(1.02) &  6.7(0.7) & 1.2521(0.2594) &   4.0(0.56) \\
11921   &  IBm     &  1678 & 19.6 & 1.7 & 2.4 & 13.49(0.05) &   3.8(0.38) &  4.3(0.4) & 0.4092(0.0968) &   6.8(0.96) \\
11944   &  Im:     &  1734 & 20.3 & 2.4 & 3.0 & 15.44(0.10) &   1.0(0.10) &  7.0(0.7) & 0.1138(0.0261) &   5.0(0.71) \\
12178   &  SABdm   &  1931 & 21.8 & 3.1 & 1.9 & 12.19(0.04) &   8.3(0.83) &  2.9(0.3) & 0.9887(0.2139) &   2.2(0.32) \\
\hline					   	          	                       	                               
   19   &  SAbc:   &  2309 & 26.1 & 3.6 & 3.6 & 11.40(0.04) &   9.2(3.23) &  1.5(0.5) & 1.4206(0.5462) &   4.0(1.46) \\
  858   &  SAb     &  2374 & 26.3 & 3.1 & 1.6 & 11.59(0.04) &  14.5(2.17) &  2.9(0.4) & 2.1483(0.4678) &   9.8(1.76) \\
  859   &  SAB0/a  &  2134 & 24.0 & 2.4 & 1.6 & 12.14(0.04) &   3.6(1.26) &  1.2(0.4) & 0.5082(0.1985) &   5.9(2.16) \\
  895   &  Sdm     &  2247 & 25.0 & 2.0 & 2.9 & 13.23(0.05) &   4.1(0.62) &  3.7(0.6) & 0.5869(0.1315) &   2.5(0.46) \\
  907   &  SAb     &  2272 & 24.0 & 5.6 & 1.3 &  9.87(0.04) &   5.8(2.03) &  0.2(0.2) & 0.7085(0.2768) &   2.2(0.81) \\
  914   &  SABcd:  &  2338 & 25.8 & 4.3 & 2.9 & 12.13(0.04) &   8.5(1.27) &  2.8(0.4) & 1.2032(0.2648) &   1.9(0.33) \\
 1211   &  Im:     &  2408 & 27.1 & 2.3 & 1.2 & 14.62(0.06) &   0.1(0.10) &  0.5(0.2) & 0.0300(0.0300) &   0.2(0.20) \\
 1554   &  SAc:    &  2101 & 23.9 & 2.8 & 2.3 & 12.11(0.04) &   4.7(1.16) &  1.5(0.4) & 0.5934(0.1809) &   1.7(0.46) \\
 3504   &  SABcd   &  2100 & 29.5 & 2.8 & 1.1 & 12.00(0.04) &  14.9(1.49) &  4.3(0.4) & 3.1049(0.5347) &   8.5(1.20) \\
 3530   &  SBcd:   &  2101 & 29.6 & 2.5 & 1.9 & 12.49(0.04) &   6.8(0.68) &  3.1(0.3) & 1.3995(0.2404) &   2.2(0.32) \\
 3522   &  S0/a    &  2132 & 30.0 & 1.8 & 1.6 & 13.99(0.05) &   1.0(0.35) &  1.8(0.6) & 0.2093(0.0787) &   0.9(0.33) \\
 3653   &  SABbc   &  2222 & 31.1 & 3.7 & 1.5 & 11.20(0.04) &  18.5(2.78) &  2.6(0.4) & 4.1224(0.8258) &   9.7(1.75) \\
 3740   &  SABc    &  2410 & 33.2 & 2.8 & 1.0 & 11.08(0.04) &  48.7(7.31) &  6.1(0.9) & 13.482(2.6248) &  24.6(4.44) \\
 3834   &  SABc:   &  2042 & 29.1 & 3.5 & 2.5 & 12.56(0.04) &   6.4(0.64) &  3.1(0.3) & 1.1374(0.1977) &   2.3(0.32) \\
 3994   &  SABab:  &  2080 & 29.5 & 2.7 & 1.9 & 12.11(0.04) &   3.4(0.84) &  1.1(0.3) & 0.6951(0.1992) &   2.1(0.56) \\
 4066   &  Scd:    &  2296 & 32.2 & 1.7 & 1.0 & 13.26(0.05) &   3.3(0.50) &  3.1(0.5) & 0.7516(0.1483) &   4.5(0.81) \\
 4260   &  Im:     &  2254 & 32.8 & 1.6 & 1.1 & 13.81(0.05) &   4.7(0.70) &  7.2(1.1) & 1.3670(0.2675) &   6.5(1.17) \\
 4273   &  SBb     &  2471 & 35.4 & 2.7 & 2.3 & 11.84(0.04) &   9.4(1.42) &  2.4(0.4) & 2.7481(0.5214) &   5.3(0.95) \\
 4270   &  SABbc   &  2479 & 35.1 & 1.6 & 1.5 & 13.28(0.05) &   0.3(0.10) &  0.3(0.2) & 0.0726(0.0268) &   0.4(0.15) \\
 4375   &  SABc:   &  2061 & 30.9 & 2.5 & 1.5 & 11.92(0.04) &   5.9(1.48) &  1.6(0.4) & 1.2345(0.3500) &   3.6(0.96) \\
 4362   &  SA0/a   &  2344 & 33.1 & 1.9 & 1.4 & 11.96(0.04) &   2.8(0.98) &  0.8(0.3) & 0.6446(0.2395) &   5.4(1.95) \\
 4393   &  SBc     &  2124 & 31.5 & 2.2 & 1.5 & 13.26(0.05) &   5.8(0.87) &  5.4(0.8) & 1.2495(0.2489) &   7.9(1.42) \\
 4390   &  SBd     &  2169 & 31.1 & 1.9 & 1.2 & 13.69(0.05) &   2.2(0.33) &  3.0(0.4) & 0.4454(0.0892) &   2.9(0.52) \\
 4444   &  SBcd    &  2081 & 31.3 & 1.5 & 1.7 & 13.31(0.05) &   3.2(0.32) &  3.1(0.3) & 0.6724(0.1113) &   8.2(1.17) \\
 4469   &  SBcd    &  2094 & 31.5 & 2.1 & 1.0 & 12.50(0.04) &  10.1(1.01) &  4.6(0.5) & 2.1677(0.3573) &  11.2(1.58) \\
 4484   &  SBb:    &  2135 & 32.1 & 2.0 & 1.5 & 11.94(0.04) &   6.0(2.10) &  1.6(0.6) & 1.3340(0.4974) &   5.0(1.82) \\
 4541   &  Sa      &  2060 & 31.4 & 3.6 & 3.0 & 11.37(0.04) &   1.5(0.36) &  0.2(0.2) & 0.2992(0.0845) &   2.0(0.54) \\
 4574   &  SBb     &  2160 & 31.1 & 2.7 & 1.7 & 11.26(0.04) &  21.1(3.17) &  3.1(0.5) & 4.2332(0.8480) &  20.9(3.76) \\
 5056   &  Sa      &  2146 & 33.4 & 2.2 & 2.0 & 11.80(0.04) &   2.6(0.91) &  0.6(0.2) & 0.6646(0.2466) &   4.0(1.44) \\
 5102   &  SABb:   &  2432 & 36.9 & 1.7 & 2.8 & 12.55(0.04) &   6.1(0.91) &  2.9(0.4) & 1.7413(0.3253) &   3.6(0.65) \\
 6517   &  Sbc     &  2491 & 38.8 & 1.7 & 1.8 & 13.19(0.05) &   2.2(0.77) &  1.9(0.7) & 0.6982(0.2553) &   3.1(1.13) \\
 7563   &  Im      &  2350 & 38.4 & 2.1 & 2.1 & 13.69(0.05) &   1.7(0.26) &  2.4(0.4) & 0.6662(0.1227) &   2.3(0.42) \\
11113   &  SABd    &  2331 & 32.0 & 1.9 & 1.1 & 13.68(0.05) &   1.7(0.25) &  2.3(0.3) & 0.4412(0.0873) &   1.6(0.28) \\
12221   &  SAd     &  2057 & 28.4 & 2.2 & 2.8 & 13.09(0.05) &   3.4(0.34) &  2.7(0.3) & 1.0863(0.1921) &   0.9(0.13) \\
12270   &  SABd    &  2116 & 24.0 & 1.8 & 1.0 & 12.65(0.04) &   3.3(1.16) &  1.8(0.6) & 0.4141(0.1618) &   3.2(1.16) \\
12294   &  SAbc    &  2194 & 25.0 & 2.3 & 2.3 & 11.29(0.04) &  29.5(4.42) &  4.4(0.7) & 4.1612(0.9322) &  14.5(2.61) \\
12343   &  SBc     &  2381 & 26.9 & 4.0 & 1.2 & 10.83(0.04) &  35.5(5.32) &  3.5(0.5) & 6.6349(1.4274) &  10.3(1.86) \\
12350   &  Sm      &  2140 & 24.3 & 2.8 & 3.1 & 13.72(0.05) &   1.5(0.22) &  2.1(0.3) & 0.2236(0.0509) &   0.4(0.07) \\
\hline						           	                       	                               
  550   &  Sb      &  2674 & 30.3 & 1.8 & 2.2 & 13.58(0.05) &   1.9(0.29) &  2.4(0.4) & 0.3924(0.0796) &   3.6(0.66) \\
 1192   &  Sb      &  2988 & 33.6 & 3.3 & 1.9 & 12.17(0.04) &   7.8(1.18) &  2.7(0.4) & 1.9408(0.3760) &   7.6(1.38) \\
 1276   &  SBdm    &  2749 & 31.3 & 1.9 & 2.1 & 13.15(0.05) &   2.7(0.41) &  2.3(0.3) & 0.6411(0.1280) &   1.5(0.27) \\
 1305   &  SAbc    &  2665 & 30.5 & 3.7 & 1.4 & 11.43(0.04) &   4.1(1.42) &  0.7(0.2) & 0.8810(0.3306) &   1.5(0.54) \\
 1313   &  SABc    &  2931 & 33.5 & 3.0 & 1.9 & 13.30(0.05) &   3.6(0.54) &  3.5(0.5) & 0.9472(0.1837) &   4.0(0.72) \\
 1378   &  SBa:    &  2935 & 37.6 & 3.5 & 2.5 & 12.15(0.04) &   1.2(0.41) &  0.4(0.2) & 1.0700(0.3923) &   0.4(0.14) \\
 1547   &  IBm     &  2640 & 30.3 & 2.2 & 1.0 & 13.25(0.05) &   2.7(0.40) &  2.5(0.4) & 0.7895(0.1600) &   1.1(0.19) \\
 2081   &  SABcd   &  2616 & 29.6 & 2.5 & 1.4 & 13.76(0.05) &   0.9(0.32) &  1.3(0.5) & 0.1673(0.0631) &   0.8(0.30) \\
 2124   &  SBa:    &  2631 & 29.7 & 2.6 & 1.0 & 11.83(0.04) &   1.9(0.65) &  0.5(0.2) & 0.3492(0.1315) &   1.9(0.68) \\
 2247   &  SBbc    &  2758 & 31.3 & 4.5 & 2.3 & 11.85(0.04) &   6.1(2.14) &  1.5(0.5) & 1.2730(0.4761) &   1.3(0.46) \\
 2603   &  Im      &  2516 & 33.7 & 1.7 & 1.2 & 14.59(0.06) &   0.6(0.21) &  1.9(0.6) & 0.3970(0.1472) &   0.7(0.25) \\
 3463   &  SABbc   &  2692 & 36.2 & 2.7 & 1.4 & 11.85(0.04) &  11.8(1.77) &  3.0(0.4) & 3.8858(0.7310) &   6.7(1.20) \\
 3701   &  SAcd:   &  2915 & 39.2 & 1.8 & 1.0 & 13.65(0.05) &   2.5(0.37) &  3.3(0.5) & 0.8725(0.1596) &   2.8(0.50) \\
 3804   &  Scd:    &  2887 & 39.0 & 1.9 & 1.5 & 12.15(0.04) &   7.1(1.07) &  2.4(0.4) & 2.2694(0.4158) &   6.6(1.18) \\
 4705   &  SBb     &  2526 & 36.4 & 2.2 & 2.0 & 12.75(0.04) &   3.9(0.59) &  2.3(0.3) & 1.0742(0.2017) &   3.4(0.61) \\
 6157   &  SAdm    &  2958 & 44.4 & 1.9 & 1.1 & 13.67(0.05) &   2.4(0.37) &  3.3(0.5) & 0.9936(0.1749) &   2.5(0.45) \\
 7308   &  Sd      &  2762 & 39.9 & 1.7 & 3.4 & 13.33(0.05) &   3.4(0.51) &  3.4(0.5) & 1.1421(0.2077) &   2.1(0.38) \\
11269   &  SABab   &  2582 & 35.0 & 2.5 & 1.9 & 12.10(0.04) &   2.6(0.92) &  0.8(0.3) & 0.7265(0.2682) &   0.9(0.34) \\
12118   &  Sab     &  2825 & 32.1 & 2.0 & 2.9 & 12.66(0.04) &   2.9(1.01) &  1.5(0.5) & 0.7141(0.2663) &   2.1(0.77) \\
12442   &  SAbc:   &  2674 & 29.8 & 2.1 & 3.7 & 12.98(0.04) &   6.0(0.90) &  4.3(0.6) & 1.2383(0.2530) &   4.7(0.84) \\
12447   &  SBbc:   &  2678 & 29.9 & 3.3 & 3.0 & 11.35(0.04) &  21.3(3.19) &  3.4(0.5) & 4.4121(0.9001) &   3.5(0.63) \\
12690   &  SBm:    &  2605 & 28.9 & 2.0 & 1.1 & 15.41(0.10) &   0.7(0.11) &  4.7(0.7) & 0.1292(0.0268) &   1.5(0.27) \\
12699   &  SBb:    &  2798 & 31.0 & 1.9 & 1.4 & 11.90(0.04) &  36.6(5.48) &  9.6(1.4) & 7.8490(1.5745) & 129.8(3.41) \\
12700   &  Im      &  2770 & 31.0 & 2.6 & 5.2 & 13.00(0.05) &   0.0(0.10) &  0.0(0.2) & 0.0000(0.0210) &   0.0(0.10) \\
12788   &  SAc     &  2956 & 32.8 & 2.2 & 1.2 & 12.51(0.04) &   9.3(1.40) &  4.3(0.6) & 2.1975(0.4300) &   8.6(1.56) \\
\hline						           	                       	                               
N2604B  &  Im      &  2104 & 31.7 & 0.7 & 1.8 & 14.71(0.06) &   0.7(0.07) &  2.4(0.2) & 0.1667(0.0274) &   3.2(0.45) \\
060-036 &  Sc      &  2115 & 32.0 & 0.9 & 4.5 & 14.39(0.06) &   2.7(0.40) &  7.0(1.1) & 0.5354(0.1059) &  13.0(2.35) \\
74.0041 &  Im      &  2160 & 31.1 & 0.5 & 1.7 & 14.37(0.06) &   0.6(0.22) &  1.6(0.6) & 0.1467(0.0549) &   4.8(1.74) \\
809+363 &  SBm     &  2471 & 35.4 & 0.4 & 2.0 & 15.02(0.10) &   0.2(0.10) &  0.7(0.2) & 0.0369(0.0180) &   1.6(0.57) \\
Shane1  &  Sm      & ~---~ & 10.9 & 0.6 & 3.0 & 17.13(0.10) &   0.5(0.10) & 15.0(1.5) & 0.0107(0.0044) &   6.3(0.89) \\
N3769A  &  SBm:    &   761 & 12.6 & 1.1 & 2.8 & 14.38(0.06) &   2.8(0.28) &  7.2(0.7) & 0.0859(0.0307) &  14.0(1.98) \\
N4810   &  Im:     &   912 & 17.7 & 1.9 & 2.4 & 14.08(0.06) &   3.6(0.36) &  7.0(0.7) & 0.2706(0.0701) &  22.0(3.11) \\
\hline												    
\end{longtable}


\section{Initial analysis of survey data}

\subsection{Individual galaxy descriptions}

\begin{figure}
   \centering
   \includegraphics[angle=0,width=5.5cm]{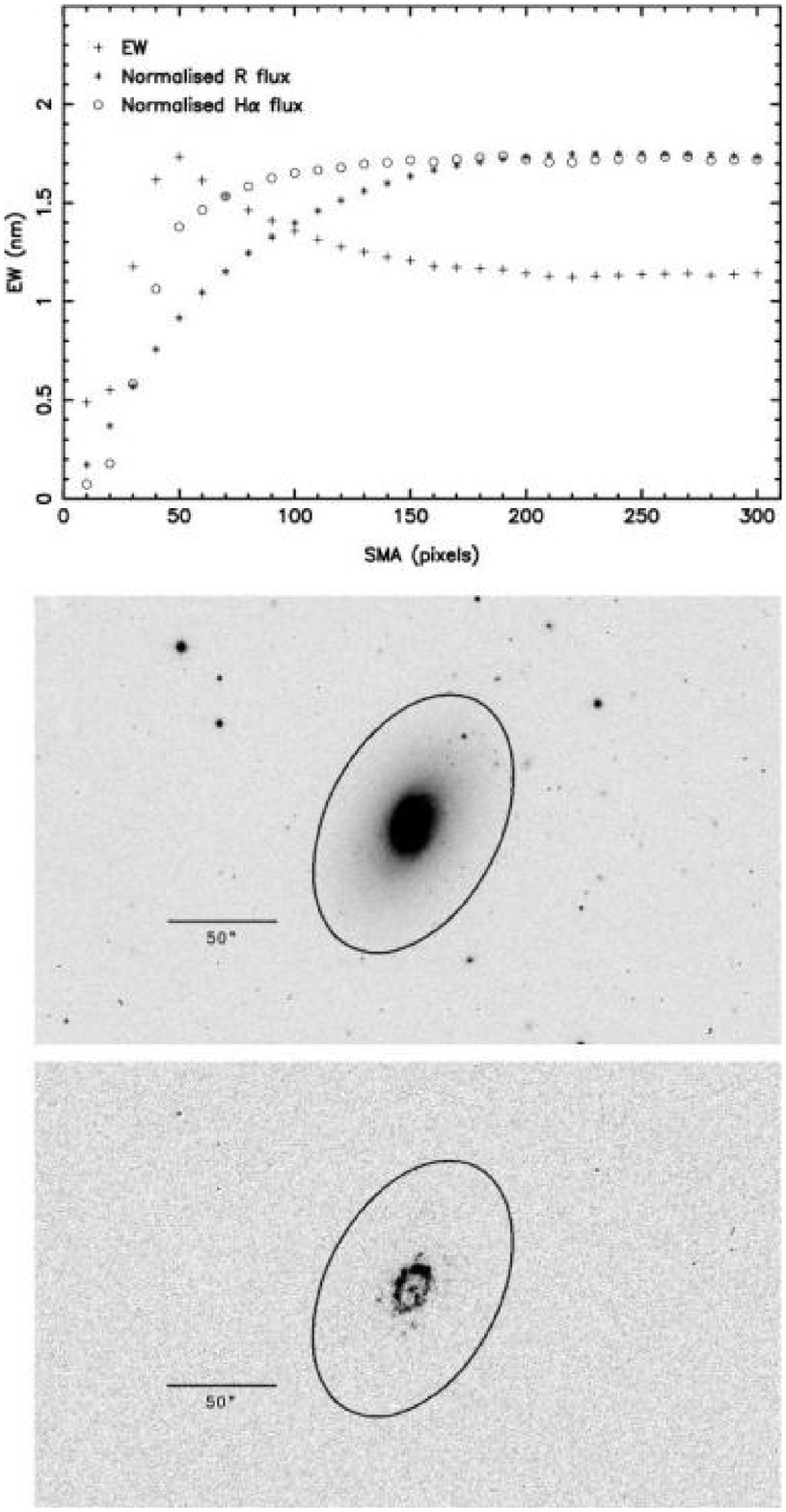}
   \includegraphics[angle=0,width=5.5cm]{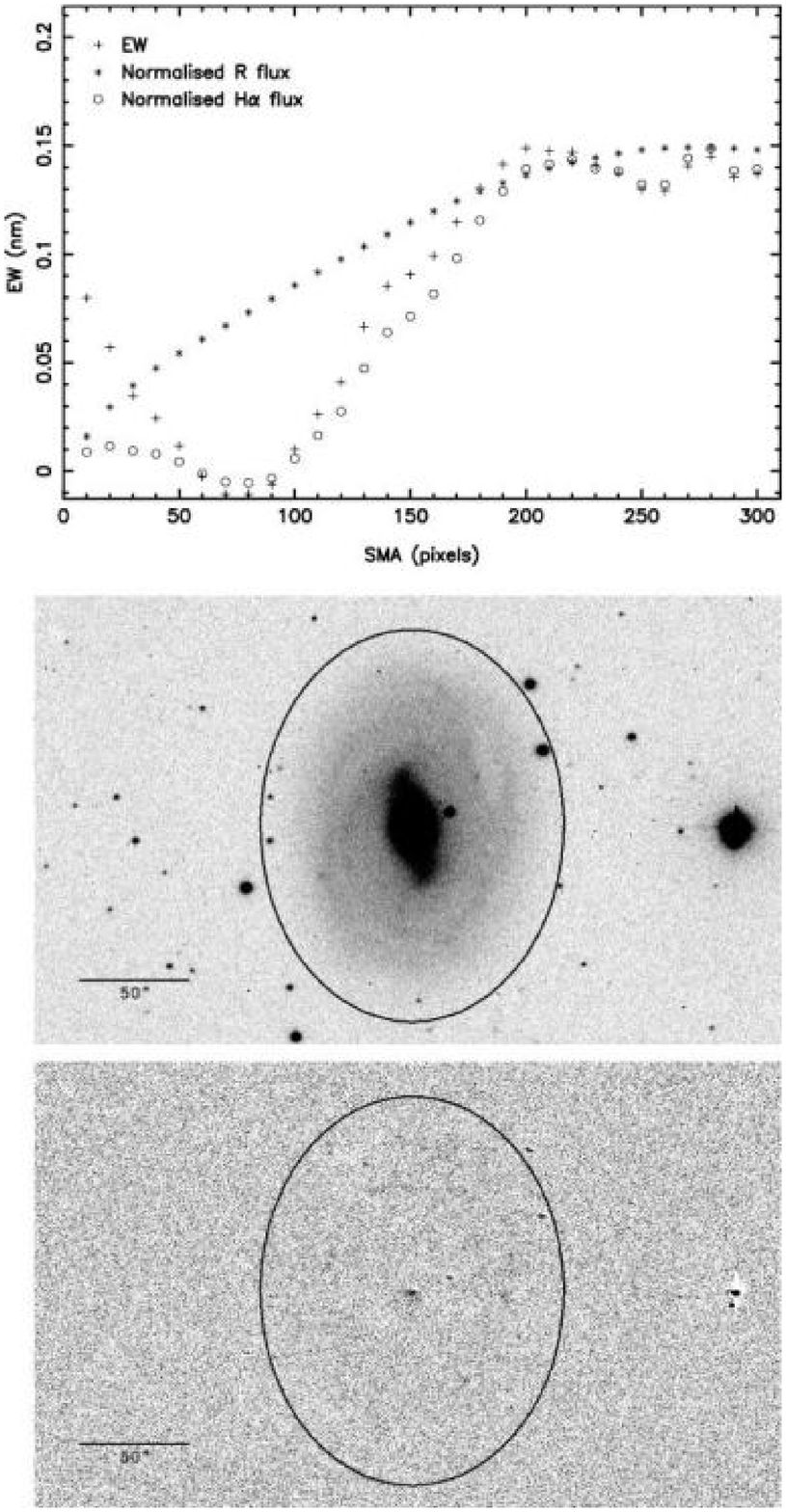}
   \includegraphics[angle=0,width=5.5cm]{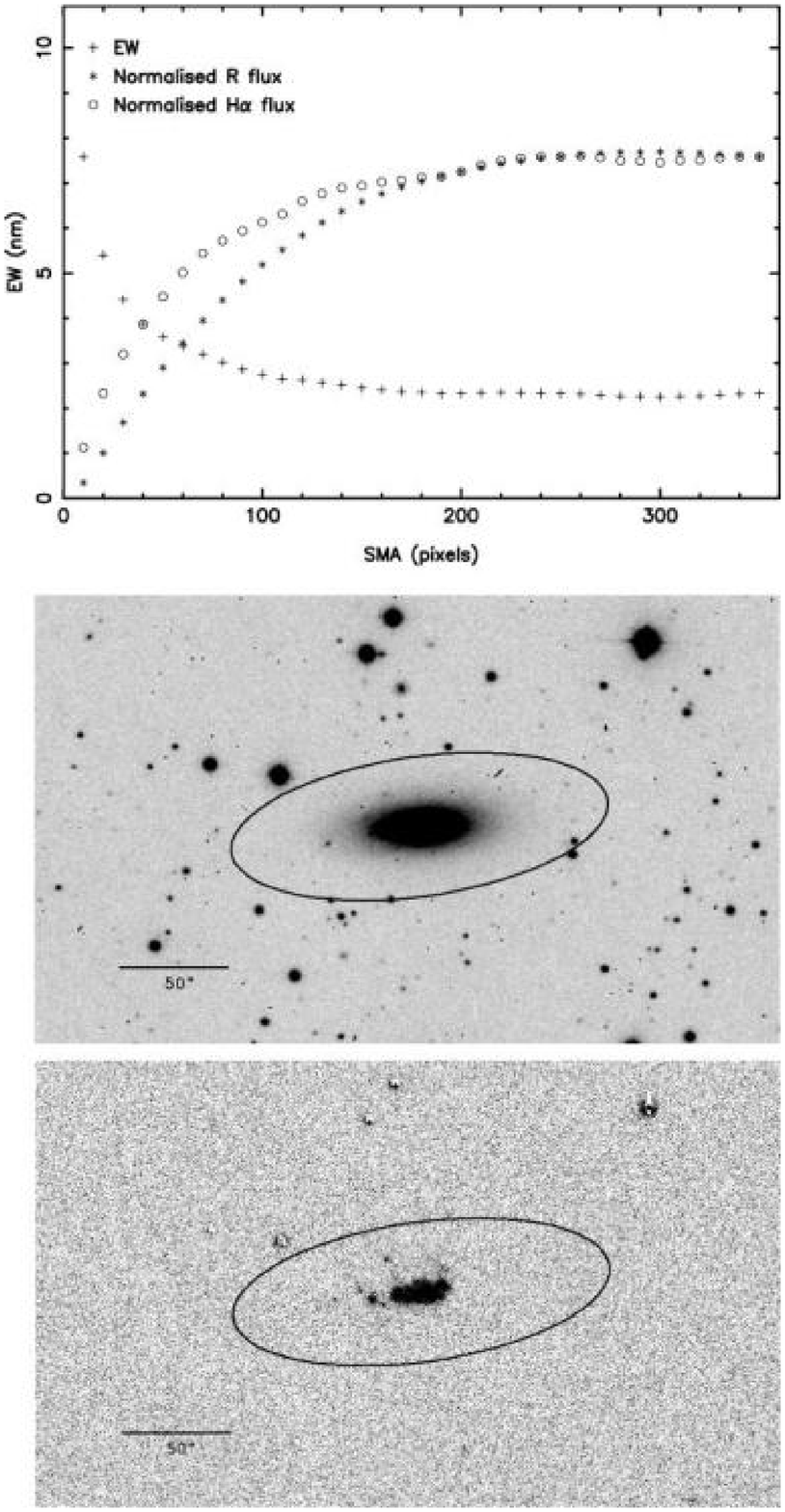}
      \caption{UGC 859, SAB0/a (left); UGC 11238, SB0/a (centre); and UGC
12043, S0/a (right).
For each galaxy, the top frame shows the H$\alpha +$[N{\sc ii}] growth
curve (circles), the $R$-band growth curve (asterisks) and the H$\alpha
+$[N{\sc ii}] EW (crosses) as a function of aperture size. The vertical
scale relates to the EW plots, and is in nm; the H$\alpha +$[N{\sc ii}]
and $R$-band fluxes are normalised arbitrarily. The
horizontal scale is the semi-major axis or radius of the apertures used,
in units of 0\farcs33 pixels. The images show the $R$-band (central
frame) and continuum-subtracted H$\alpha +$[N{\sc ii}] images (bottom
frame), with superposed ellipses to show the apertures used for total
line fluxes and $R$ magnitudes. } 
\label{Figa} 
\end{figure}

\begin{figure}
   \centering
   \includegraphics[angle=0,width=5.5cm]{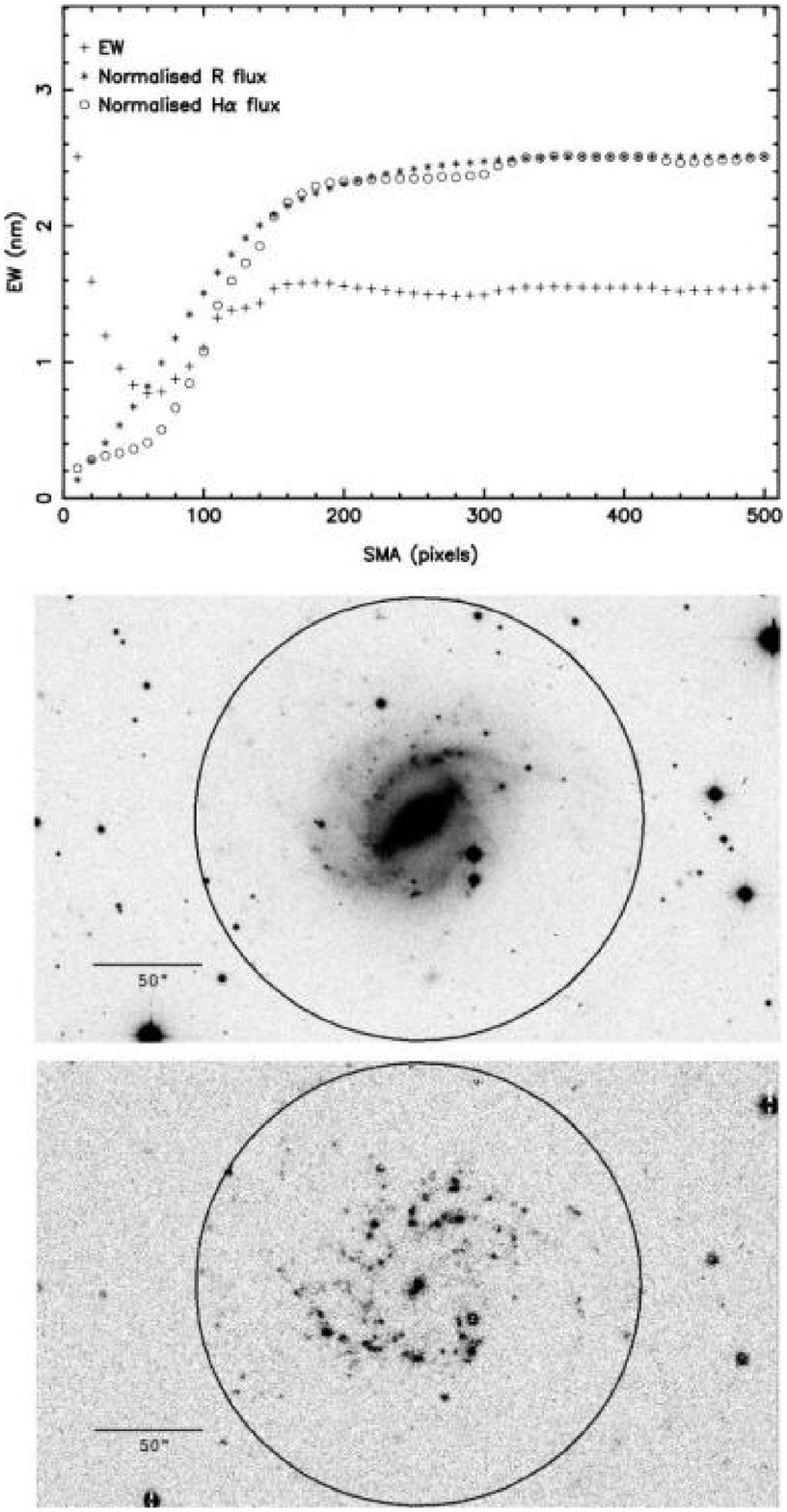}
   \includegraphics[angle=0,width=5.5cm]{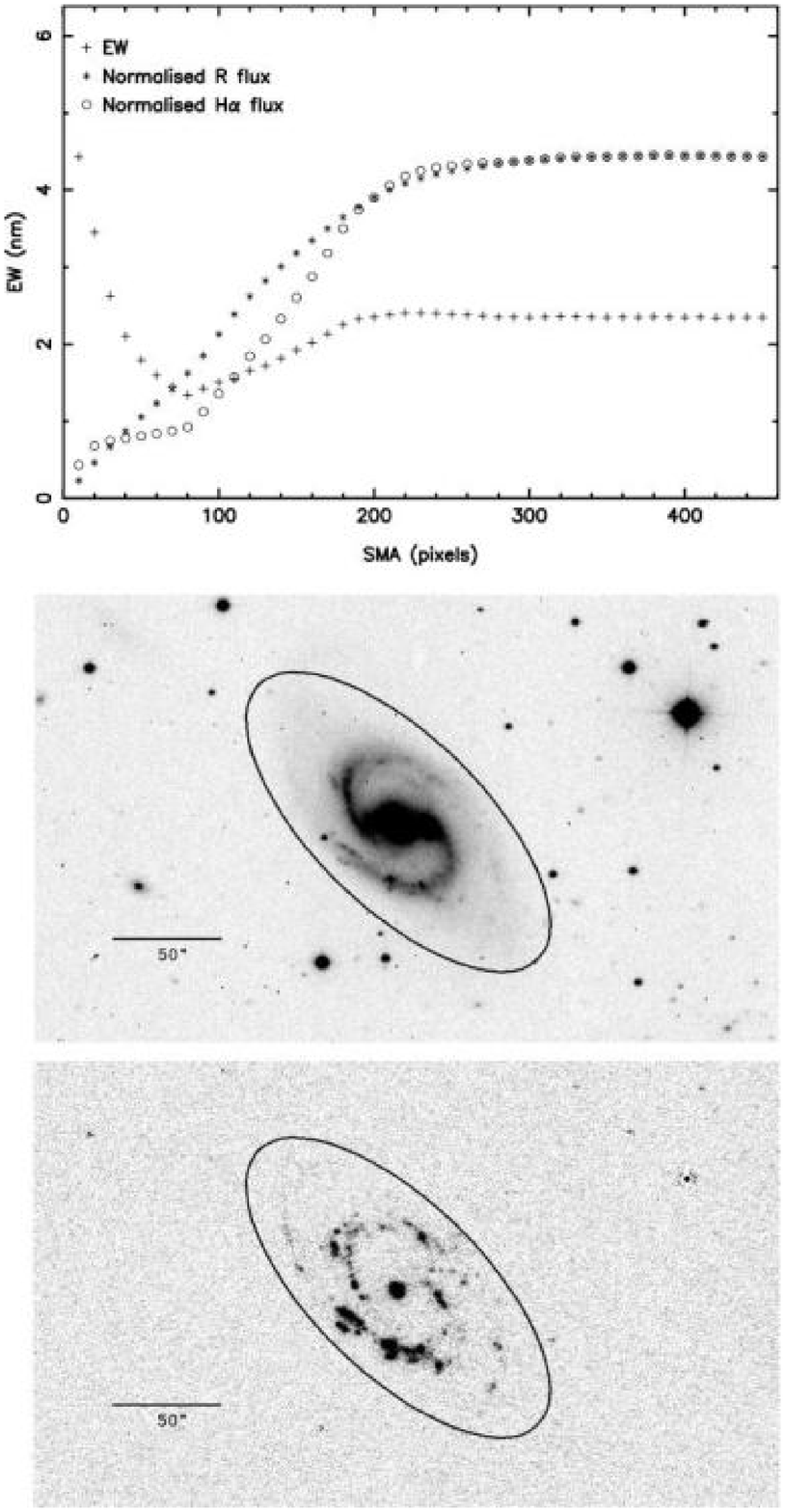}
   \includegraphics[angle=0,width=5.5cm]{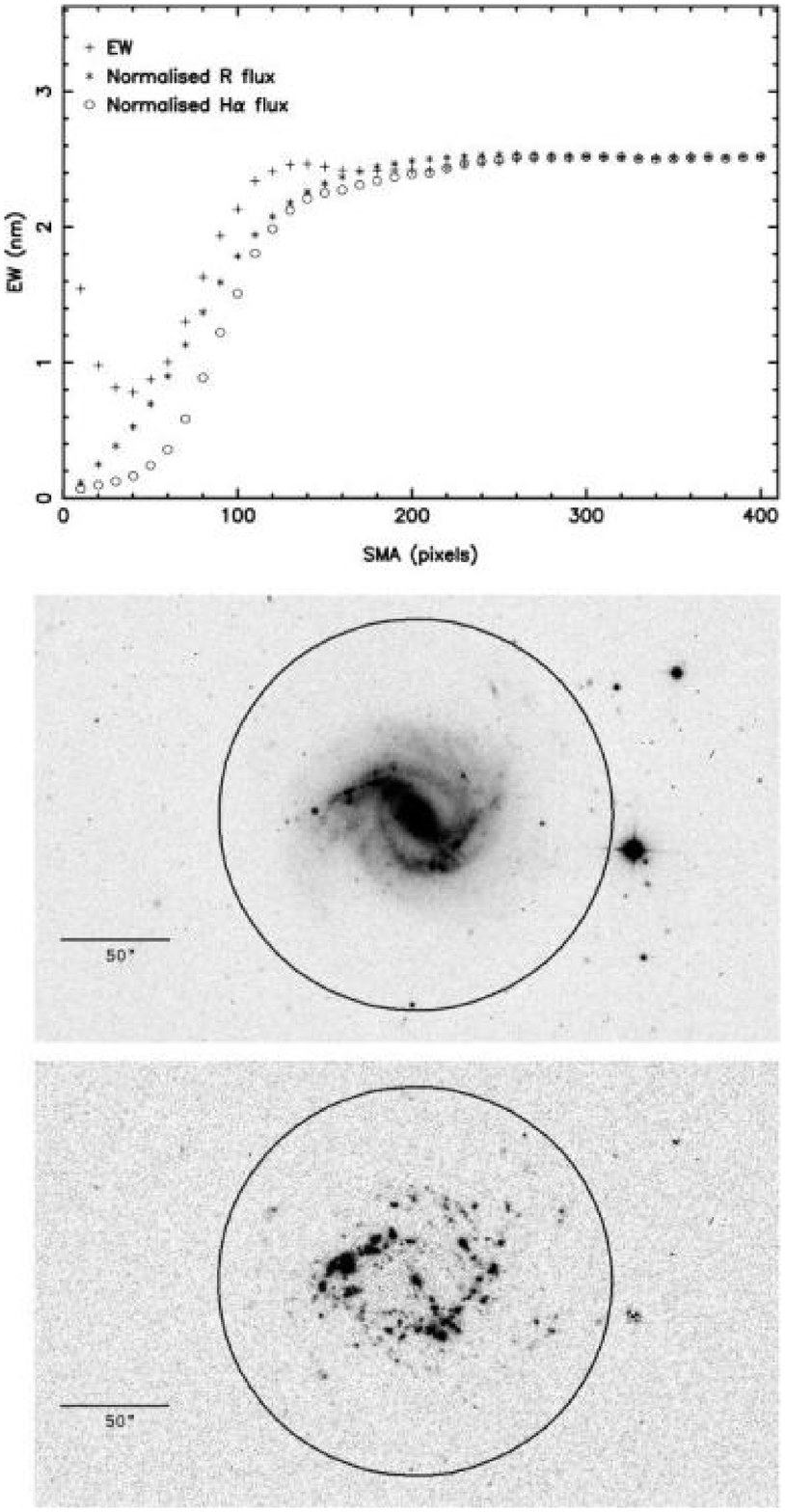}
      \caption{UGC~3685, SBb (left): UGC~4273, SBb (centre); and
UGC~6077, SBb (right).  See Fig.~\ref{Figa} for details.
              }
         \label{Figb}
   \end{figure}

\begin{figure}
   \centering
   \includegraphics[angle=0,width=5.5cm]{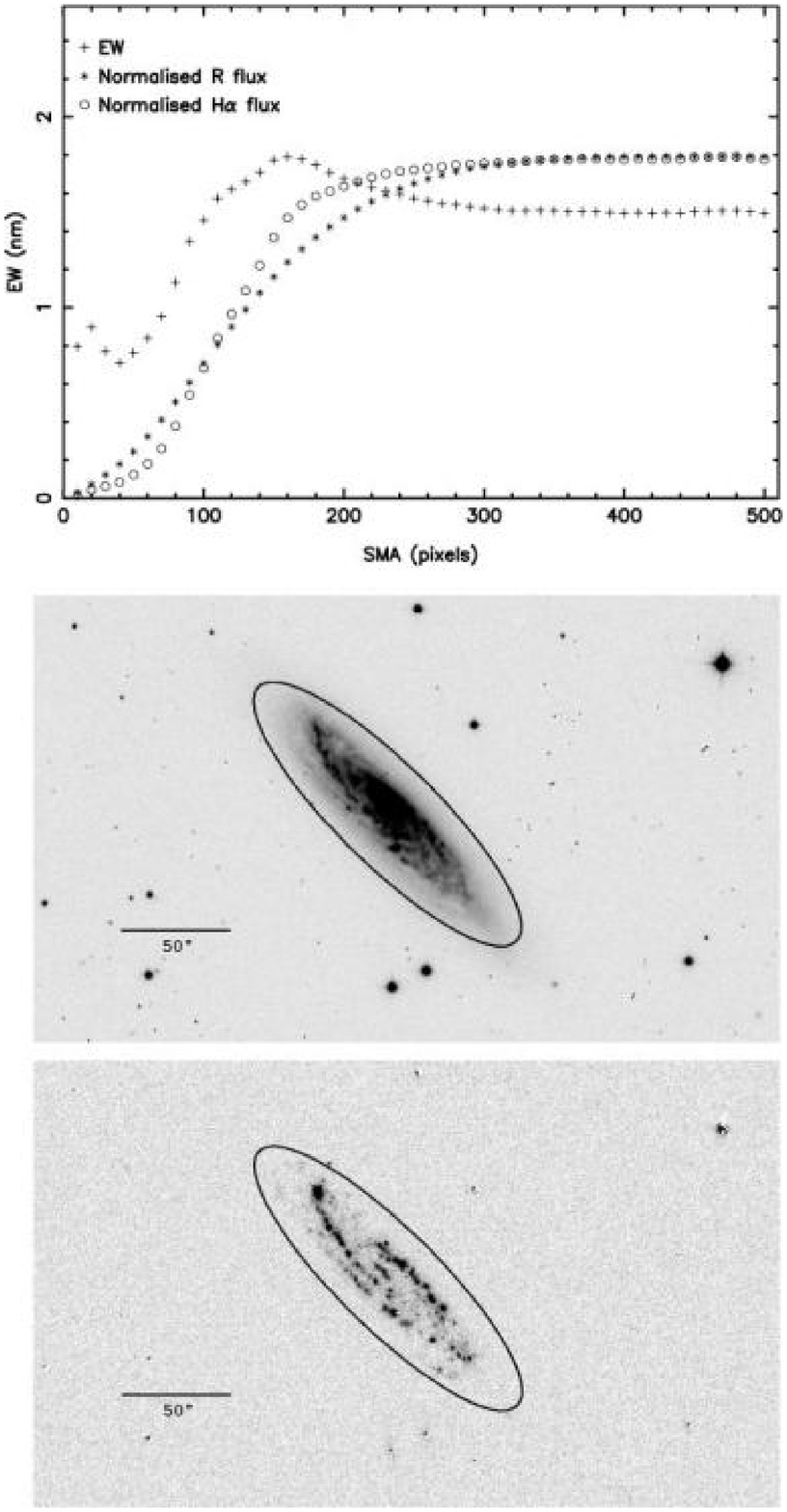}
   \includegraphics[angle=0,width=5.5cm]{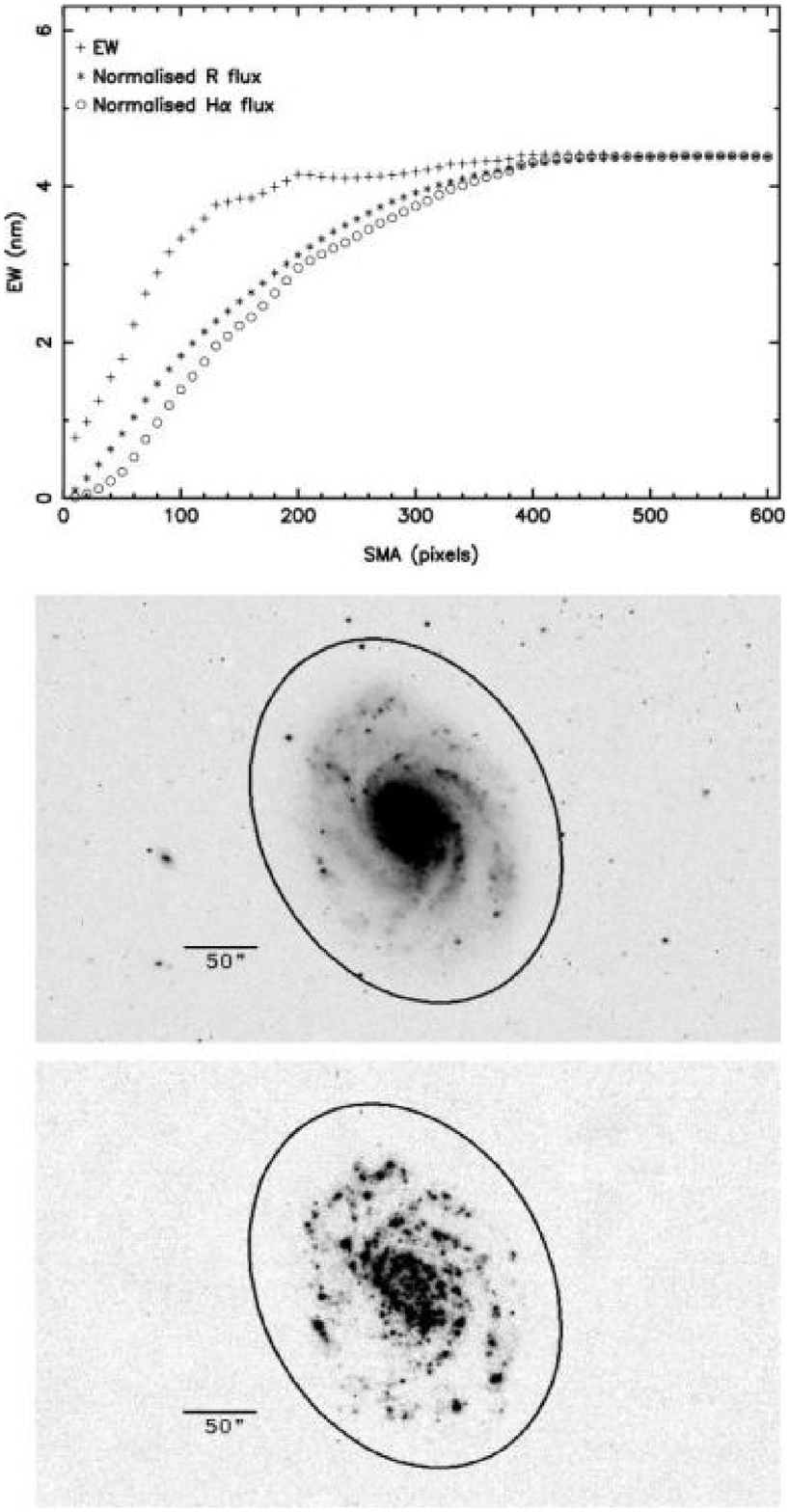}
   \includegraphics[angle=0,width=5.5cm]{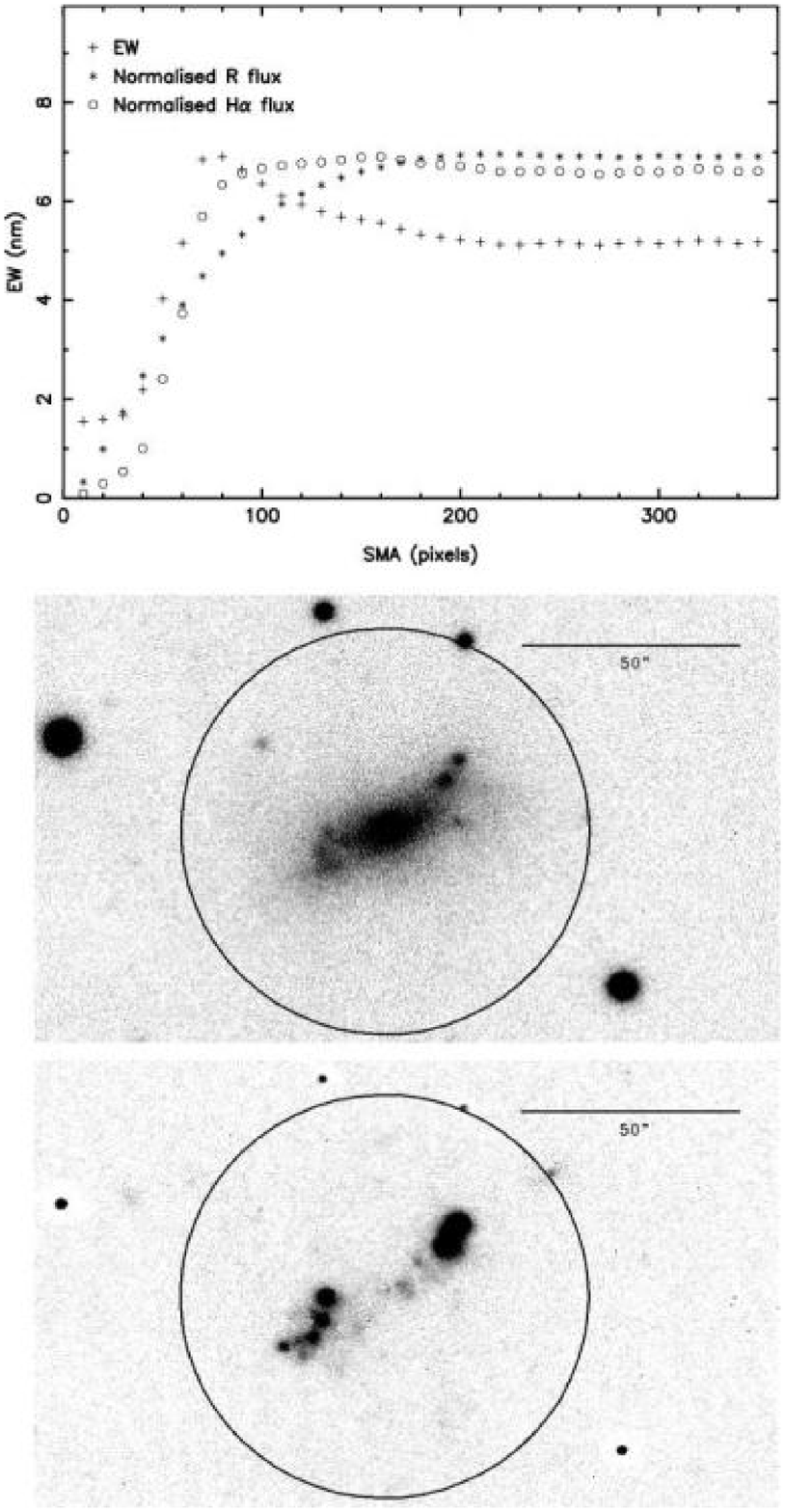}
      \caption{UGC~19, Sbc (left); UGC~6644, Sc (centre); and UGC~3711,
IBm (right).  See Fig.~\ref{Figa} for details.
              }
         \label{Figc}
   \end{figure}

\begin{figure}
   \centering
   \includegraphics[angle=0,width=5.5cm]{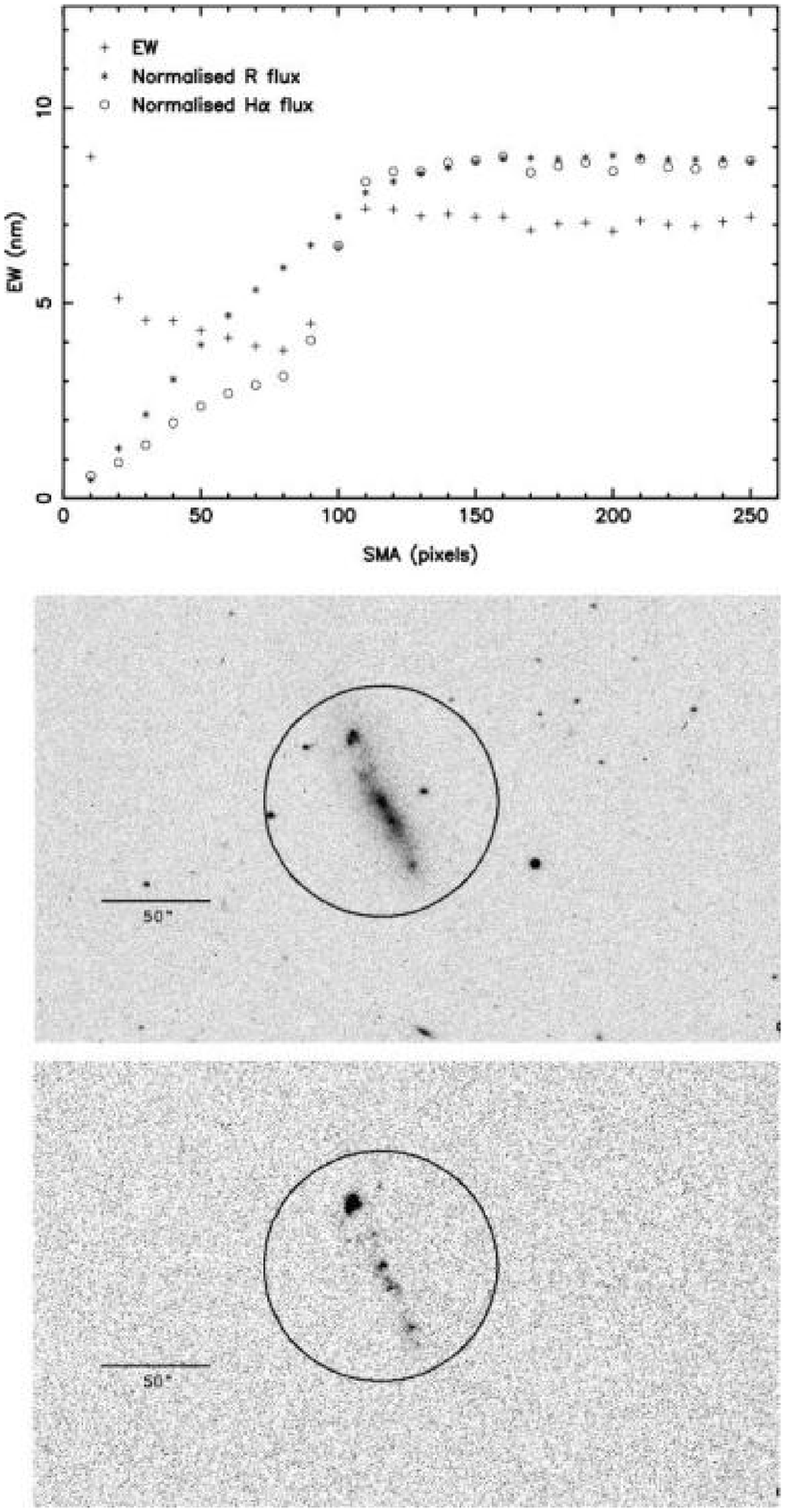}
   \includegraphics[angle=0,width=5.5cm]{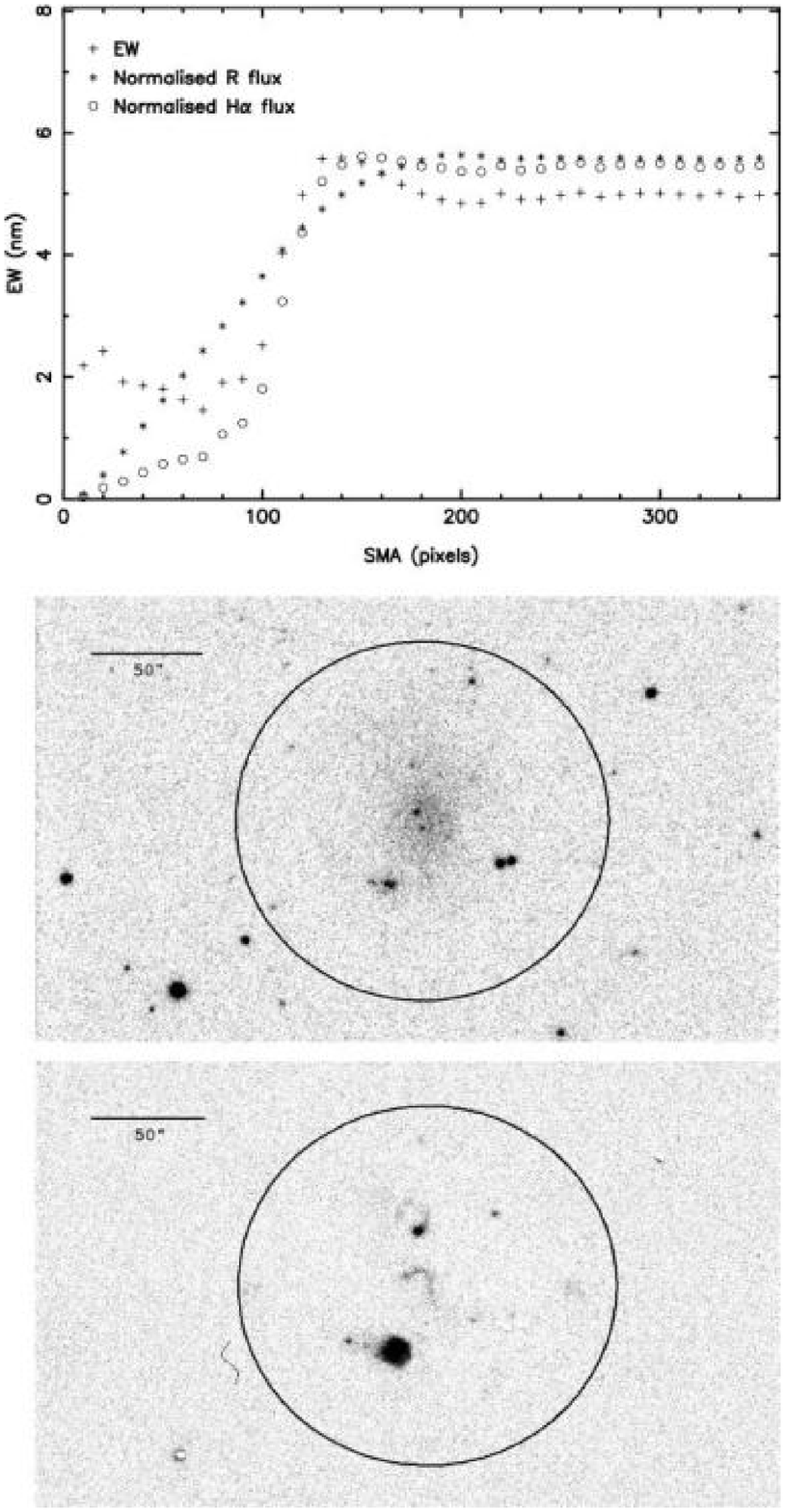}
   \includegraphics[angle=0,width=5.5cm]{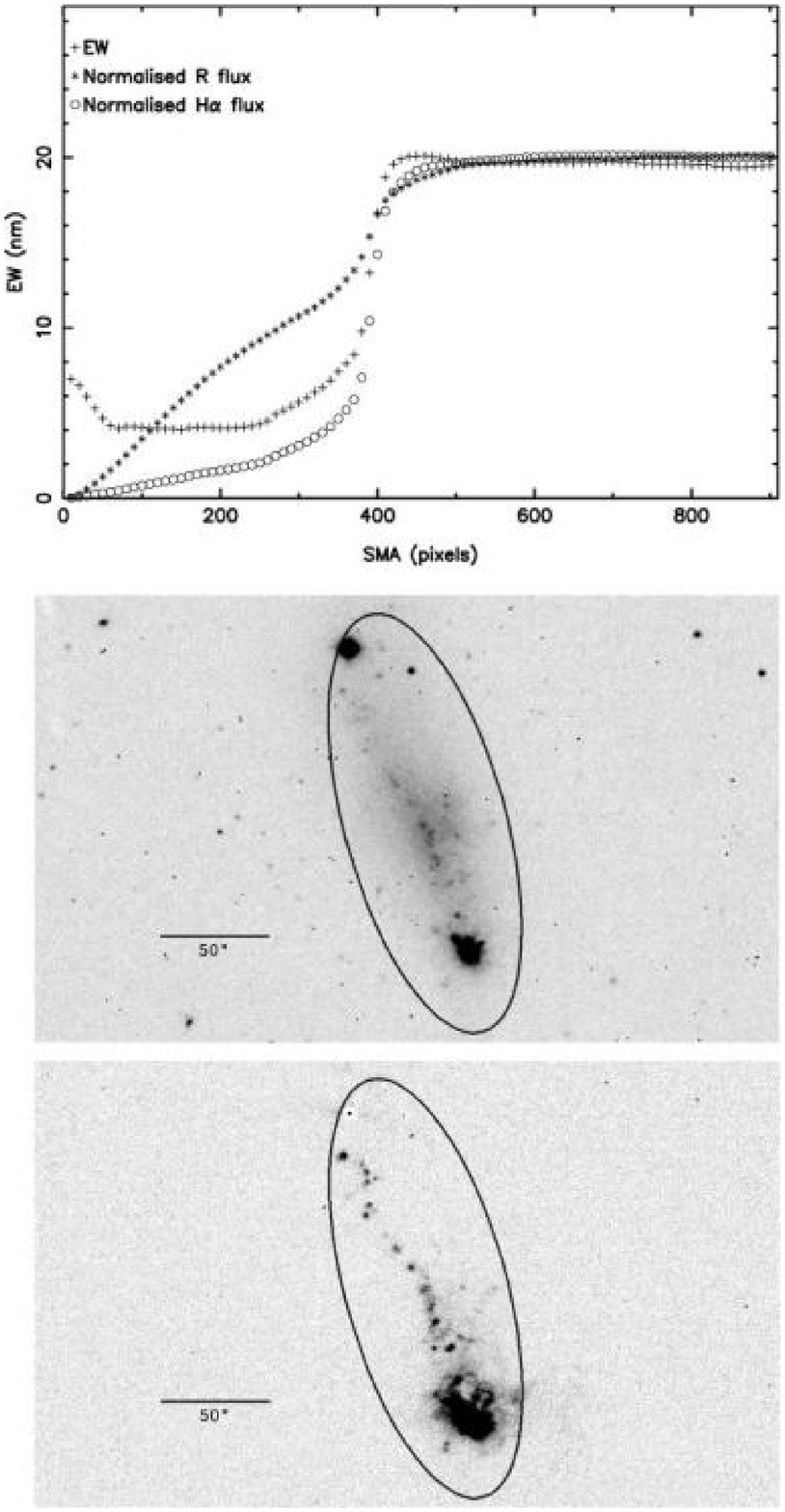}
      \caption{UGC~7326, Im (left); UGC~7866, IABm (centre); and
UGC~8098, SBm (right).  See Fig.~\ref{Figa} for details.
              }
         \label{Figd}
   \end{figure}

\begin{figure}
   \centering
   \includegraphics[angle=0,width=5.5cm]{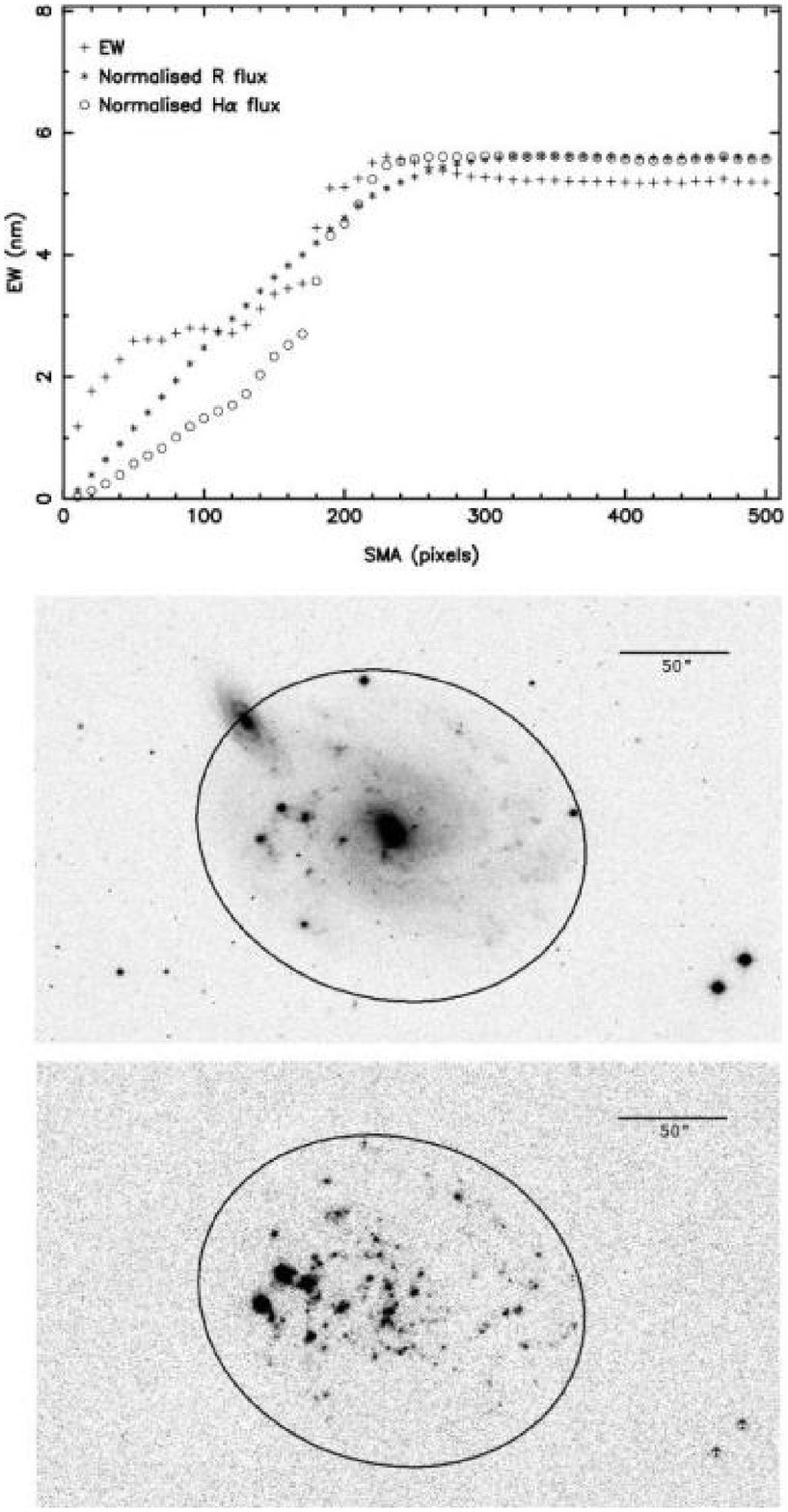}
   \includegraphics[angle=0,width=5.5cm]{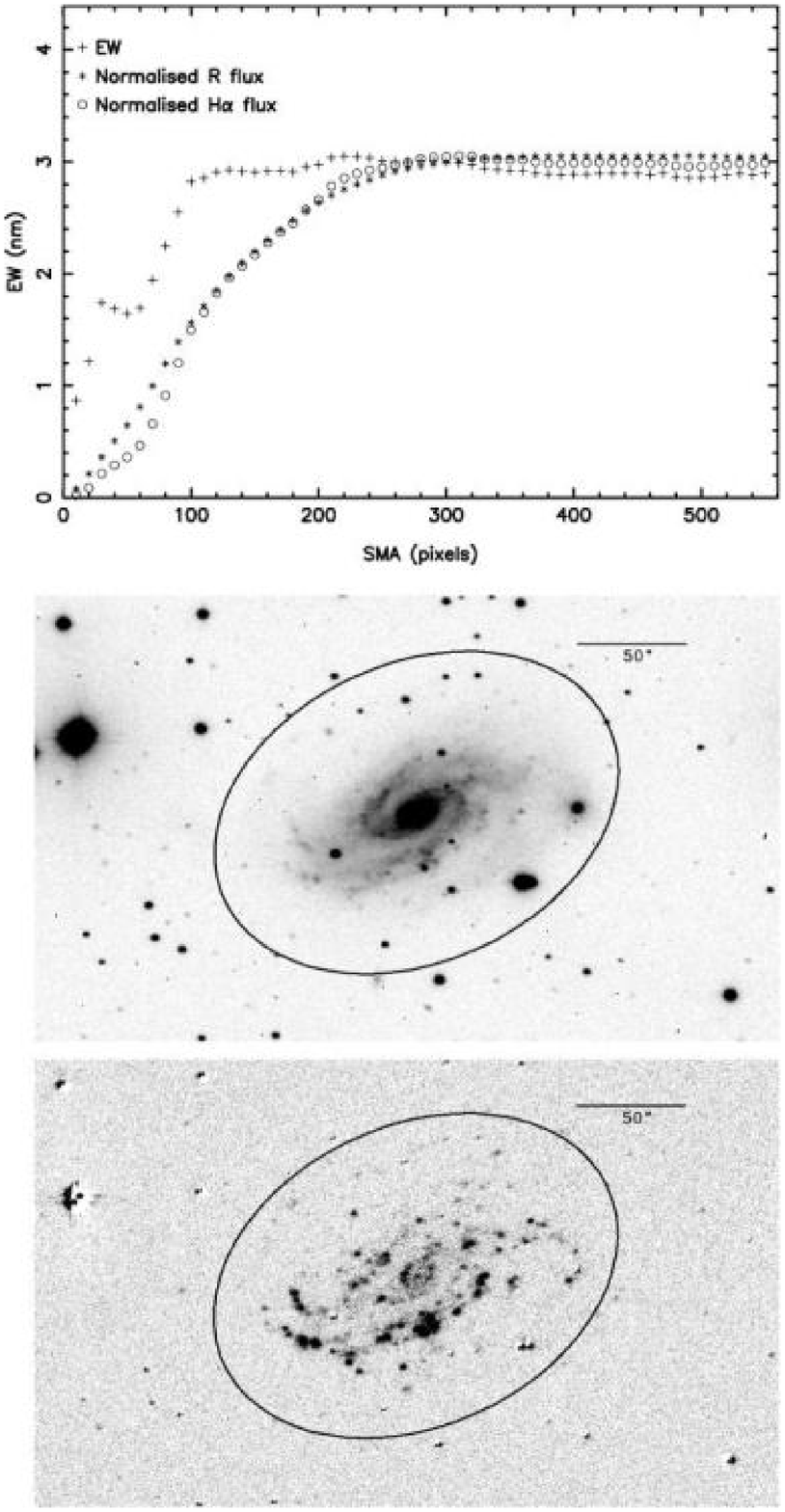}
   \includegraphics[angle=0,width=5.5cm]{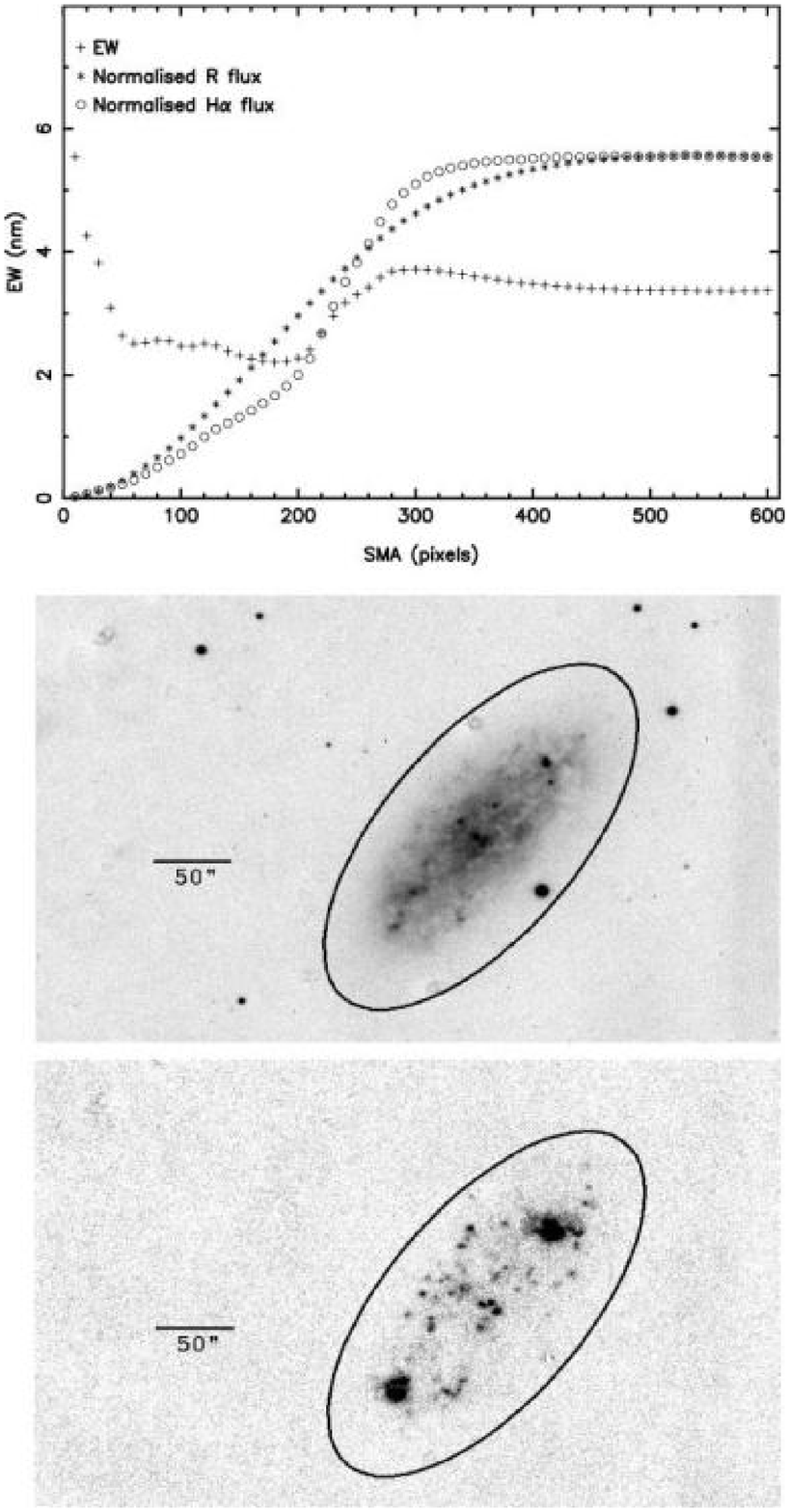}
      \caption{UGC~806, SABcd (left); UGC~3463, SABbc (centre); and
UGC~5221, Sc (right).  See Fig.~\ref{Figa} for details.
              }
         \label{Fige}
   \end{figure}

\begin{figure}
   \centering
   \includegraphics[angle=0,width=5.5cm]{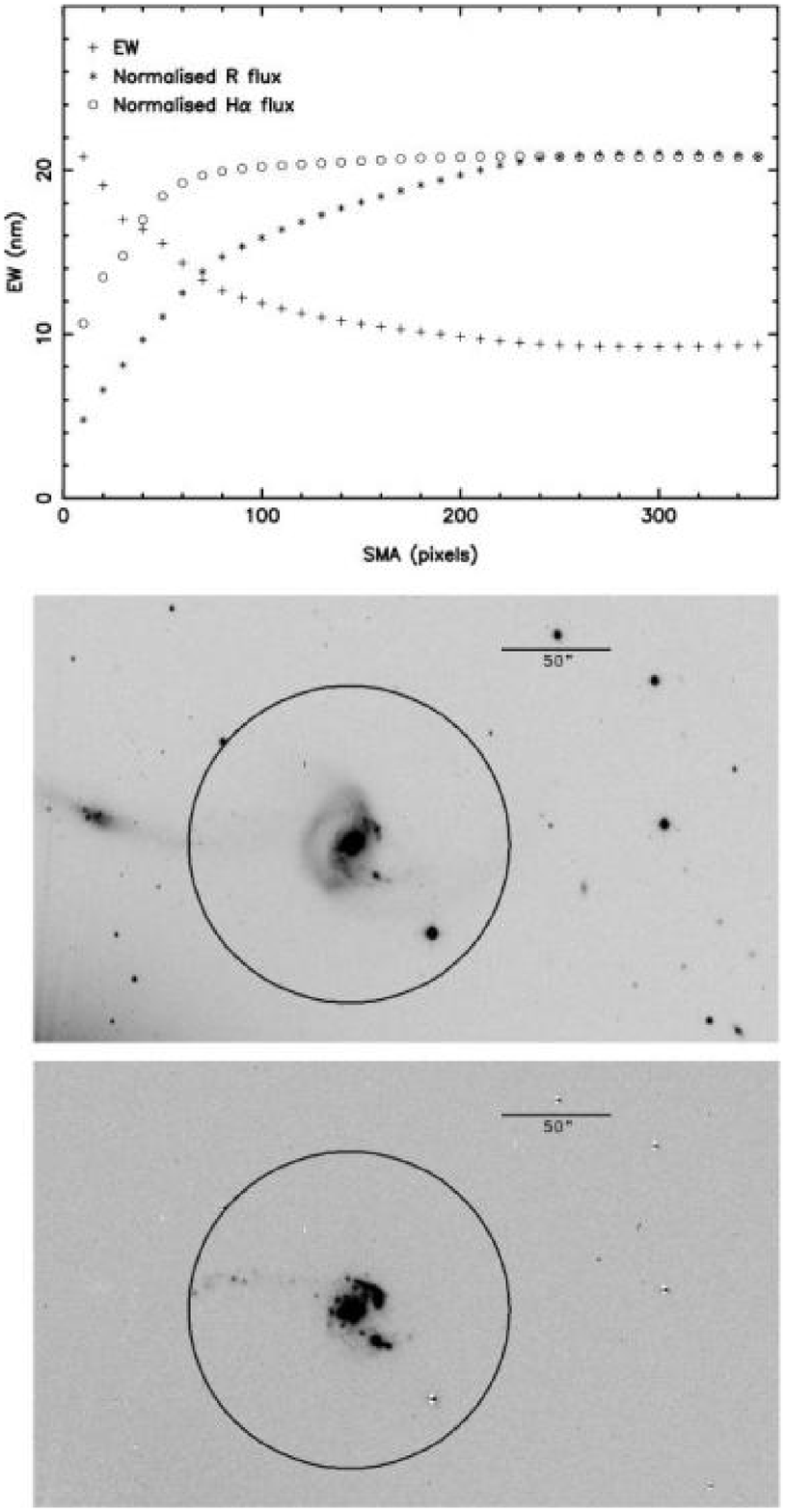}
   \includegraphics[angle=0,width=5.5cm]{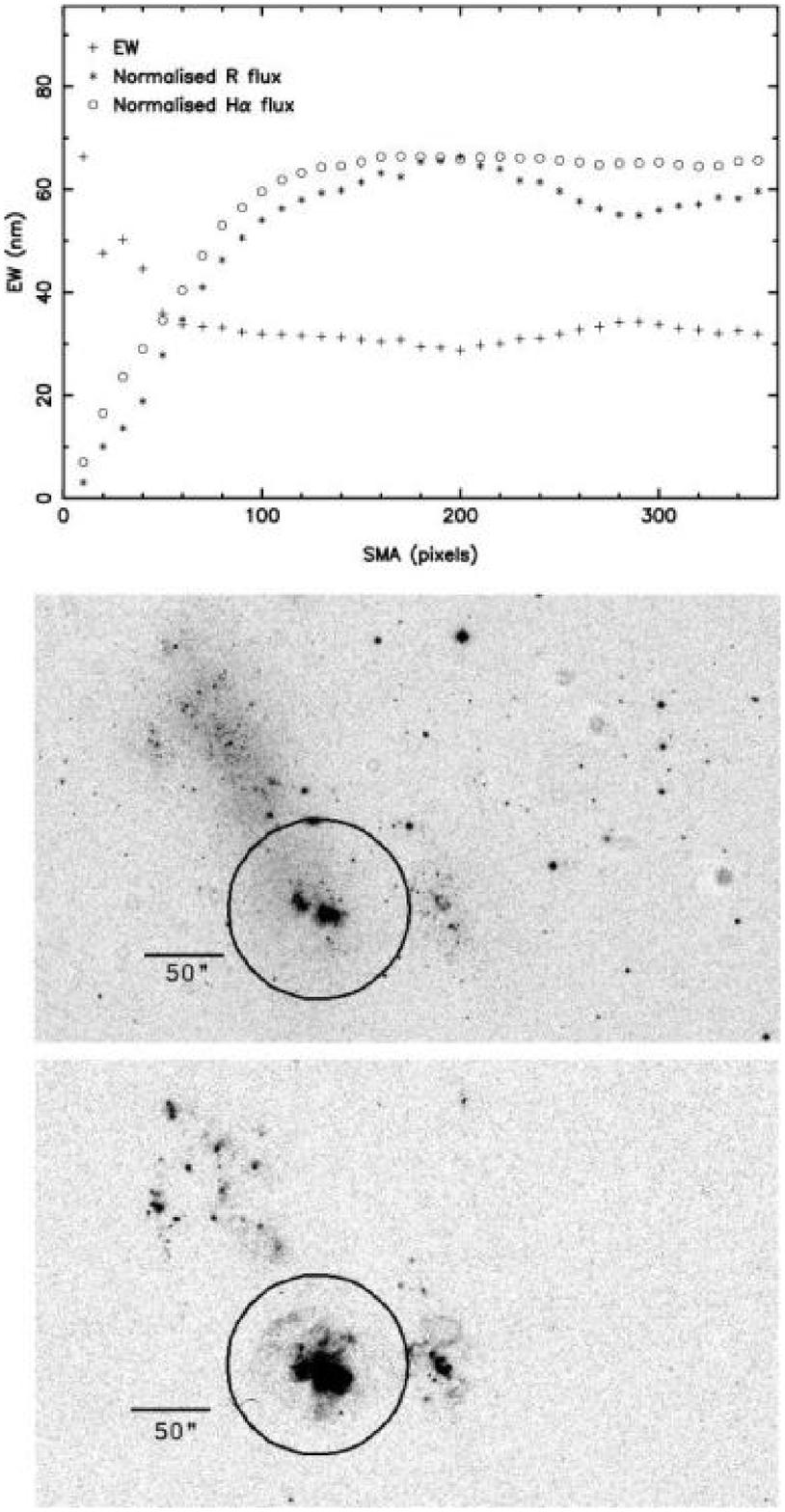}
   \includegraphics[angle=0,width=5.5cm]{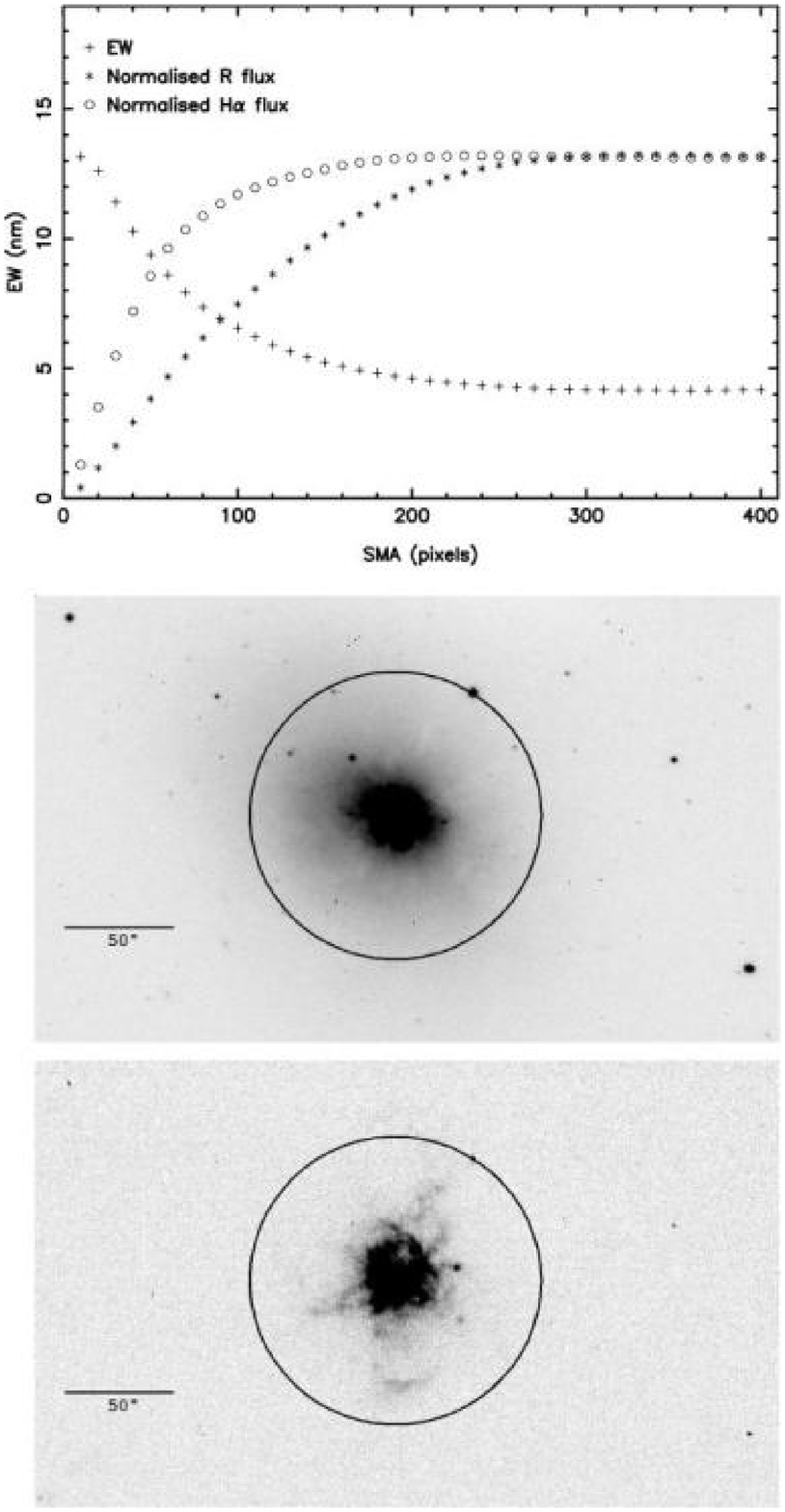}
      \caption{UGC 12699, SBb pec (left); UGC 3847, Irr (centre);  and
UGC 5398, I0 (right).  See Fig.~\ref{Figa} for details.
              }
         \label{Figf}
   \end{figure}

We will now consider some representative examples of the galaxies in
this sample in more detail, to illustrate the information which can be
gleaned from this dataset.  Future papers will explore the issues
raised here in greater depth.  In the accompanying Figs.
(\ref{Figa}--\ref{Figf}), the graph in the top frame shows the
H$\alpha +$[N{\sc ii}] growth curve (circles), the $R$-band growth curve
(asterisks) and the H$\alpha +$[N{\sc ii}] EW (crosses) as a function of aperture
size. The vertical scale relates to the EW plots, and is in nm; the
H$\alpha +$[N{\sc ii}] and $R$-band fluxes are normalised arbitrarily to fit in
these plots, but the calibration can be derived from the total
H$\alpha +$[N{\sc ii}] fluxes and $R$ magnitudes given in Table 3, which correspond
to the `plateau' levels in these plots. The horizontal scale is the
semi-major axis (or radius, for circular apertures) of the apertures
used, in units of 0\farcs33 pixels. The images show the $R$-band
(central frame) and continuum-subtracted H$\alpha +$[N{\sc ii}] images (bottom
frame), and are oriented with North upwards and East to the left.
Galaxies were chosen for these figures to illustrate both `typical'
examples of the galaxy types and data quality of this survey, and some
of the interesting or extreme objects.

Figure~\ref{Figa} shows three galaxies classified as extreme early types.
Despite their similar classifications and optical appearances, these
galaxies show widely differing star formation morphologies. UGC~859 shows
strong star formation in a very regular ring, around the edge of the
bulge (also described by Pogge \& Eskridge \cite{pogg}).  This is
reflected in the EW curve, which exhibits a deep central dip due to the
strong continuum from the central bulge, a peak at the radius of the
star-forming ring, and a steady drop at larger radii due to the presence
of only old stars outside the ring, to an overall value of about 1~nm.
This is typical of moderately star-forming bright spiral galaxies.
UGC~12043 shows more centrally-concentrated star formation, with just a
central peak in the EW curve and a monotonic decline with radius, at
least at the resolution of the present data. The overall EW of $\sim$2~nm
is high for such an early type galaxy, and both UGC~859 and UGC~12043 are
much stronger emission-line sources than were found in the early-type
galaxy samples studied by Caldwell et al. (\cite{cal91},\cite{cal94}).
UGC~11238, in contrast, has barely detectable line emission, and is
included here to highlight the fact that there is no selection bias in
favour of star-forming galaxies in the sample, and any statistical
analysis will include undetected or barely-detected galaxies like this
one.  This figure illustrates how well the continuum subtraction removes
the light from the old stellar population in the galaxy images.

The galaxies illustrated in Fig.~\ref{Figb}, UGC~3685, UGC~4273 and
UGC~6077 are all classified as SBb, and all show qualitatively similar
H$\alpha +$[N{\sc ii}] distributions, with a significant central peak, a star
formation `desert' in the region swept out by the bar, and substantial
star formation in H{\sc ii} regions scattered around the disk.  This results
in an EW curve with a strong central peak, a broad dip, and a gentle
outer rise to the plateau level at 1.5--3~nm.  The frequency of
central peaks in strongly-barred spiral galaxies may reflect the bar-driven 
feeding of gas into the central regions of galaxies predicted by
several authors (e.g., Arsenault \cite{arsenau}; Quillen et al.
\cite{quil}).  This feeding has been used to explain the high star
formation rates generally found in strongly-barred galaxies
(e.g., Hawarden et al. \cite{hawa}; Dressel \cite{dres}; Huang et al.
\cite{huan}; Martinet \& Friedli \cite{mart}) via nuclear starbursts.

Figure~\ref{Figc} shows UGC~19, which is an Sbc with a marked ring of
star formation, and is a good example of how this is revealed by a local
maximum in the EW at intermediate radii (100-200 pixels semi-major axis,
or 4--8~kpc at the adopted distance for this galaxy). UGC~6644, on the
other hand, exhibits a monotonically increasing EW curve, showing that
the stellar population becomes increasingly dominated by young stars with
radius.  This is to be expected in the central regions, where the old
bulge population is bright in the $R$ band and has little associated line
emission, but in UGC~6644 there is evidence that the star formation
regions are more widely distributed than the old stellar population even
in the disk. This question of the relative distributions of young and old
stars, including the effects of inclination on these distributions, will
be studied systematically across all Hubble types in a later paper in
this series.

The sample contains a large number of Magellanic irregular galaxies, and these
show an interesting variety in star formation morphologies.  UGC~7326
(classified Im), UGC~7866 (an IABm) and UGC~8098 (an SBm, also named
NGC~4861; see also Conselice et al. \cite{cons} who present
ground-based and Hubble Space Telescope imaging of this galaxy) all
show highly asymmetric or `cometary' morphologies (Fig.~\ref{Figd})
in their continuum-subtracted H$\alpha +$[N{\sc ii}] images, with extremely intense
star formation centres which are displaced very significantly from the
centres of the old stellar distributions.  This is reflected in long
central troughs in the EW curves, with sharp rises to very high values
when the star formation centres are reached (all
growth curves are centred on the peaks of the old stellar light in the
$R$-band images). The
decoupling between the smooth old stellar distribution and the highly
clumped and asymmetric sites of ongoing star formation is particularly
marked in these galaxies. The overall EW values for these three
galaxies lie in the range 5--20~nm, much higher than is typical for
even late-type spiral galaxies.  Local EW values in apertures centred on the
H{\sc ii} complexes lie in the range 50-100~nm for these galaxies. Similarly
intense star formation is found in the strongly-barred IBm UGC~3711
(Fig.~\ref{Figc}), but in this case the star formation complexes are
found in two off-nuclear regions, which appear to be associated with
the bar ends.

Significant asymmetry in H$\alpha +$[N{\sc ii}] distributions is also apparent in
some of the spiral galaxies.  A particularly striking example of this
is UGC~806 (Fig.~\ref{Fige}), classified SABcd, which contains 3
extremely bright star formation complexes to the east of the bulge.
Schaerer et al. (\cite{scha}) refer to this object as a Wolf-Rayet
galaxy, and note that it may be part of NGC~450 (UGC~806) which
appears likely given the close agreement in recession velocity
(1693~km~s$^{-1}$ for the star formation complexes c.f.
1761~km~s$^{-1}$ systemic).  However, even with these three bright
regions removed, the H{\sc ii} region distribution is significantly
asymmetric, with a substantial deficit to the north and west.  A less
marked asymmetry is shown by the SABbc galaxy UGC~3463
(Fig.~\ref{Fige}), where most of the bright H{\sc ii} regions lie
south of the nucleus.  However, in this case the line of asymmetry
coincides with the major axis of this inclined spiral, and it is
tempting to ascribe the asymmetry to differential extinction effects,
possibly due to systematic offsets between the dust lanes and the star
formation centres in spiral arms.  The underlying galaxy appears
highly symmetric, as shown by the near-IR K-band image of UGC~3463
presented by Peletier et al.  (\cite{pele}), with no obvious
differences between the upper and lower spiral arms in the light
emitted by the old stellar population.  The question of extinction in
arms and asymmetries in H$\alpha$ morphologies will be examined
further in a later paper of this series.

UGC~5221 (Fig.~\ref{Fige}) is an Sc spiral, possibly in the M81
group, with an unusual star formation morphology in that the bulk of
the H$\alpha +$[N{\sc ii}] emission comes from two regions symmetrically placed
along the major axis of the galaxy.  If the galaxy elongation is
interpreted as an inclined disk (and we note that, for example,
Bronkalla \& Notni (\cite{bron}) cite UGC~5221 as a prime example of a
`pure-disk' galaxy), this morphology must be seen as a
coincidental alignment, but it does suggest the possibility that the
main body of UGC~5221 should be seen as a bar. In this case, the star
formation is occurring at the bar ends, as is commonly seen in
strongly barred galaxies (e.g., UGC~3711, UGC~4273 and UGC~4708 in this
paper).

The sample includes some interacting galaxies, including UGC~12699
(Fig.~\ref{Figf}), better known as the prototypical starburst
(Weedman et al. \cite{weedman}) NGC~7714 or Arp~284. As expected, our
data show extremely intense central star formation in NGC~7714, but
interestingly none at all from the small companion NGC~7715 (as was
also noted by Gonz\'alez-Delgado et al. \cite{gonz} from their
H$\alpha$ imaging), although there are some H$\alpha +$[N{\sc ii}] knots from the
tail apparently joining the two galaxies.  This is despite the
classification of Im for NGC~7715 and its very disturbed, knotty
appearance, which may be due at least in part to foreground stars.
Another possible interaction which is worthy of comment is UGC~3847
and UGC~3851 (NGC~2363 and NGC~2366, shown in Fig.~\ref{Figf}), both
irregular galaxies, with the 2 highest H$\alpha +$[N{\sc ii}] EWs of
all 334 galaxies observed in this study. Drissen et al. (\cite{dris})
consider UGC~3847 to be an outlying H{\sc ii} region of UGC~3851, and note
that UGC~3847 is the highest surface brightness H{\sc ii} region in the
entire sky.  UGC~3851 is not actually part of the statistical
sample, since it is too large to satisfy the selection criteria, but
is included in this discussion of individual objects due to its
proximity to UGC~3847.

Finally, we include UGC~5398 (Fig.~\ref{Figf}) as a cautionary
example of the difficulties in interpreting H$\alpha$ emission in a
uniform manner. The morphology of the line emission in UGC~5398 is
very different to that of most of the sample, showing plumes and
possible bubble structures.  This galaxy was interpreted by Martin
(\cite{martin}) as an example of the `supershell' phenomenon, where
large numbers of supernovae in a short time interval create
large-scale galactic winds, which blow shell structures of this types.
However, this morphology is also similar to the Extended Nuclear
Emission-line Regions (ENERs) found by Hameed \& Devereux
(\cite{hameed}) in H$\alpha$ imaging of the central regions of early
type galaxies, which they ascribe to gas that is either shock-ionized
or photoionized by UV radiation from bulge post asymptotic giant
branch stars.  This, and the likely contribution of AGN line emission
from the central regions of at least some of our sample galaxies, are
reminders that not all line emission need be related to star
formation.  These effects will be studied in more detail in later
papers in this series.

\subsection{Global properties of the sample}

The full sample of 334 galaxies provides an excellent database for
quantifying star formation activity as a function of galaxy type and
luminosity.  Some of the most fundamental correlations and
distributions are illustrated in the remaining figures in this section.  
  
   \begin{figure}
   \centering
   \includegraphics[angle=-90,width=10cm]{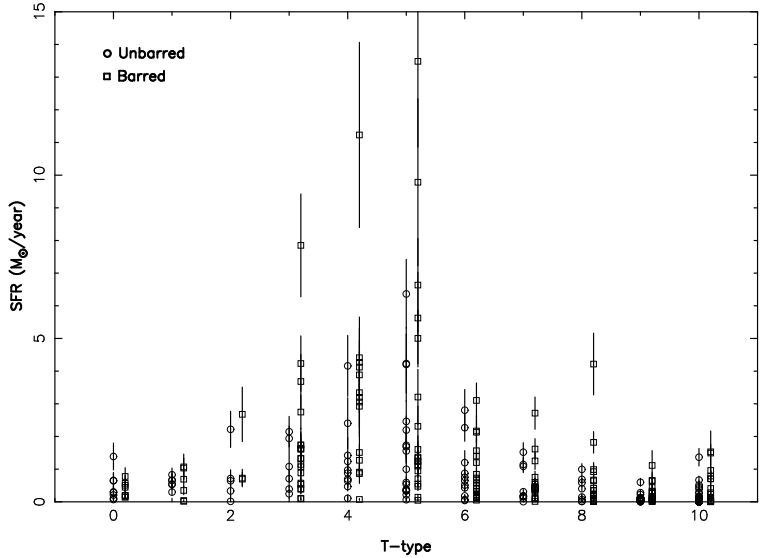}
      \caption{Distribution of star formation rates as a function of 
      Hubble T type.
              }
         \label{sfrvt}
   \end{figure}

Figure~\ref{sfrvt} shows the distribution of individual galaxy star
formation rates as a function of Hubble T-type.  The main result from
these distributions is that, as expected, spiral galaxies of types Sbc
and Sc have the highest individual star formation rates. Barred (SAB and
SB types) and unbarred distributions are shown separately, showing that
bars appear to be related to moderately higher star formation rates, and
the 5 galaxies with the highest star formation rates are all barred.
This is not surprising as we have already found that bars can induce
circumnuclear star formation (see Fig.~\ref{Figb}).  For all Hubble
types, the modal values of the distributions lie close to zero star
formation rate, i.e. very low star formation rates are extremely common
for all types.

   \begin{figure}
   \centering
   \includegraphics[angle=-90,width=10cm]{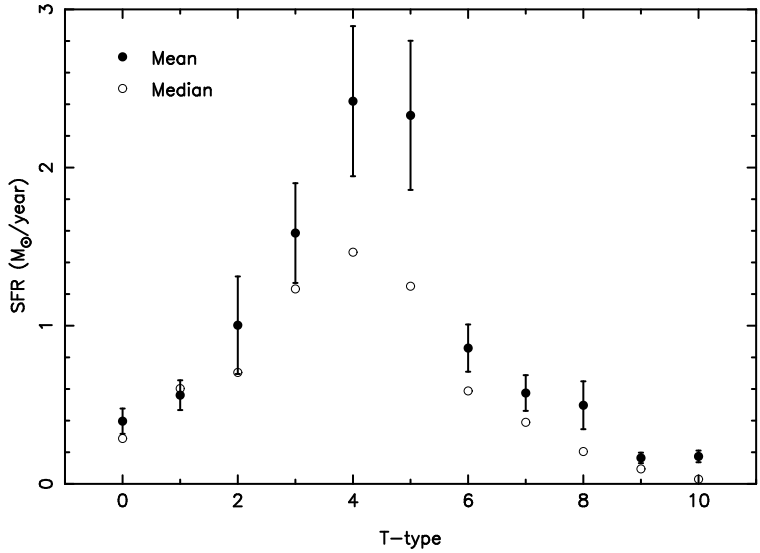}
      \caption{Mean and median star formation rates as a function of 
      Hubble T type.
              }
         \label{msfrvt}
   \end{figure}

Figure~\ref{msfrvt} shows mean star formation rates (filled points) as
a function of Hubble type, with error bars showing the standard error
on the mean of the distribution for each type. The open
circles show the median star formation rates for each type, to
demonstrate that the trends are not dominated by the few outliers at
high star formation rates seen in Hubble types Sb--Sc inclusive.

Figures~\ref{sfrvt} and \ref{msfrvt} can be compared directly with the
results presented by Kennicutt (\cite{k83}), where we have converted the
latter from an assumed Hubble constant of H$_0 =$ 50 to our assumed value
of 75~km~s$^{-1}$Mpc$^{-1}$. The majority of Kennicutt's sample (59/97)
are of type Sc, and the mean star formation rate for these 59
galaxies is 2.3~M$_{\odot}$~yr$^{-1}$, with a standard error on the mean
of 0.3~M$_{\odot}$~yr$^{-1}$, in excellent agreement with the value shown
in fig.~\ref{msfrvt}.  For the 13 Sbc galaxies in Kennicutt's study, the
mean star formation rate is 3.1~M$_{\odot}$~yr$^{-1}$, standard error
0.4~M$_{\odot}$~yr$^{-1}$, slightly larger than the mean we find for this
type. For the 9 Sb galaxies in Kennicutt's study, the mean star formation
rate is 1.5~M$_{\odot}$~yr$^{-1}$, standard error
0.3~M$_{\odot}$~yr$^{-1}$, in excellent agreement with our value.  All
the other Hubble types have 5 or fewer galaxies in Kennicutt's study,
ruling out a statistically meaningful comparison on a type-by-type basis,
but the overall agreement with our fig.~\ref{sfrvt} is essentially
perfect. For example, every one of Kennicutt's 97 measured star formation
rates lies within the range of rates determined in the present study for
galaxies of that T-type.

Less good agreement is found with the star formation rates in early-type
spirals presented by Caldwell et al. (\cite{cal91}, \cite{cal94}).  These
authors find uniformly low star formation rates in the 5 S0/a galaxies
they study (the highest being 0.004~M$_{\odot}$~yr$^{-1}$, 2 galaxies
having measured rates of 0.001~M$_{\odot}$~yr$^{-1}$ and the remaining 2
upper limits only), while for 6 Sa galaxies the mean rate was found to be
0.17~M$_{\odot}$~yr$^{-1}$, with a standard error of
0.06~M$_{\odot}$~yr$^{-1}$ and a maximum of 0.4~M$_{\odot}$~yr$^{-1}$.
For reasons that are not clear, the present study find examples of S0/a
and Sa galaxies with much higher star formation rates; for example
UGC~859, a type S0/a spiral shown in Fig.~\ref{Figa}, has a prominent
ring of star formation and a star formation rate of
0.5~M$_{\odot}$~yr$^{-1}$.

   \begin{figure}
   \centering
   \includegraphics[angle=-90,width=10cm]{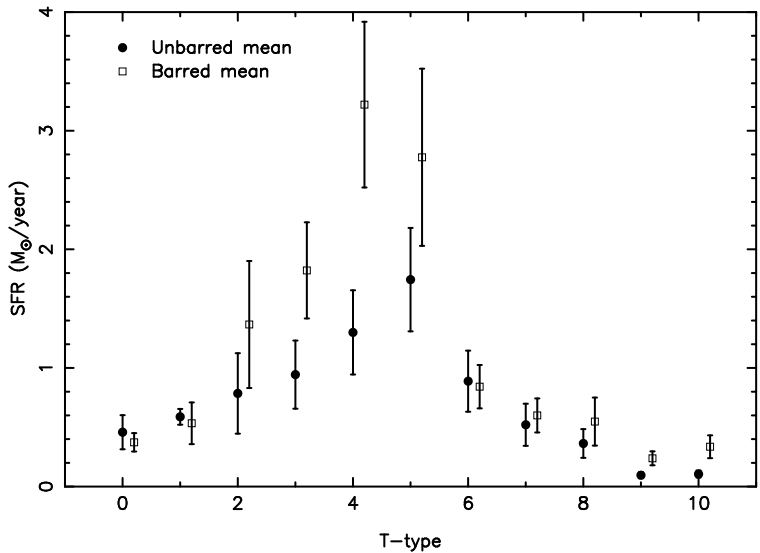}
      \caption{Mean star formation rates as a function of 
      Hubble T type, separated into barred and unbarred types.
              }
         \label{msfrvt_ub}
   \end{figure}

Figure~\ref{msfrvt_ub} again shows mean star formation rates as a
function of Hubble type, but this time separated into barred and
unbarred types.  Again it is clear that bars can enhance star
formation, although the effect is only significant in the intermediate
Hubble types, and possibly in the extreme late-type galaxies.

   \begin{figure}
   \centering
   \includegraphics[angle=-90,width=10cm]{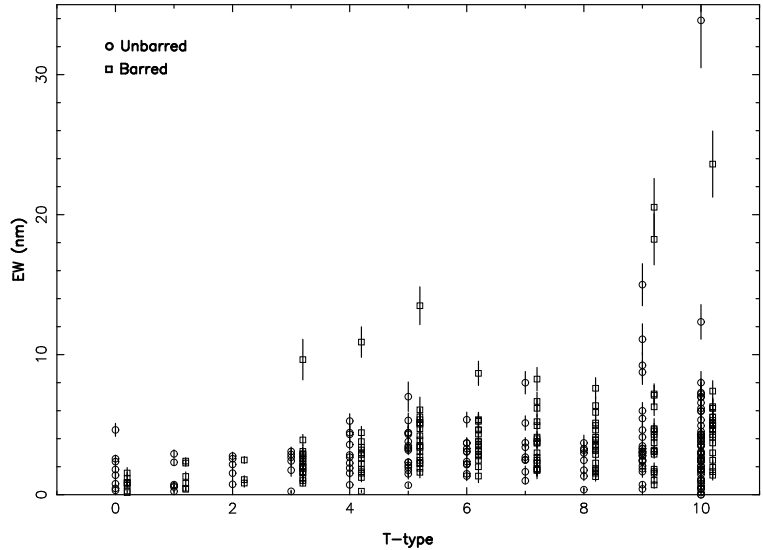}
      \caption{Distribution of H$\alpha +$[N{\sc ii}] EWs as a function of 
      Hubble T type.
              }
         \label{ewvt}
   \end{figure}

   \begin{figure}
   \centering
   \includegraphics[angle=-90,width=10cm]{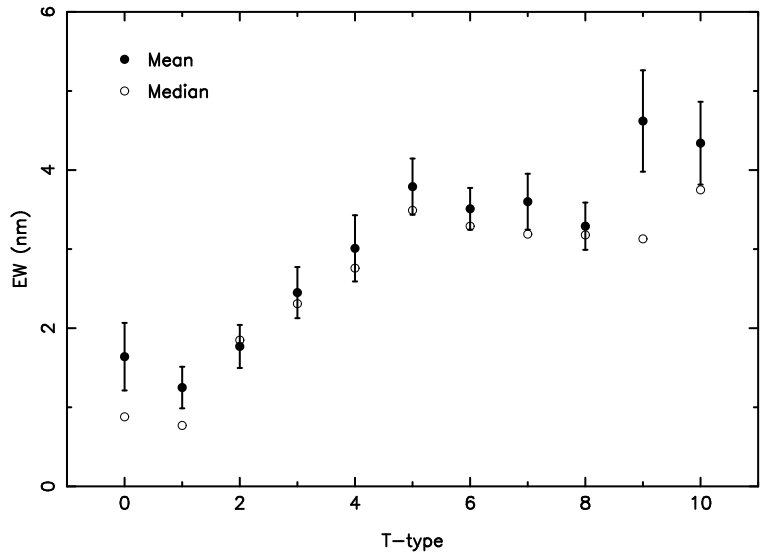}
      \caption{Mean and median H$\alpha +$[N{\sc ii}] EW as a function of 
      Hubble T type.
              }
         \label{mewvt}
   \end{figure}

Equivalent width values provide a measure of star formation
efficiency, being normalised by the luminosity of the older stellar population
of the galaxy. Figures~\ref{ewvt} and \ref{mewvt} show the
distribution of individual galaxy H$\alpha$ $+$ [N{\sc ii}] EWs, and the mean
values of this parameter, as a function of Hubble type.  It is clear
that the highest EWs are seen in the late type spiral galaxies
and the irregular galaxies, with the five highest values belonging to Sm or
Im galaxies.  However, the apparent dependence of mean EW on Hubble
type is surprisingly weak for types later than Sbc.  It should also be
noted that the EW values for the earliest Hubble types may be depressed
by the H$\alpha$ absorption from the stars in the dominant bulges in
these galaxies.

Figure~\ref{ewvt} can be compared with Fig. 3 of Kennicutt (\cite{ken98}),
which gives H$\alpha$ $+$ [N{\sc ii}] EWs for a large sample of nearby
spiral galaxies, with the majority of the data being taken from Kennicutt
\& Kent (\cite{ken83}).  For the later-type spirals (Sc--Sm), the agreement in
the distributions of EW is good, with marginal evidence for higher EWs in
the present sample ($\sim$3.5~nm average over these types, cf. 2.9~nm for
the same types in Kennicutt \& Kent \cite{ken83}), consistent with the
measurement offset noted in the comparison with Kennicutt's data in
section 4.2.  However, for the earliest types (S0/a--Sab), the agreement
is less good, with the present sample containing several early-type
galaxies with H$\alpha$ $+$ [N{\sc ii}] EWs several times larger than
those found in Kennicutt's compilation.  This is most marked for the S0/a
type galaxies, where all 11 of Kennicutt's galaxies lie well below 0.5~nm
EW, compared to a {\em mean} of 1.7~nm for the 16 galaxies in the present
sample.  Even for the Sb type galaxies, the present study finds a mean EW
approximately twice that of Kennicutt (\cite{ken98}).  Apart from this
discrepancy, the overall pattern of EW with increasing T-type is very
similar in the two studies, with Kennicutt (\cite{ken98}) also finding
mean EW to increase from type S0/a to type Sc, and then to remain
approximately constant for the later types.

Loveday et al. (\cite{love99}) present histograms of H$\alpha$ EW for
large numbers of galaxies from the Stromlo-APM redshift survey.  Since
their data are spectroscopic with a resolution of $\sim$5\AA, they have
resolved the H$\alpha$ and [N{\sc ii}] lines and present EWs for H$\alpha$
only.  Over all spiral types, their median H$\alpha$ EW is 1.06~nm, and for
irregulars it is 1.69~nm. These values are significantly lower than those
for both the present study and Kennicutt (\cite{k83}). The median
H$\alpha$ $+$ [N{\sc ii}] EWs for the present study are 2.80~nm for
spirals and 3.75~nm for irregulars, which reduce to 2.2 and 3.5~nm
respectively after application of the standard Kennicutt correction for
[N{\sc ii}] contamination, still approximately double the EW values found
by Loveday et al. (\cite{love99}).  The median H$\alpha$ $+$ [N{\sc ii}]
EWs for the 114 spirals observed by Kennicutt (\cite{k83}), and for the
110 spirals observed by Romanishin (\cite{rom90}), are both 2.4~nm, or
1.8~nm after correction for [N{\sc ii}] emission.  The reasons for the
lower values found by Loveday et al. (\cite{love99}) are not clear at
present, but seem unlikely to be due entirely to differences in galaxy
classifications, since all types later than Sab in the present study have
median EW values higher than the overall mean found by Loveday et al.
(\cite{love99}).  Other factors that may play a part are different surface
brightness selection criteria between the APM survey and the nearby
galaxy catalogues used by the other studies, and possibly aperture
effects, although all of these studies present EWs which should be close
to total values.

   \begin{figure}
   \centering
   \includegraphics[angle=-90,width=10cm]{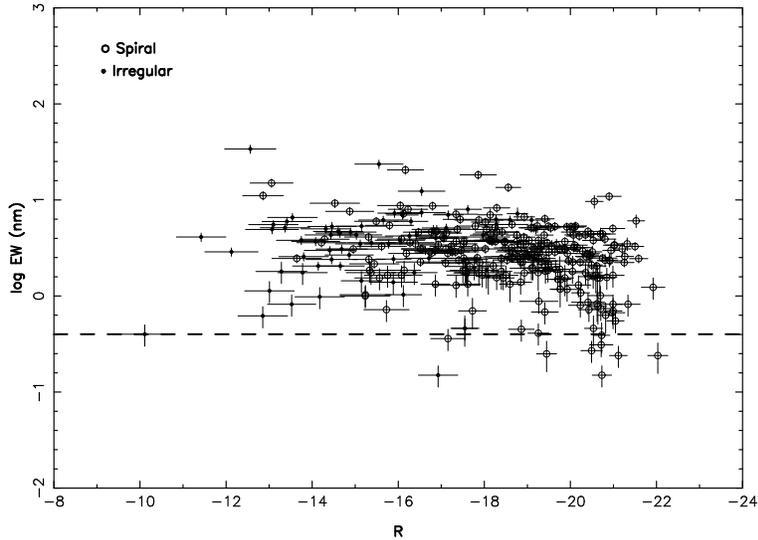}
      \caption{Log H$\alpha +$[N{\sc ii}] EW as a function of 
      total $R$-band absolute magnitude. The dashed line shows the
      EW completeness limit of 0.4~nm.
              }
         \label{ewvr}
   \end{figure}

Figure~\ref{ewvr} shows the distribution of individual galaxy H$\alpha$
$+$ [N{\sc ii}] EWs, where the decimal logarithm is plotted to better
show separate points, as a function of total $R$ magnitude.  This again
shows a surprising lack of dependence of EW on the properties of the
underlying galaxy.  The diagram mainly shows a large scatter in EW at any
given luminosity.  Any trend in EW with luminosity is due to a population
of very low EW galaxies seen in the brightest spiral galaxies only (and
these are the galaxies most likely to have EWs reduced by H$\alpha$
absorption), but apart from this the distribution is completely flat as a
function of luminosity.  This lack of correlation has been found by
previous studies, but over a smaller range in absolute magnitude.  For
example, Kennicutt \& Kent (\cite{ken83}) find no luminosity dependence
in the H$\alpha$ $+$ [N{\sc ii}] EW of Sc spirals over a 5~mag range in
M$_B$.  Tresse and Maddox (\cite{tres98}) find no correlation in an
identical plot for 138 galaxies out to $z=$0.3 from the Canada-France
Redshift Survey (CFRS), over a 6~mag range in M$_B$.  However, Tresse et al.
(\cite{tres02}) do find a trend in the closely related parameter,
Log(L(H$\alpha$)/L$_B$), when plotted against M$_B$, for CFRS galaxies
between redshift 0 and 1.1.  This trend is in the sense that the most
luminous galaxies have higher H$\alpha$-to-continuum flux ratios than the
lower luminosity galaxies, although the strength of the correlation
depends on the reddening correction applied, and the interpretation of
this result is complicated by the fact that the most luminous galaxies
also tend to be the most distant.

\begin{figure}
   \centering
   \includegraphics[angle=-90,width=10cm]{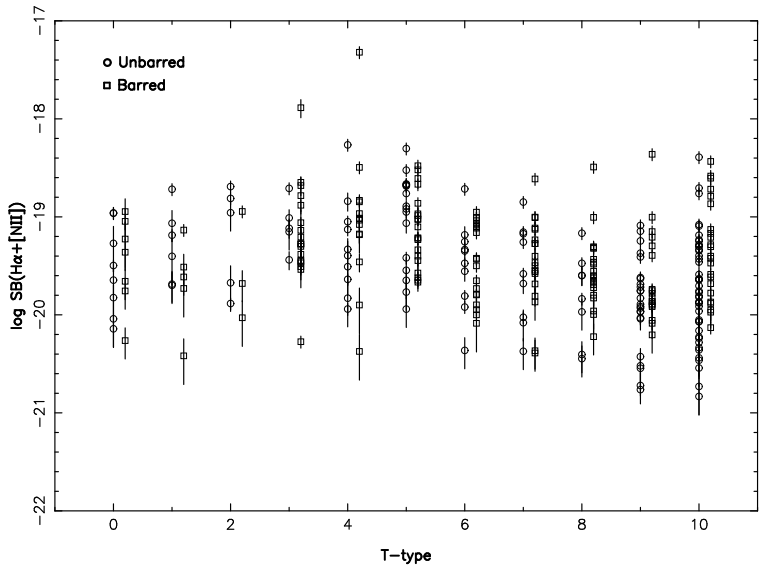}
      \caption{H$\alpha +$[N{\sc ii}] surface brightness as a function
      of galaxy type.
              }
         \label{sbvt}
   \end{figure}

\begin{figure}
   \centering
   \includegraphics[angle=-90,width=10cm]{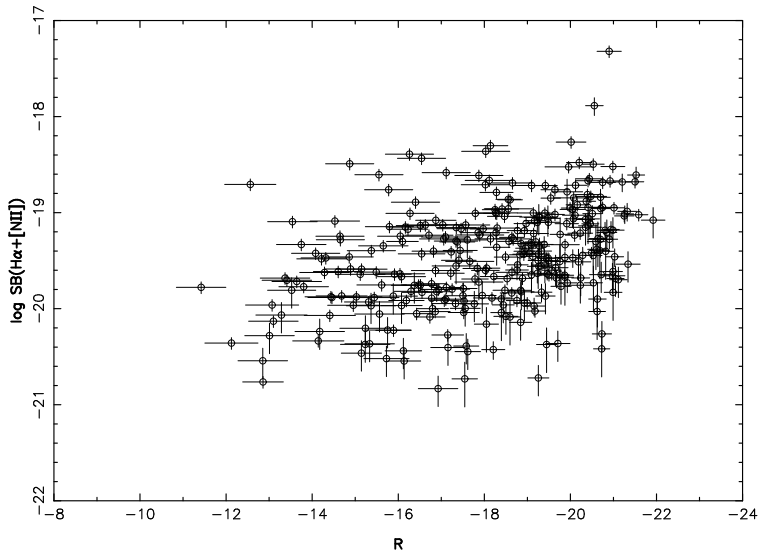}
      \caption{H$\alpha +$[N{\sc ii}] surface brightness as a function of 
      total $R$-band absolute magnitude. 
              }
         \label{sbvr}
   \end{figure}

An alternative way to prevent size effects from masking trends in star
formation rate is to normalise by galaxy surface area, giving an
H$\alpha +$[N{\sc ii}] surface brightness, or equivalently a star formation rate
per unit surface area.  Figures~\ref{sbvt} and \ref{sbvr} present such data for the
present galaxy sample, where the surface brightness used is a hybrid
value, plotted in arbitrary units.  This hybrid surface brightness is
based on the total detected H$\alpha +$[N{\sc ii}] flux, divided by the square of
the $R$-band Petrosian radius of the galaxy.  This latter radius is
defined to be independent of galaxy distance and photometric errors,
and is basically the radius at which the local surface brightness is a
given factor lower than the mean surface brightness within that radius.
We adopt the definition of Petrosian radius used by Shimasaku et al.
(\cite{shim}), who adopt a value of 0.2 for the $\eta$ parameter used
to define the Petrosian radius (see Petrosian \cite{petr} for the
original explanation of this parameter).
This radius is much better defined for the smooth $R$-band profiles than
for the H$\alpha +$[N{\sc ii}] images, hence the use of the hybrid surface
brightnesses.  Figure~\ref{sbvt} shows the distributions of this surface
brightness parameter as a function of galaxy Hubble type, showing that
the distributions are essentially independent of galaxy type, although
the range in surface brightness is broad at all Hubble types. In
particular, it is noteworthy that the irregular galaxies (T$=$10) have
a very similar distribution to the spiral galaxies, even though their overall
star formation rates are much lower.

This conclusion is emphasised by Fig.~\ref{sbvr}, which shows
H$\alpha +$[N{\sc ii}] surface brightness as a function of total $R$-band galaxy
magnitude.  Again the distribution is essentially flat, with the only
obvious outliers being two bright spiral galaxies with high H$\alpha +$[N{\sc ii}] surface
brightnesses.

\section{Conclusions and future work}

We have presented new narrow-band H$\alpha +$[N{\sc ii}] line and $R$-band imaging
data photometry, of a sample of 334 galaxies carefully selected to
span all star-forming types in the local Universe.  The galaxies were
taken from the UGC, and hence are subject to the selection criteria of
that catalogue, and in addition we have required a measured recession
velocity less than 3000~km~s$^{-1}$, D$_{25}$ diameters between 1\farcm7 and
6\farcm0, and Hubble types between S0/a and Im.

The new data have
been used to quantify the distributions of star formation activity as
a function of galaxy type and luminosity.  We find a strong
correlation between galaxy Hubble type and total star formation rate,
in the expected sense that the most strongly star-forming galaxies are
those of Hubble types Sbc and Sc, although even within these types
there is a wide range in star formation rates.  H$\alpha +$[N{\sc ii}] EW
values, averaged over entire galaxies, give a measure of star
formation activity normalised by the stellar luminosity of the galaxy;
these also show a trend with galaxy type, with early-type spiral galaxies
having low EW values, and irregular galaxies the highest EWs, at least in the
mean.  However, the distribution of EW as a function of galaxy
luminosity is surprisingly flat, both in terms of median EW and the
scatter in this parameter.

In this paper we have only skimmed the surface of the science to be
undertaken in this project.  A key area for further study concerns the
corrections which need to be applied to H$\alpha$ fluxes derived from
narrow-band imaging to account for [N{\sc ii}] line contamination and
extinction internal to the galaxies under study, since these dominate the
uncertainties in the derived star formation rates (see also the detailed
examination of this question by Charlot et al. \cite{char02}, and work by
Buat et al. \cite{buat02} and Rosa-Gonz\'alez et al. \cite{gonz02}).  The
spatial dependence of the [N{\sc ii}] correction will be investigated,
since the [N{\sc ii}] line strength depends on metallicity and requires a
harder radiation field than H$\alpha$, suggesting that the relative
strengths of the two lines may be very different in the disk and
circumnuclear environments. We will also look at the type dependence of
the H$\alpha$ extinction correction in detail, building on the work of
Kennicutt \& Kent (\cite{ken83}), since our sample contains many
late-type and low-luminosity galaxies which were not well represented in
their study.

Sect.  6.1 of the present paper illustrates that the H$\alpha$ images
contain substantial information on the distribution of line-emitting
gas {\em within} galaxies, as well as on the total quantity.  We will
use this information to compare the distributions of star formation
regions with the old stellar populations (as represented by $R$-band
light) in galaxy disks and in irregular galaxies, with a key question
being whether new stars are produced with a distribution similar to
that of the old stars, or whether there is any evidence for disks and
irregular galaxies being constructed `inside-out' or `outside-in'.
The distributions of both new and old stellar populations will be
quantified for this analysis by concentration indices.

Whilst the present H$\alpha$ sample has been constructed to sample the
field population, i.e. avoiding the centres of rich galaxy clusters,
it still samples a range of galaxy environments, including completely
isolated galaxies, galaxies in known pairs or groups, and some
interacting systems.  We will thus study the effects of environment
(local galaxy density) and interactions with very close neighbours on
star formation rate, surface density and distribution in spiral and
irregular galaxies, for comparison with results obtained by Kennicutt et
al. (\cite{ken87}), Loveday et
al. (\cite{love99}), Carter et al. (\cite{cart01}) and Koopman et
al. (\cite{koopman}).

A final aim for this study is to determine the total star formation
rate in the local Universe, by correcting the observed galaxy sample for
the known selection effects to represent a volume-limited sample.  This
will yield a galaxy luminosity function, from which we can estimate the
total star formation rate per unit volume of the local Universe, and the
contribution to this total from different galaxy luminosities and Hubble
types.  We believe this to be a complementary approach to previous
studies in this area, e.g. those by Gallego et al. (\cite{gallego95}) who
looked at 176 star-forming galaxies out to $z\sim$0.045, and Tresse \&
Maddox (\cite{tres98}), 138 galaxies to $z\sim$0.3. Our study is much more local than
these, raising concerns about how representative a volume it may sample,
but intensively samples that volume given the large sample size. The
proximity of the galaxies also enables us to include the faintest
star-forming dwarfs, and gives us excellent spatial resolution within the
galaxies, thus disentangling nuclear and disc contributions to the total
star formation rate.

\begin{acknowledgements}

We thank the Isaac Newton Group support scientists, telescope operators
and technical staff for volunteering their willing assistance throughout
the many observing nights of this programme. We are grateful for the full
allocation of time to this project by the Panel for the Allocation of
Telescope Time of the UK Particle Physics and Astronomy Research Council.
This research has made use of the NASA/IPAC Extragalactic Database (NED)
which is operated by the Jet Propulsion Laboratory, California Institute
of Technology, under contract with the National Aeronautics and Space
Administration.  The referee is thanked for many helpful suggestions
which improved the presentation of this paper.

\end{acknowledgements}

\end{document}